\documentclass[11pt]{article}
\pdfoutput=1

\usepackage{pdflscape} 
\usepackage{times}
\usepackage{graphicx}
\usepackage{amsmath, amssymb}
\usepackage[round, sort&compress, authoryear]{natbib}
\usepackage{authblk}
\usepackage{color}
\usepackage{appendix}

\newcommand{\swallow}[1]{ }
\newtheorem{theorem}{Theorem}
\newtheorem{lemma}{Lemma}
\newcommand{\beas}{\begin{eqnarray*}}
\newcommand{\enas}{\end{eqnarray*}}
\newcommand{\beginsupplement}{%
        \setcounter{table}{0}
        \renewcommand{\thetable}{S\arabic{table}}%
        \setcounter{figure}{0}
        \renewcommand{\thefigure}{S\arabic{figure}}%
     }

\makeatletter
\DeclareRobustCommand*\cal{\@fontswitch\relax\mathcal}
\makeatother

\usepackage{subfigure}
\usepackage{epstopdf}

\usepackage{url}

\begin{document}

\title{Inference of Markovian Properties of Molecular Sequences from NGS Data and Applications to Comparative Genomics}
\author[1]{Jie Ren}
\author[2]{Kai Song}
\author[2]{Minghua Deng}
\author[3]{Gesine Reinert}
\author[4,5]{Charles H. Cannon}
\author[1,6, ]{Fengzhu Sun  \footnote{to whom correspondence should be addressed; fsun@usc.edu}}

\affil[1]{\footnotesize Molecular and Computational Biology
Program, University of Southern California, Los Angeles,
California, USA;}
\affil[2]{School of Mathematical Sciences, Peking University, Beijing, China;}
\affil[3]{Department of Statistics, University of Oxford, 1 South Parks Road, Oxford OX1 3TG, UK;}
\affil[4]{Department of Biological Sciences, Texas Tech University, TX 79409-3131, USA;}
\affil[5]{Xishuangbanna Tropical Botanic Garden, Chinese Academy of Sciences, Yunnan, China;}
\affil[6]{Centre for Computational Systems Biology, School of
Mathematical Sciences, Fudan University, Shanghai, China.}

\renewcommand\Authands{ and }

\normalsize

\date{}

\maketitle

\begin{abstract}
Next Generation Sequencing (NGS) technologies generate  large
amounts of short read data for many different organisms. The fact
that NGS reads are generally short makes it challenging to assemble
the reads and reconstruct the original genome sequence. For
clustering genomes using such NGS data, word-count based
alignment-free sequence comparison is a promising approach, but for
this approach, the underlying expected word counts are essential.

A plausible model for this underlying distribution of word counts is
given through modelling the DNA sequence as a Markov chain (MC). For
single long sequences, efficient statistics are available to
estimate the order of MCs
 and the transition probability matrix for the sequences.
As NGS data do not provide a single long sequence, inference methods
on Markovian properties of sequences based on single long sequences
cannot be directly used for NGS short read data.

Here we derive a normal approximation for such word counts. We also show that the
traditional Chi-square statistic has an approximate gamma
distribution, using the Lander-Waterman model for
physical mapping.
We propose several methods to estimate the order of the MC
based on NGS reads and evaluate them using simulations.
We illustrate the applications of our results by clustering genomic
sequences of several vertebrate and tree species based on NGS reads
using alignment-free sequence \textcolor{black}{dissimilarity measures}. We find that
the estimated order of the MC has a considerable effect on the
clustering results, and that the clustering results that use a MC of
the estimated order give a plausible clustering of the species.

Our implementation of the statistics developed here is available as
R package ``NGS.MC'' at
\url{http://www-rcf.usc.edu/~fsun/Programs/NGS-MC/NGS-MC.html}.

\end{abstract}


\section{Introduction}
\label{section:Introduction}

NGS technologies generate  large amounts of overlapping short read data for many different organisms;
 for example a read is a subsequence of less than 400 bps for
Illumina and 700 bps for 454 sequencing technologies, and can
sometimes be much shorter. The fact that  NGS reads are generally
short makes it challenging to reconstruct the original genome
sequence.

\textcolor{black}{Recently several word-count based alignment-free
sequence comparison methods have been applied to infer the
relationship among different species
\citep{yi2013co,song2013alignment} and metagenomic samples
\citep{jiang2012comparison,wang2014comparison,behnam2014amordad,hurwitz2014modeling}
based on NGS reads without assembly. Our alignment-free sequence
dissimilarity measures, $d_2^*$ and $d_2^S$
\citep{song2013alignment,song2014new}, and their variants
\citep{liu2011new,behnam2013geometric,ren2013multiple} have shown
promise. }These methods require the knowledge about the approximate
distribution of word counts in the underlying sequences. While a
model which assumes that all letters in the sequence are equally
likely is relatively straightforward to analyse, see
\cite{reinert2009alignment}, a Markov model for the underlying
sequences is more realistic.

Markov chains (MC) have been widely used to model molecular
sequences \citep{almagor1983markov} with many applications including
the study of dependencies between the bases
\citep{blaisdell1985markov}, the enrichment and depletion of certain
word patterns \citep{pevzner1989linguistics}, prediction of
occurrences of long word patterns from short patterns
\citep{hong1990prediction,arnold1988mono}, and the detection of
signals in introns \citep{avery1987analysis}. \cite{narlikar2013one}
studied the effect of the order of MCs on several biological
problems including phylogenetic analysis, assignment of sequence
fragments to different genomes in metagnomic studies, motif
discovery, and functional classification of promoters. These
applications  showed the importance of accurate specification of the
order of MCs. Reliable estimators for the order of the MC and
 the transition probability
matrix based on the sequence data are crucial.

Based on relatively long molecular sequences, for a general finite
state MC sequence of letters from a finite alphabet ${\mathcal A} =
\{1, 2, \cdots, L\}$ of size $L$, \cite{hoel1954test} showed, under
the hypothesis that the long sequence follows a $(k-2)$-th order MC,
that twice the log-likelihood ratio of the probability of \textcolor{black}{the sequence}
under a $(k-1)$-th order MC versus that under the $(k-2)$-th order
MC model follows approximately a $\chi^2$-distribution with ${df}_k
=
 (L-1)^2 L^{k-2}$ degrees of freedom under general conditions. He also approximated the
log-likelihood ratio by the Pearson-type statistic
 \begin{equation}
  S_k = \sum_{\mathbf{w} \in \mathcal{A}^k } \frac{ \left(  N_{\mathbf{w}} - E_{\mathbf{w}} \right)^2 }{  E_{\mathbf{w}} },
  \label{S2_k equation}
  \end{equation}
which is also approximately $\chi_{}^2$-distributed with the same
degrees of freedom. Here, $\mathbf{w} = {w}_1 {w}_2\cdots {w}_k$
denotes a $k$-word formed of  letters $w_i \in {\mathcal{A}}$,
$^{-}\mathbf{w} = w_2\cdots {w}_k$, $\mathbf{w}^{-} = {w}_1
{w}_2\cdots {w}_{k-1}$, and $^-\mathbf{w}^- = {w}_2 \cdots
{w}_{k-1}$; $ N_{\mathbf{w}}$ denotes the  count of the word
$\mathbf{w}$ in the sequence, and $E_{\mathbf{w}} =
\frac{N_{^{-}\mathbf{w} }
N_{\mathbf{w}^{-}}}{N_{^{-}\mathbf{w}^{-}}}$ is the estimated
expected count of ${\mathbf{w}}$ if the sequence is generated by a
MC of order $k-2$. Here $k \ge 3$; see also \cite{avery1999fitting}
for  a detailed study,
\cite{billingsley1961statisticalb,billingsley1961statisticala} for
an an excellent exposition of statistical issues related to MCs, as
well as
\cite{waterman1995introduction,reinert2000probabilistic,reinert2005statistics,ewens2005statistical}
for applications to  sequence analysis.

The Chi-square statistic \eqref{S2_k equation} and the log-likelihood ratio statistics can be used to  test the order of a MC, using all  $k$-words $\mathbf{w} \in \mathcal{A}^k$. When a particular order of MC is rejected, we
can  identify particular word patterns that are exceptional, through  the approximate distribution of
$N_\mathbf{w}$. The  approximate distributions of $N_\mathbf{w}$ in
long sequences is well understood, see for example
\cite{waterman1995introduction,reinert2005statistics,reinert2000probabilistic}.
In particular, suppose that the sequence follows a stationary
($k-2$)-th order MC and let
\begin{align*}
 \hat{\sigma}^2_\mathbf{w}
&= E_\mathbf{w} \left ( 1 - \frac{N_{^-\mathbf{w}}}{N_{^-\mathbf{w}^-}} \right ) \left ( 1 - \frac{N_{\mathbf{w}^-}}{N_{^-\mathbf{w}^-}} \right ).
 \label{eq:variance}
 \end{align*}
For
\begin{equation}
 Z_\mathbf{w} =  \frac{N_\mathbf{w} - E_\mathbf{w}}{\hat{\sigma}_\mathbf{w}} ,
 \label{individual_word}
\end{equation}
 Theorem 6.4.2 in \cite{reinert2005statistics} gives that, as
 sequence length goes to infinity,  for all real values $x$,
$
 \mathbb{P} ( Z_\mathbf{w} \le x) \rightarrow \Phi(x),
$
where $\Phi$ denotes the cumulative distribution function of a standard normal variable.
We also say that $ Z_\mathbf{w} $ converges to the standard normal distribution $N(0,1)$ {\it{in distribution}}.
This asymptotic result can then be used to find exceptional words in long sequences.

Given an NGS short read sample, it is tempting to use the test
statistic $S_k$ defined in (\ref{S2_k equation}) to test
the order of a MC by simply counting the number of the occurrences
of words in short read data. However, as the short reads from NGS
data are sampled randomly from the genome, some parts of the genome
are possibly not sampled and some parts are possibly sampled
extensively. The sampling process introduces additional randomness
to the statistic, and makes $S_k$ deviate from its traditional
$\chi^2$ -distribution. Similarly, the approximate distribution of
$Z_\mathbf{w}$  given in \eqref{individual_word} will  be different
from the standard normal distribution.

 In this paper, we study these
approximate distributions,  both theoretically and by simulations.
First we extend the statistics $S_k$ and $Z_\mathbf{w}$ for a MC
sequence to $S_k^R$ and $Z_\mathbf{w}^R$ for the NGS read data. Our
underlying model for the distribution of reads along the genome is
the potentially inhomogeneous Lander-Waterman model for physical
mapping \citep{lander1988genomic}. We discover that for a set of
short reads sampled from a $(k-2)$-th order MC sequence, the
statistic $S_k^R$ follows approximately a gamma distribution with
shape parameter ${df}_k/2$ and scale parameter $2d$, where $d$ is a
factor related to the distribution of the reads along the genome. We
also show that, with the same factor $d$, the distribution of the
single word statistic $Z_\mathbf{w}^R/\sqrt{d}$ tends to the
standard normal distribution. Based on the theoretical results, we
introduce an estimator for the order of the MC using NGS data.  For
practical purposes, we also give  an estimator for the factor $d$
when the underlying reads sampling distribution is unknown. 
To the best of our knowledge,
this is the first study of the Markovian properties of molecular
sequences based on NGS read data.

To illustrate our theoretical results and our estimators, we first
carry out a simulation study based on transition probability
matrices which are estimated from cis-regulatory module (CRM) DNA
sequences, and insert repeats. We simulate different read lengths,
numbers of reads, inhomogeneous sampling, as well as sequencing
errors, and we include a regime where the sampling rate depends on
the GC content. \textcolor{black}{If the GC bias is not very strong or the
sequencing depth is not very low, then the simulation results agree with our theoretical predictions despite the theoretical assumptions being slightly violated. }

Next we apply our methods to cluster 28 vertebrate species using our
alignment-free dissimilarity measures $d_2^*$ and $d_2^S$ under
different MC models which are estimated from NGS read samples. The
estimated orders based on NGS data without assembly are found to be
consistent with those inferred directly from the long genome
sequences. The clustering performs best when using MCs around
the estimated order. Applying the same analysis to 13 tropical tree
species whose genomes are unknown, based on their NGS read samples,
the most plausible clustering is achieved when using a MC model of
order close to the one estimated from the NGS reads.

The paper is organized as follows. The ``Methods" section contains
the probabilistic models of generating the MC sequence and sampling
the short reads, as well as  the theorem for the approximate
distributions of $S_k^R$ and $Z_\mathbf{w}^R$ for NGS data. This
theorem is used to derive our estimators for the order of the MC and
for the factor $d$. In the ``Results" section, we first provide
extensive simulation studies including the comparison of the
theoretical approximate distributions and the simulated results for
$S_k^R$ and $Z_\mathbf{w}^R$, the effect of inhomogeneous sampling
and sequencing errors, the efficiency of the estimator of the factor
$d$, and the evaluations of the methods for estimating the MC order.
Second, we estimate the orders of the MCs for 28 vertebrate species
based on the simulated whole genome NGS samples. We then use our
dissimilarity measures $d_2^*$ and $d_2^S$ to
cluster the NGS samples of the 28 species under different MC orders
to see the effect on the performance of the clustering. The
applications show that our new methods are effective for the
inference of relationships among sequences based on NGS reads.
Finally, we use our methods to study the relationships among 13 tree
species whose complete genomic sequences as well as their
phylogenetic relationships are unknown. Our clustering results are
consistent with the physical characteristics of the tree species.
The paper concludes with some discussion of the study.


\section{Methods}
\label{methods}

\subsection{Probabilistic modeling of a MC sequence and random sampling of the reads using NGS}
\label{model}

In NGS,  a large number of reads are randomly sampled
from the genome. Hence two random processes are involved in the
generation of the short read data: the generation of the
underlying genome sequence and the random sampling of
the reads.

We use an $r$-th order homogeneous ergodic MC to model the
underlying genome sequence with each letter taking values in a
finite alphabet set ${\mathcal A}$ of size $L$. Since our study is
based on genomic sequences, $L=4$. As in
\cite{lander1988genomic,zhang2008modeling,zhai2012normal,daley2013predicting,simpson2014exploring},
we assume that the genome is continuous and that the
 distribution of reads along the genome follows a potentially
\textit{inhomogeneous} Poisson process with rate $c(x)$ at position
$x$. If $c(x) = c$ for all $x$, we refer to the sampling of the reads as
\textit{homogeneous}. We assume that all sampled reads have the same
length of $\beta$ bps. A total of $M$ reads are independently
sampled from the genome of length $G$ bps.

We extend the statistics $S_k$ and $Z_\mathbf{w}$ in 
(\ref{S2_k equation}) and (\ref{individual_word}) to NGS short read
data accordingly. Let $N_\mathbf{w}^R$ be the number of occurrences
of the $k$-word $\mathbf{w}$ in the short read data, where the superscript $R$ refers to the ``read'' data, and define

 \begin{equation}
  S_k^R = \sum_{\mathbf{w} \in \mathcal{A}^k } \frac{ \left(  N_{\mathbf{w}}^R - E_{\mathbf{w}}^R \right)^2 }{  E_{\mathbf{w}}^R },
  \label{S2_kR equation}
  \end{equation}
\begin{equation}
 Z_\mathbf{w}^R =  \frac{N_\mathbf{w}^R - E_\mathbf{w}^R}{\hat{\sigma}_\mathbf{w}^R},
 \label{individual_wordR}
 \end{equation}
where $$E_{\mathbf{w}}^R = \frac{N_{^{-}\mathbf{w} }^R N_{\mathbf{w}^{-}}^R}{N_{^{-}\mathbf{w}^{-}}^R} \ \text{and}\
(\hat{\sigma}^R_\mathbf{w})^2 = E_\mathbf{w}^R \left ( 1 - \frac{N_{^-\mathbf{w}}^R}{N_{^-\mathbf{w}^-}^R} \right ) \left ( 1 - \frac{N_{\mathbf{w}^-}^R}{N_{^-\mathbf{w}^-}^R} \right ).$$

We have the following theorem on the approximate
distributions of $S_k^R$ and $Z_\mathbf{w}^R$; the proof is given in
the Supplementary Materials. Note that for each read we discard the
last $k-1$ positions as they would lead to words of length less than
$k$; the error made with this approximation is asymptotically
negligible when $k$ is small relative to $\beta$.

\begin{theorem}
Assume that the underlying genome follows a ($k-2$)-th order MC which assigns non-zero probability to every $k$-word $\mathbf{w}$.  Let
$S_k^R$ and $Z_{\mathbf{w}}^R$ be defined as in 
(\ref{S2_kR equation}) and (\ref{individual_wordR}), respectively.
Suppose that the genome of length $G$ can be divided into (not necessarily contiguous) regions with constant coverage {\color{black}{$r_i$}} for the $i$-th region, so that  every base is covered exactly $r_i$ times, based on the first
$\beta-k+1$ positions of the reads. Let $G_i$
be the length of the $i$-th region
that changes with $G$ in a way such that $\lim_{G \rightarrow \infty} G_i/G= f_i > 0$ for the $i$-th region, $i = 1, 2, \cdots$.
\swallow{{\color{black}{and that $\lim_{G \rightarrow \infty}\frac{Mk}{\sqrt{G}} \rightarrow 0$.}}}
 Let
\begin{equation} \label{effcov} d = \frac{\sum_{i} r_i^2 f_i}{\sum_{i} r_i f_i}.
\end{equation}   Then, as $G \rightarrow \infty$,
    \begin{enumerate}
    \item[a)] For each $k$-word $\mathbf{w}$, in distribution, $Z_{\mathbf{w}}^R/\sqrt{d} \rightarrow N(0,1)$.
    \item[b)] The statistic $S_k^R/d$ has an approximate $\chi^2$-distribution
    with ${df}_k = (L-1)^2 L^{k-2}$ degrees of freedom; equivalently, the
    statistic $S_k^R$ has an approximate gamma distribution with shape
    parameter ${df}_k/2$ and scale parameter $2d$.
    \end{enumerate}
\label{main-theorem}
\end{theorem}

If the $M$ reads are sampled homogeneously along the genome with
coverage $c$ based on the first $\beta - k+1$ positions of the reads
along the genome, i.e. $c = \frac{M(\beta-k+1)}{G-k+1}$, the
Lander-Waterman formula \citep{lander1988genomic} shows that the
fraction of genome covered $r_i = i$ times is $f_i = \exp(-c)
c^i/i!$. Under this assumption,  we obtain
$$ d = \frac{\sum_{i} i^2 f_i}{\sum_{i} i f_i} = \frac{c^2 + c}{c} = c + 1.$$
The results in  Theorem \ref{main-theorem}  continue to
hold when taking $d = c+1$.

In the Lander-Waterman model for physical mapping
\citep{lander1988genomic}, the factor $c = \frac{M\beta }{G}$ is the
coverage of the genome.
 Hence  we  refer to $d$ from \eqref{effcov} as  the  {\it effective coverage}
 of the reads along the genome based on the first $\beta-k+1$ positions of each read.

\subsection{Estimating the order of the MC based on NGS reads} \label{five methods}

Based on Theorem \ref{main-theorem}, we can estimate the order
{\color{black}{$r$}} of a MC sequence using NGS reads. First, we
test the null hypothesis that the sequence follows an  independent
identically distributed (i.i.d; MC order = 0) model.  For a test at
significance
 level $\alpha$, if $S_2^R/d$ is higher than the
$1-\alpha$ quantile of the $\chi^2$-distribution with $df = (L-1)^2$
degrees of freedom, the i.i.d hypothesis is rejected. If this null
hypothesis is rejected, then here we  propose an estimator for the
order of a MC; it is an analog
 of a corresponding established estimator of MC orders
based on long sequences that has  been shown to be effective. In the
Supplementary Materials we present four related estimators as well
as estimators based on the AIC and BIC information criteria; the one
presented here has the best performance in simulation studies.

We  assume that the word length $k \ge 2$ and that the assumptions
of  Theorem \ref{main-theorem} are satisfied. Then, for $k \geq
r+2$, $S_k^R/d$ has approximately a $\chi^2$-distribution with $
(L-1)^2 L^{k-2}$ degrees of freedom.  If $ k < r+2$,   then
$S_k^R/d$ will typically be larger than expected from this
$\chi^2$-distribution. For $k \geq r+2$,  the law of large numbers
gives that ${\frac{S^R_{k+1}}{L S^R_k}} \rightarrow 1$ for $G
\rightarrow \infty$; if $k< r+2$ then the ratio will be much  larger
than 1 in the limit.   Therefore we can estimate $r$ as follows:
\begin{equation}
\hat{r}_{S_k} = \mathrm{argmin}_k \left \{\frac{S^R_{k+1}}{ S_{k}^R}
\right \} - 1. \label{eq:r-Sk-estimate}
\end{equation}
In general, we want the value of $\mathrm{min}_k\left
\{\frac{S^R_{k+1}}{ S_{k}^R} \right \}$ to be very small, e.g, less
than 0.01.

Using the law of large numbers it can be shown that under our
assumptions this estimator is consistent, in the sense that
$\hat{r}_{S_k}$ tends to $r$ in probability as $G$ tends to
infinity.

\subsection{Estimating the effective coverage $d$}
\label{estimate d}

Often the effective coverage $d$ is not known and we would like to estimate the effective coverage $d$ using NGS short read
data. From Theorem \ref{main-theorem}, we can see that, under the
general conditions stated in the theorem, $(Z^R_{\mathbf{w}})^2/d$
follows a $ \chi^2$-distribution with one degree of freedom. Since
the median of the $\chi^2$-distribution with one degree of freedom
is about 0.456, we can use the scaled median as a robust estimator
for  $d$;
\begin{equation}
\hat{d} = \mathrm{median}\{ (Z^R_{\mathbf{w}})^2, ~~ \mathbf{w} \in {\mathcal A}^k\}/0.456.
\label{eq:d-estimator}
\end{equation}

When we assume that the underlying long sequence
follows a MC of order at most $m$,  we use $(m+2)$-words to estimate
$d$ using (\ref{eq:d-estimator}).

Note that  for an i.i.d. model sequence, the set of 2-words would
not yield meaningful results as there are only 16 different 2-words
and the median based on  16 numbers is generally not reliable. As an
underlying genome sequence following an $r$-th order MC can also be
seen as an $(r+1)$-th, $(r+2)$-th, \dots , and higher order MC
sequence, we can use $k$-words with relatively large $k$ ($\geq
r+2$) to estimate the factor $d$, if the maximum order of a MC is
unknown beforehand.

\subsection{Simulation study}

For the simulation study, we first generate MCs of different orders.
For realistic parameter values,  the transition probability matrices
of the MCs are based on real cis-regulatory module (CRM) DNA
sequences in mouse forebrain from \cite{blow2010chip}.
\textcolor{black}{We use CRM sequences here because CRM sequences are often used to study the
effectiveness of alignment-free sequence dissimilarity
measures \citep{goke2012estimation,song2014new,ren2013multiple}.} To
take into consideration that in real genomic sequences, many repeat
regions are present, we insert repeats into the generated MCs. We
simulate NGS data by sampling a varying number of reads of different
lengths from the MC, varying genome length as well as coverage.

We include homogeneous and inhomogeneous sampling of the reads as
well as sequencing errors. We also let the sampling rate of the
reads depend on the GC content of the fragments  based on data from
the current sequencing technologies
\citep{benjamini2012summarizing}. 
We set the sequencing
error rate at 10\%, which is relatively high compared to the true
sequencing error rate in real sequencing in order to clearly distinguish  among the estimators with regards to their robustness to sequencing
errors. When a sequencing error occurs at a position, the nucleotide
base is changed to one of the other three nucleotides with equal probability.

Once the NGS reads are generated,
we calculate the statistics $S_k^R$ and $Z_\mathbf{w}^R$ for each
word $\mathbf{w}$, the order estimator $\hat{r}_{S_k}$ and the
estimator for effective coverage $d$ based on 
(\ref{eq:d-estimator}); each procedure is repeated 1000 times.
In each repeat experiment, we let the order estimator choose the
model from 1st, 2nd, $\cdots$, 5th order MCs; we estimate the
effective coverage $d$ by \eqref{eq:d-estimator}, using 3-tuples for
a first order MC,  and 4-tuples for a second order MC. The details
are given in the Supplementary Materials.

\subsection{Applications to the study of relationships among
organisms} \label{subsubsection:applications}

We test our methods on real and simulated NGS data from 28
vertebrate species whose complete genomic sequences are available
and that are comprehensively studied in
\citep{miller200728,karolchik2008ucsc}. We download the genomes of
the 28 vertebrate species from UCSC Genome Browser, and then use
MetaSim \citep{richter2008metasim} to simulate reads from each of
the 28 vertebrate species. In simulations the accuracy of the order estimation increases with read coverage. To reflect a worst-case scenario, we set the read coverage to be 1 as a lower
bound for the performance although the current sequencing technology
can generate data with very high read coverage. 
\textcolor{black}{We set MetaSim to generate reads of
length 62bp under the error rate which is estimated by Illumina in our simulations.}

To estimate the order of MC based on the NGS sample for each of the
28 species, we  apply the order estimator $\hat{r}_{S_k}$ in
 (\ref{eq:r-Sk-estimate}); there is no sharp ratio
transition found over $k=2, \cdots, 14$. Given that real genomes consist of multiple types of regions (coding,
non-coding and regulatory regions) and each type may fit to
different MC models, the result indicates that no suitable MC model
can adequately fit all the patterns in the genome. Instead, we fit
the data with a MC model that can explain the majority (say 80\%) of
the word patterns in the genome. Motivated by the normal
approximation of a particular word statistic in Theorem
\ref{main-theorem}, we study the fraction of $k$-words whose
occurrences can be explained using the statistic
$(Z_{\mathbf{w}}^R)^2/d$ by comparison to a $\chi^2$-distribution
with one degree of freedom with type I error 0.01. We estimate the
order of MC to be the smallest $k-2$ under which more than 80\% of
$k$-words can be explained by the $(k-2)$-th order MC.

To cluster the organisms, we use the inferred MC models to estimate
the expected number of occurrences of word patterns and then study
the relationships among the organisms using our dissimilarity measures $d_2^*$ and $d_2^S$. We briefly present their
definitions below, please see \cite{song2013alignment,song2014new} for details. Then
we apply a similar approach to study the relationships among 13 tree
species with NGS reads, for which neither  the complete genome
sequences nor their relationships are known. To estimate the unknown
effective coverage $d$ using $k$-words by
(\ref{eq:d-estimator}), we let $k$ to be relatively large and use
$\hat{d}$ as the value at which the estimated $d$ stabilizes as $k$
increases.

\subsection{\textcolor{black}{Alignment-free sequence comparison dissimilarity measures}}

Consider two sets of NGS reads from two genomes. We use superscripts
$(1)$ and $(2)$ to denote the first and the second read set,
respectively. Suppose that $M^{(i)}$ reads of length $\beta^{(i)}$
are in the $i$-th data set. 
Since the reads can come from either the
forward strand or the reverse strand of the genome in NGS, we
supplement the observed reads by their complements and refer to the
joint set of the reads and the complements as the read set.

Let $N_{\mathbf{w}}^{(i)}$ be the count
of the word $\mathbf{w}$ in the $i$-th data set. 
We define
$EN_{\mathbf{w}}^{(i)}$ to be the expected number of occurrences of
word $\mathbf{w}$ based on either the i.i.d model or a Markov model,
$ EN_{\mathbf{w}}^{(i)} = M^{(i)} (\beta^{(i)} - k + 1) ( p_\mathbf{w}^{(i)} + p_{\bar{\mathbf{w}}}^{(i)} ), $
where $M^{(i)} (\beta^{(i)} - k + 1)$ is the total number of
$k$-word in the $i$-th sample,  $\bar{\mathbf{w}}$ is the complement
of word $\mathbf{w}$, and $p_\mathbf{w}^{(i)}$ is the probability of
 word $\mathbf{w}$ in the $i$-th genome under a specific
model. 
Then we define $D_2^*$ and $D_2^S$ as follows,

$$D_2^* = \sum \limits_{\mathbf{w}\in \mathcal{A}^{k}}\frac{\tilde{N}_\mathbf{w}^{(1)}\tilde{N}_\mathbf{w}^{(2)}}{\sqrt{EN_\mathbf{w}^{(1)} EN_\mathbf{w}^{(2)}}},  \text{ and} \
D_2^S =  \sum \limits_{\mathbf{w}\in \mathcal{A}^{k}}\frac{\tilde{N}_\mathbf{w}^{(1)}\tilde{N}_\mathbf{w}^{(2)}}{ \sqrt{\big(\tilde{N}_\mathbf{w}^{(1)}\big)^2+\big(\tilde{N}_\mathbf{w}^{(2)}\big)^2} },
$$

where $\tilde{N}_\mathbf{w}^{(i)}= N_\mathbf{w}^{(i)} - EN_\mathbf{w}^{(i)}$, $i=1, 2$.
Further, the dissimilarity measures $d_2^*$ and $d_2^S$, ranging from 0 to 1, are defined as,

\begin{align*}
d_{2}^{*} &= \frac{1}{2} \left ( 1 - \frac{D_{2}^{*}}{\sqrt{\sum \limits_{\mathbf{w}\in\mathcal{A}^{k}} \frac{ \left(\tilde{N}_\mathbf{w}^{(1)} \right)^2} {EN_\mathbf{w}^{(1)}} } \sqrt{\sum \limits_{\mathbf{w}\in\mathcal{A}^{k}} \frac{\left(\tilde{N}_\mathbf{w}^{(2)}\right)^2} {EN_\mathbf{w}^{(2)}}}  } \right ),   \text{ and} \\
d_{2}^{S} &= \frac{1}{2} \left ( 1 - \frac{D_{2}^{S}}{\sqrt{\sum \limits_{\mathbf{w}\in\mathcal{A}^{k}} \frac{ \big( \tilde{N}_\mathbf{w}^{(1)} \big)^2 }{  \sqrt{\big(\tilde{N}_\mathbf{w}^{(1)}\big)^2+\big(\tilde{N}_\mathbf{w}^{(2)}\big)^2}}}
\sqrt{\sum \limits_{\mathbf{w}\in\mathcal{A}^{k}} \frac{\big( \tilde{N}_\mathbf{w}^{(2)} \big)^2}{ \sqrt{\big(\tilde{N}_\mathbf{w}^{(1)}\big)^2+\big(\tilde{N}_\mathbf{w}^{(2)}\big)^2}} }} \right ). 
\end{align*}

For comparison, we also use a simplistic dissimilarity measure based
on the non-centered correlation of the word frequencies defined as
$d_2  = \frac{1}{2} \left( 1 - \frac{\sum \limits_{\mathbf{w}\in
\mathcal{A}^{k}}N_{\mathbf{w}}^{(1)}N_{\mathbf{w}}^{(2)}}{{\sqrt{\sum \limits_{\mathbf{w}\in\mathcal{A}^{k}} \left(N_{\mathbf{w}}^{(1)} \right)^{2}}\sqrt{\sum \limits_{\mathbf{w}\in\mathcal{A}^{k}}\left( N_{\mathbf{w}}^{(2)} \right)^{2}}}}
\right). $

\swallow{
Let $N_{\mathbf{w}}^{(i)}$ be the count
of the word $\mathbf{w}$ in the $i$-th data set. Let
$EN_\mathbf{w}^{(i)} = M^{(i)}(\beta^{(i)}-k+1)p_\mathbf{w}^{(i)} $
be the expected count of $\mathbf{w}$, where $p_\mathbf{w}^{(i)}$ is
the probability of the word $\mathbf{w}$ under a specific MC model.
 Then $D_2^*$ and $D_2^S$ for the two NGS samples are defined as,

$D_2^*=\sum \limits_{\mathbf{w}\in \mathcal{A}^{k}}\frac{\tilde{N}_\mathbf{w}^{(1)}\tilde{N}_\mathbf{w}^{(2)}}{\sqrt{EN_\mathbf{w}^{(1)} EN_\mathbf{w}^{(2)}}}$ ,
$D_2^S=\sum \limits_{\mathbf{w}\in \mathcal{A}^{k}}\frac{\tilde{N}_\mathbf{w}^{(1)}\tilde{N}_\mathbf{w}^{(2)}}{ \sqrt{\left(\tilde{N}_\mathbf{w}^{(1)}\right)^2+\left(\tilde{N}_\mathbf{w}^{(2)}\right)^2} },$
where $\tilde{N}_\mathbf{w}^{(i)}= N_\mathbf{w}^{(i)} - EN_\mathbf{w}^{(i)}$, $i=1, 2$.
Further, the dissimilarity measures $d_2^*$ and $d_2^S$ ranging from 0 to 1 are defined as,

$d_{2}^{*}=\frac{1}{2} \Bigg ( 1 - \frac{D_{2}^{*}}{\sqrt{\sum \limits_{\mathbf{w}\in\mathcal{A}^{k}} \left(\tilde{N}_\mathbf{w}^{(1)} \right)^2 / EN_\mathbf{w}^{(1)}}\sqrt{\sum \limits_{\mathbf{w}\in\mathcal{A}^{k}}\left(\tilde{N}_\mathbf{w}^{(2)}\right)^2 / EN_\mathbf{w}^{(2)}}} \Bigg )$,

$d_{2}^{S}=\frac{1}{2} \Bigg ( 1 - \frac{D_{2}^{S}}{\sqrt{\sum \limits_{\mathbf{w}\in\mathcal{A}^{k}} \frac{ \left( \tilde{N}_\mathbf{w}^{(1)} \right)^2 }{  \sqrt{\left(\tilde{N}_\mathbf{w}^{(1)}\right)^2+\left(\tilde{N}_\mathbf{w}^{(2)}\right)^2}}}
\sqrt{\sum \limits_{\mathbf{w}\in\mathcal{A}^{k}} \frac{\left( \tilde{N}_\mathbf{w}^{(2)} \right)^2}{ \sqrt{\left(\tilde{N}_\mathbf{w}^{(1)}\right)^2+\left(\tilde{N}_\mathbf{w}^{(2)}\right)^2}} }}  \Bigg).$

For comparison, we also use a simplistic dissimilarity measure based
on the non-centered correlation of the word frequencies defined as
$d_2  = \frac{1}{2} \left( 1 - \frac{\sum_{\mathbf{w}\in
\mathcal{A}^{k}}N_{\mathbf{w}}^{(1)}N_{\mathbf{w}}^{(2)}}{{\sqrt{\sum_{\mathbf{w}\in\mathcal{A}^{k}} \left(N_{\mathbf{w}}^{(1)} \right)^{2}}\sqrt{\sum_{\mathbf{w}\in\mathcal{A}^{k}}\left( N_{\mathbf{w}}^{(2)} \right)^{2}}}}
\right). $
}


\section{Results}
\label{results}

\subsection{Summary of simulation results}
Due to page limitations, we summarize the simulation results here;
details are given in the Supplementary Materials. Our extensive
simulations show that the simulated mean, standard deviation and
distributions of $S_k^R$ and $Z_{\mathbf{w}}^R$ are very close to
their corresponding theoretical approximations given by Theorem
\ref{main-theorem}. Both the effective coverage and the MC order can
be estimated accurately under the parameter settings of the current
sequencing technologies.


\subsection{The relationship among 28 vertebrate species}
Table S4 shows the estimated orders of MCs for a group of 28 vertebrate species
that are studied in \citep{miller200728,karolchik2008ucsc}  based on
simulated NGS short reads. 
For each of the 28 species, we compute the
fraction of the $k$-words that have $(Z_{\mathbf{w}}^R)^2/\hat{d}$
within the 99\% of a $\chi^2$-distribution with one degree of
freedom, for $k=8, 9, \dots, 14$. Using 80\% as a threshold, we
estimate the order of MC for each species to be the smallest $k-2$
under which the fraction of words that can be explained by the
$(k-2)$-th order MC is greater than the threshold.

Comparing our results with the results  in \cite{narlikar2013one},
where the order of MCs for a selection of vertebrate genomes was
estimated by  AIC and BIC criteria using whole genome sequences, we
find that the estimated order based on NGS read data are almost the
same as that estimated based on the whole genome sequences in
\cite{narlikar2013one}. Our  proposed methods of estimating the
order of MC based on short reads of NGS data achieve the same
accuracy as that based on whole genome sequences.

For a given value of $k$, we compute $d_2^*$ and $d_2^S$ 
using an $r$-th
order MC, $r=0 \ (\text{i.i.d model}), \dots, (k-2)$ for each pair of
species, yielding a $28 \times 28$ pairwise dissimilarity matrix
under each MC model. To evaluate the dissimilarity measures, we use
the pairwise distance matrix obtained from Figure S1 in
\cite{miller200728} as the gold standard for the dissimilarity
between each pair of the 28 species;  the matrix is given as Table
S5 in the Supplementary Materials. Note that the estimated orders of
the 28 species range from 7 to 11, and the average order is 10. To
study the performance of the dissimilarity measures under different
orders of MC, we choose $k=14$ such that we can study the results
under the MC model with orders up to $12$.

Table \ref{28species_correlation} shows Spearman's rank correlation
coefficient (SPCC)
 between the standard distance and the dissimilarity estimated by the $d_2$-type
 measures under MC models of various orders; higher SPCC indicates better performance.
 \textcolor{black}{Both measures, $d_2^*$ and $d_2^S$, achieve their best results of SPCC=0.92
 when using a MC of order 12. Note that using a simplistic dissimilarity measure $d_2$ only gives SPCC=0.08.}

In general both $d_2^*$ and $d_2^S$ obtain higher SPCC with the
standard matrix as the order of MC increases, except for $d_2^S$ at
order 9. In particular, the measure $d_2^*$ has negative correlation
coefficient with the standard distance under the i.i.d model. The
SPCC becomes stable when the order of the MC used for the analysis
is close to 11, the maximum estimated MC orders over the 28 species.
Here $d_2^S$ is less affected by the order of the MC than $d_2^*$.
When the appropriate order of MC is used, $d_2^*$ and $d_2^S$
perform similarly and much better than $d_2$.

\begin{table}[h]
\centering
\setlength{\tabcolsep}{2pt}
\begin{tabular}{c | c | c | c | c | c  |  c }
$d_2$-type   & order=0 &  order=5 & order=9 & order=10 & order=11 &
order=12 \\ \hline
$d_2^*$ & -0.21   &  -0.16  & 0.85    & 0.89     & 0.90     & 0.92     \\
$d_2^S$ & 0.86    &  0.87 &  0.85    & 0.88     & 0.90     & 0.92    \\
\end{tabular} \\
\caption{The Spearman's rank correlation coefficient (SPCC) between
the true distance matrix and the dissimilarity matrix by $d_2$-type
dissimilarity measures under MC models with order 0 (i.i.d), 5, 9, 10, 11 and
12. The length of the $k$-tuple word is 14. }
\label{28species_correlation}
\end{table}

\subsection{The relationship among 13 tropical tree species with unknown reference genomes}

We also apply our method to the 13 tree species based on the NGS
shotgun read data sets in \cite{cannon2010assembly}. The reference
genome sequences for the 13 tree species are unknown. Our objective
is to cluster these tree species using $d_2^*$ and $d_2^S$ with MCs
for the sequences.

The estimated order of the MC for all the 13 tree species is 8. We
use the dissimilarity measures $d_2^*$ and $d_2^S$
under various orders of MC as the background model to cluster the 13
tree species from their NGS reads. We choose $k=11$ so that we
explore the MC with order up to 9. We use the Unweighted Pair Group
Method with Arithmetic Mean (UPGMA) to cluster the tree species.

The 13 trees species can be generally classified into two groups: 5
tree species from \textit{Moraceae} and 8 tree species from
\textit{Fagaceae}. The two \textit{Moraceaes}, \textit{Ficus
altissima} and \textit{Ficus microcarpa}, should cluster together
because they are known to be closely related and are both large
hemiepiphytic trees while the other three \textit{Moraceae} species
are small dioecious shrubs. Within the \textit{Fagaceae} group, the
two \textit{Castanopsis} species should cluster together, and the
five \textit{Lithocarpus} species should also form a subgroup.
\textit{Trigonobalanus doichangensis (Fagaceae)} is an ancestral
genus that is very divergent from the rest of the family and has
undergone considerable sequence evolution. It should not group
within the class of \textit{Castanopsis} and \textit{Lithocarpus} in
\textit{Fagaceae}.

Figure \ref{13tree_d2star} shows the clustering results of the 13
tree species using $d_2^*$ under MCs of order 0 (\text{i.i.d}), 4, 8 and 9. The
trees are built based on all the reads. From the results we can see,
under the i.i.d model, \textit{Lithocarpus} mixes up with
\textit{Castanopsis}; \textit{T. doichangensis} can not be separated
from the rest of \textit{Fagaceae}, while under the MC of order
greater than 4,  \textit{T. doichangensis} is successfully separated
from the rest of the \textit{Fagaceae}. Moreover, within the
\textit{Moraceae} group, \textit{Ficus fistulas} and \textit{Ficus
langkokensis} form a subgroup under the i.i.d model, and they are
separated under the MC with order greater than 4. While
\textit{F. langkokensis} is the closest \textit{Maraceae} to the
\textit{Fagaceae} under 4th order MC, \textit{F. fistulosa}
becomes the closest species to the \textit{Fagaceaes} under 8th and
9th order MCs.

In order to see whether the clustering of the tree species can be
correctly inferred using only a portion of the shotgun read data, we
randomly sample 10\% of the total read data for each tree species to
cluster them. To study the variation of the clusters due to random
sampling of the reads, we repeat the sampling process 30 times and
calculate the frequencies of each internal branch of the clustering
using all the reads occurring among the 30 clusterings. The number
on the branch refers to the frequency of the branch occurring among
the 30 clusterings based on random sampled 10\% reads. 
Three branches of the tree under MC of order 9 have
frequencies of occurrence less than 30. When using the MC of a very
high order, the clustering becomes unstable.

\begin{figure}[ht]
\centering
\begin{subfigure}[k11, order0, $d_2^*$]{\includegraphics[height=4.5cm,width=9cm,angle=0]{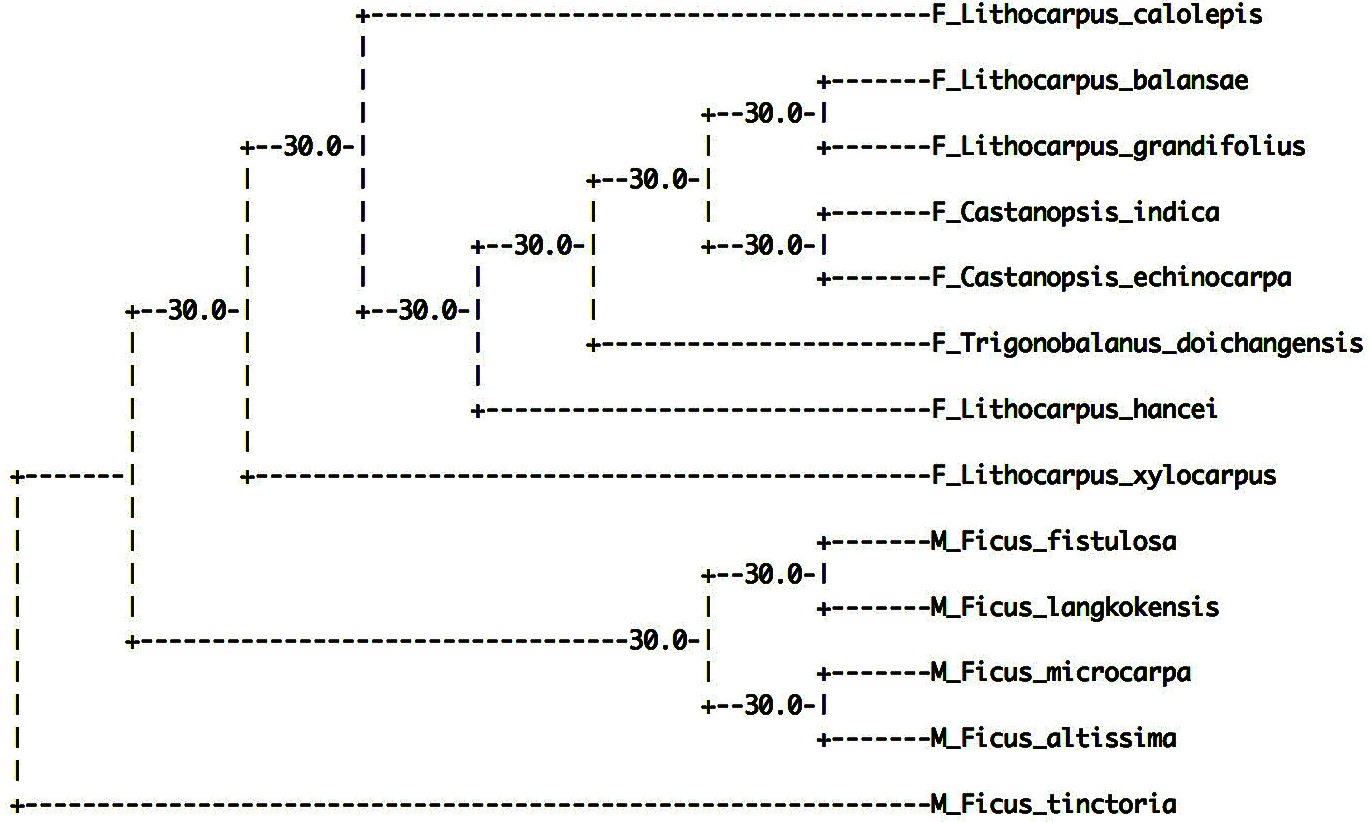}
\label{fig:subfig1}
 }%
\end{subfigure}
 \begin{subfigure}[k11, order4, $d_2^*$]{\includegraphics[height=4.5cm,width=9cm,angle=0]{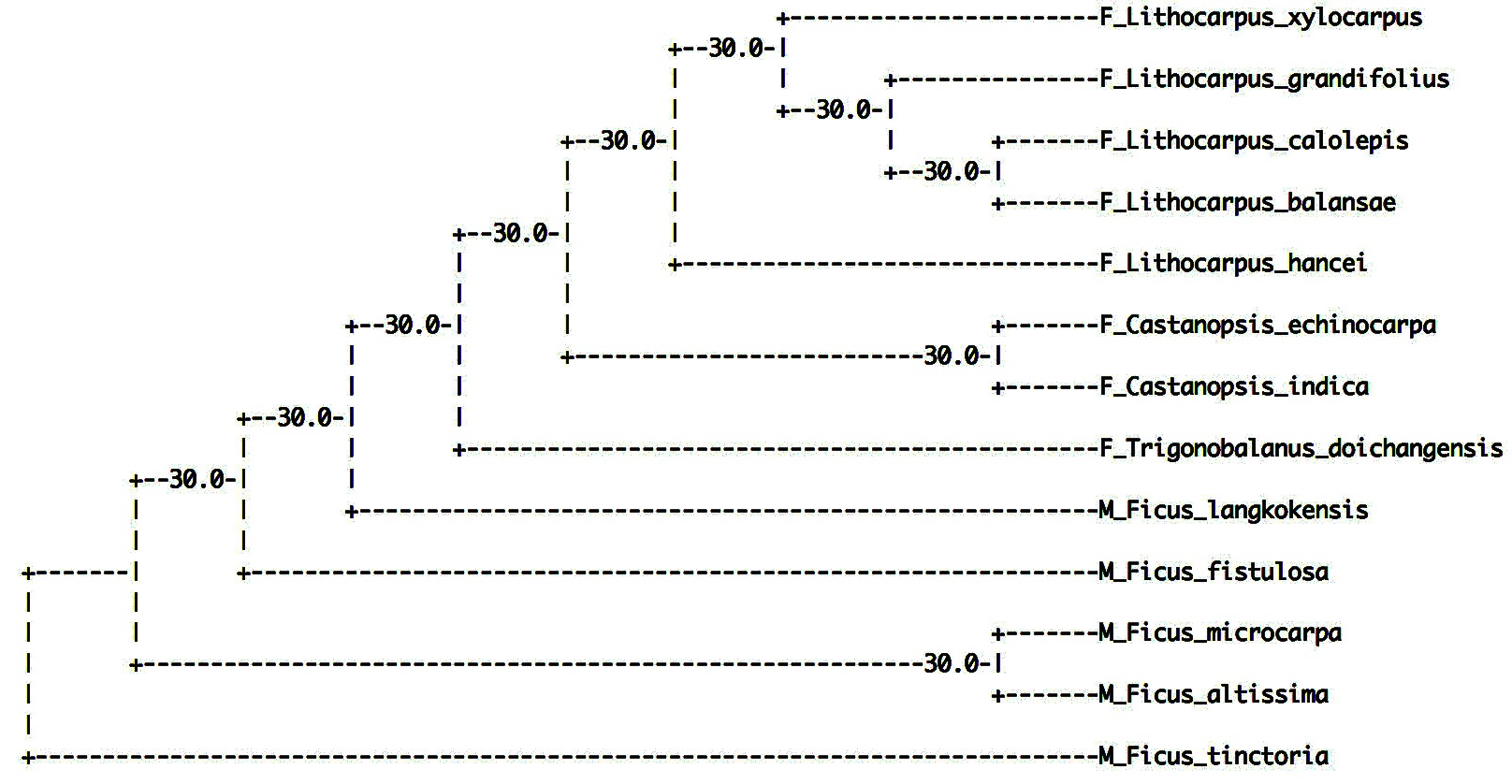}
   \label{fig:subfig2}
 }%
\end{subfigure}
 \begin{subfigure}[k11, order8, $d_2^*$]{\includegraphics[height=4.5cm,width=9cm,angle=0]{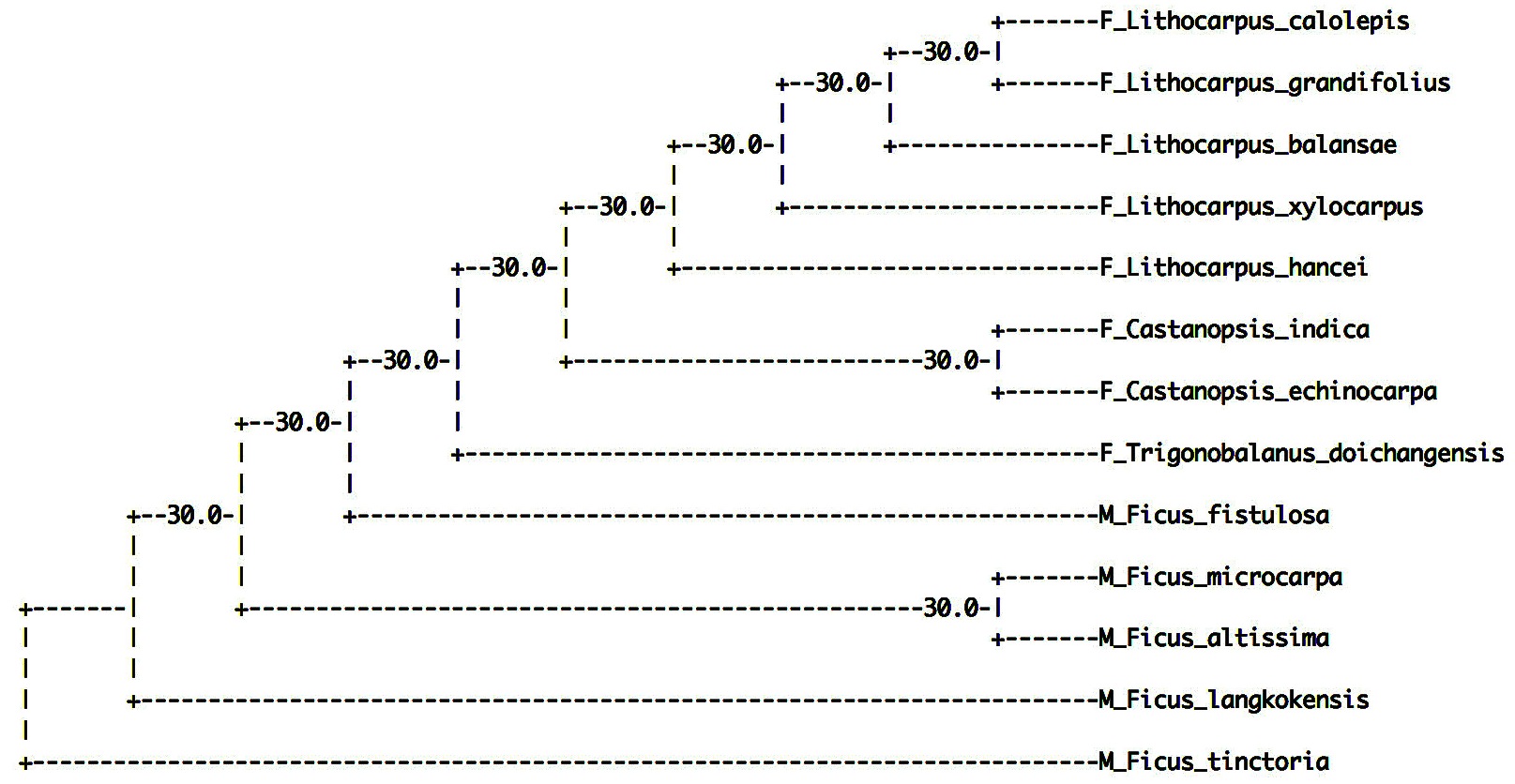}
   \label{fig:subfig3}
 }%
\end{subfigure}
 \begin{subfigure}[k11, order9, $d_2^*$]{\includegraphics[height=4.5cm,width=9cm,angle=0]{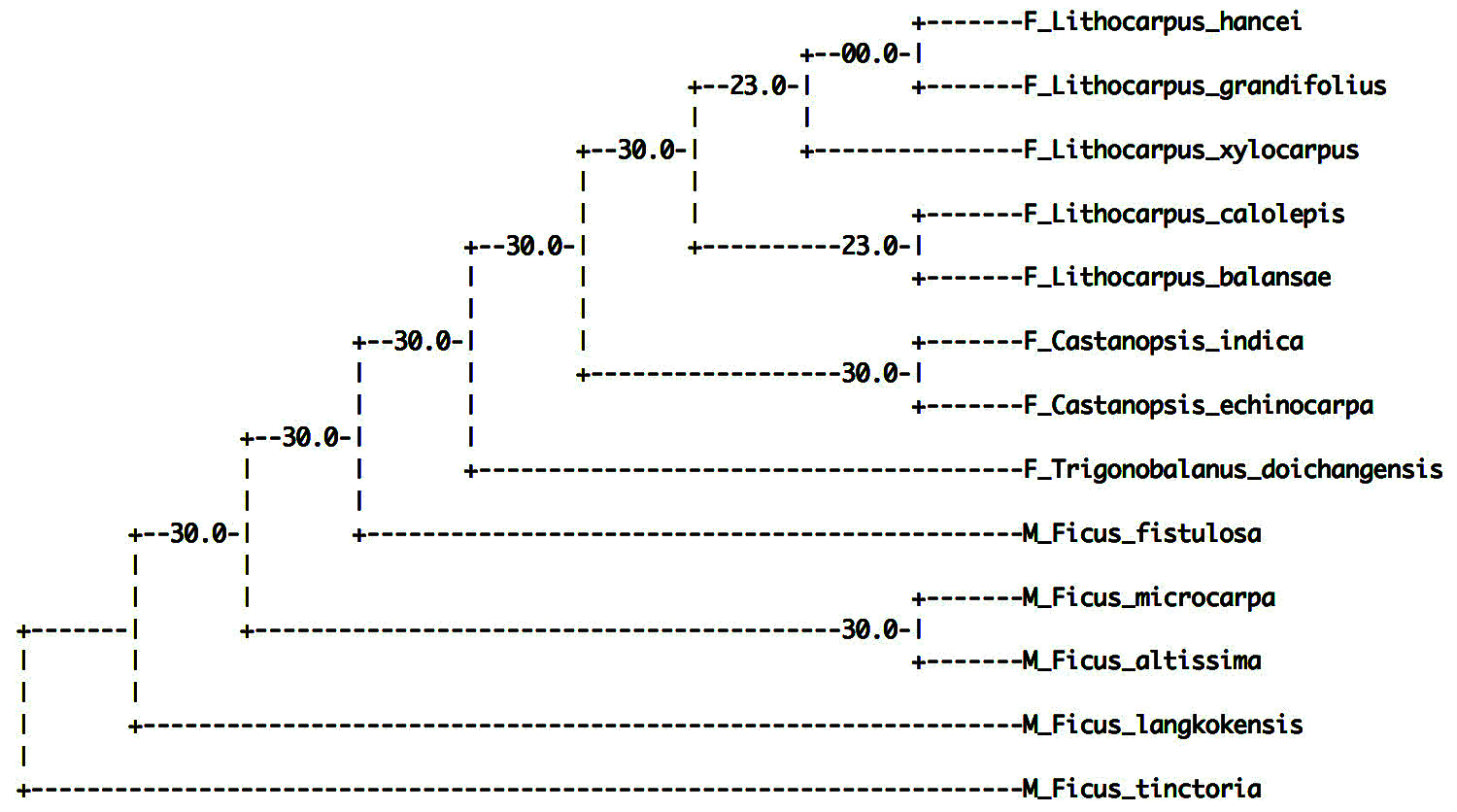}
   \label{fig:subfig4}
 }%
\end{subfigure}
\caption{The clustering of the 13 tropical tree species using
$d_2^*$ under MC with order 0 \ (\text{i.i.d}), 4, 8 and 9. The number on
the branch refers to the frequency of the branch occurring among the
30 clusterings based on random sampled 10\% reads. {The letter `F'
at the beginning of the names represents \textit{Fagaceae};
similarly the letter `M' represents \textit{Maraceae}. }}
\label{13tree_d2star}
\end{figure}

For the clustering results using $d_2^S$, see Figure S7. Under MC
with all four orders, the two \textit{Castanopsis} and the five
\textit{Lithocarpus} species are grouped separately, and \textit{F.
altissima (Moraceae)} and \textit{F.microcarpa (Moraceae)} are
clustered together. Under the i.i.d model, \textit{T.doichangenesis
(Fagaceae)} is successfully separated from \textit{Lithocarpus}, but
it is not the most outside species in the \textit{Fagaceae} group.
When the MC order is greater than 4, \textit{T.doichangenesis
(Fagaceae)} gets separated from the rest of the \textit{Fagaceaes}.
It can also be seen that when using the  i.i.d model, or a  MC with
order 8 or greater, some of the branches becomes unstable.

In general, the results show that the clustering becomes more
accurate as the order of MC increases using both $d_2^*$ and
$d_2^S$. Under the i.i.d model, the clustering based on $d_2^*$ does
not correctly separate \textit{Castanopsis} from
\textit{Lithocarpus}, while the clustering based on $d_2^S$ groups
the two types separately. With higher order MCs, $d_2^*$
successfully separates \textit{Castanopsis} from
\textit{Lithocarpus}. The general clustering structure among
\textit{Lithocarpus}, \textit{Castanopsis}, \textit{Trigonobalanus}
and \textit{Ficus} stays correct when order is greater than 4 for
both measures. 
\textcolor{black}{When using the MC with order higher than the
estimated order, the clustering is unstable 
and indeed the branch for {\it{L.Hancei (Fagaceae)}} is not supported on the last tree when using only 10\% of the data. 
With a large number of parameters to estimate, 10\% of the data does not suffice to capture the information in the data.
The best clustering is achieved under a MC of order 8
and 9.
}

\section{Discussion} \label{discussion}
Next generation sequencing technologies provide large amount of data
in the form of short reads. Assembly of the millions of short reads
to recover the long sequence is challenging, because the relative
short length of the reads makes it difficult to resolve the repeat
regions, not all regions may be covered, and assembly is
time consuming. While multiple sequence alignment may be
prohibitive, we can use word-count based
dissimilarity measures to cluster the underlying species. These
measures require an underlying probability model for the
sequences; Markov chains are a reasonable model for such sequences.
While transition probabilities can be estimated directly from count
data, estimating the order of a MC here is not straightforward.

Methods for estimating the order of a MC of a long sequence have
been developed since the 1950s, but estimating the order of a MC
directly from a set of short reads without assembly has not been
studied yet. In this paper, we develop two statistics $S_k^R$ and
$Z_\mathbf{w}^R$ and show  that  both $S_k^R$ and $Z_\mathbf{w}^R$
have surprisingly  simple approximate distributions with only two
parameters, one of them depending on the order of the original long
MC sequence, and the other one depending on the distribution of the
reads along the sequence. Intriguingly, one of these parameters is
$d = c + 1$  under homogeneous sampling, where $c$ is the coverage
of the reads along the genome based on the first $\beta - k + 1$
positions of each read.

Based on the property of $S_k^R$ and $Z_\mathbf{w}^R$, we develop an
estimator for the order of a MC as well as an estimator for the
parameter $d$ based on NGS data. Extensive simulation studies are
carried out to verify the theorem and evaluate the estimator.
\swallow{We find that the estimators $\hat{r}_{S_k}$,
$\hat{r}_{p_k}$, $\hat{r}_{Z_k}$ and $\hat{r}_{PS}$ provide the most
accurate and robust estimation of the order of a MC, regardless of
the homogeneous and inhomogeneous regime.}

  Finally, we apply the estimation methods to two NGS data sets.
Since the real genome sequences consist of coding, non-coding and
various regulatory regions, single standard MC models do not fit the
data well. Moreover, some enriched patterns, such as the motif
sequences, are widespread throughout the genomes and violate the
simple MC model for the whole genome sequence. Hence studying the
fraction of $k$-words whose occurrences can be explained using the
statistic $(Z_\mathbf{w}^R)^2/d$ by comparison to a $\chi^2_{1}$
distribution is a more realistic way to determine the order of the
MC for a real genome sequence. The estimated orders are  consistent
with the orders estimated directly from the full genome sequences using
BIC methods.

Our primary motivation for this study is alignment-free genome
comparison using NGS data. Further, we cluster the 28 species based
on the NGS data using MC models with various orders. The results
show that the clustering performs best and gives stable results when
using a MC model with order on and above the estimated order. In
addition, we apply the same analysis to 13 tropical tree species
whose reference genomes are unknown; again  the best clustering is
achieved under a MC with the order within the estimated range.

When the sequence length is short or the sequencing depth is low,
 the numbers of occurrences of some $k$-words become small or even
zero. Then the assumption of non-zero variance for all word counts
which underlies the gamma approximation for $S_k^R$  no longer holds
and the gamma approximation  may not work well.  In such a situation
an exact test for the order of MCs in the spirit of
\cite{besag2013exact} could be very helpful. 
In this paper we have only made a start on the Markov chain modelling of NGS data. An exhaustive study of errors in the data, in the form of power studies, could help to further understand the application range of our results.  
Finally, in this work
we take the estimation of the transition probabilities for granted,
once the order of the MC is determined. While   the estimation of
the transition probabilities of the MC model of a long sequence has
been studied by \cite{anderson1957statistical} and
\cite{baum1966statistical}, it would be interesting to extend these
methods to NGS data.

\section*{Acknowledgments}
The authors would like to thank anonymous referees for helpful comments on this work and on previous related work.
We thank Dr. Xiaohui Xie from UCI for explanation of the 28 vertebrate species.
The research is supported by National Natural Science Foundation of China (No.10871009, 10721403),
and National Key Basic Research Project of China (No.2009CB918503). FS is partially supported by
US NIH P50 HG 002790 and NSF DMS-1043075 and
OCE 1136818. GDR is partially supported by EPSRC EP/K032402/1.

{\it Conflict of Interest}: : None declared.

\bibliography{document_sun.bib}
\bibliographystyle{natbib}

\newpage
\clearpage

\section*{Supplementary Materials}
\appendix

\makeatletter
\DeclareRobustCommand*\cal{\@fontswitch\relax\mathcal}
\makeatother

\beginsupplement  

\section{Proof of Theorem 1}

\label{proof}

Here we assume that the conditions of Theorem 1 prevail. In order to prove Theorem 1, we introduce some notations.
Let
 $N_{\mathbf{w}}(i)$ denote the number of occurrences of  $\mathbf{w}$ in the $i$-th region, $i = 1, 2, \cdots$;
 $E_{\mathbf{w}}(i) = \frac{N_{^{-}\mathbf{w} }(i) N_{\mathbf{w}^{-}}(i)}{N_{^{-}\mathbf{w}^{-}}(i)}$ be the estimated expected number of occurrences of $\mathbf{w}$ in the $i$-th region under the $(k-2)$-th order MC model;
and $P_{\mathbf{w}}$ be the probability of $\mathbf{w}$ assuming that the MC starts from the stationary distribution. Similarly, let
 $N_{\mathbf{w}}$, $E_{\mathbf{w}}$, and $\hat{\sigma}^2_\mathbf{w}$ be the observed, expected, and variance of the number of occurrences of $\mathbf{w}$ along the long genome sequence.
 The same notations with superscript ``R" indicate the corresponding quantities based on the short read data.
We assume that $k$ is small compared to $\beta$, and hence the edge effects are small. Then we have the following lemma.


\begin{lemma}
\begin{itemize}
\item[a)]  {{For any $k$-word $\mathbf{w}$, for any region $i= 1, \ldots, B$,}} in distribution,
\begin{equation}
\lim_{G \rightarrow \infty} \frac{N_\mathbf{w}^R(i) - E_\mathbf{w}^R(i)}{\hat{\sigma}_\mathbf{w}^R} = N \left (0, \frac{r_i^2 f_i}{\sum_j r_j f_j} \right).
\label{eq:i-th-interval}
\end{equation}

\item[b)] For any $k$-word $\mathbf{w}$, in probability,
\begin{equation}
\lim_{G \rightarrow \infty} \frac{E_\mathbf{w}^R - \sum_{i} E_\mathbf{w}^R(i)}{\sqrt{E_\mathbf{w}^R}} = 0.
\label{eq:expectation_difference}
\end{equation}
\end{itemize}
\label{lemma:expectation_difference}
\end{lemma}

{\bf Proof of Lemma \ref{lemma:expectation_difference}.}
To prove part a),  {{note that under the null model that the MC is $(k-2)$-th order,
 Theorem 6.4.2 in \cite{reinert2005statistics} gives that, as
 sequence length goes to infinity,  in distribution,
$$
\frac{N_\mathbf{w}(i) - E_\mathbf{w} (i) }{\hat{\sigma}_\mathbf{w} (i)} \rightarrow N(0, 1).
$$
Here we use that  the Markov chain assigns non-zero probability to every $k$-word $\mathbf{w}$ and hence the variance ${\sigma}_\mathbf{w}$ would be non-zero.
As $N^R_\mathbf{w}(j) = r_j N_\mathbf{w}(j)$ the corresponding result for $N^R_\mathbf{w}(j) $ follows; it remains to identify the asymptotic variance. We have for any region $i$ }}
\begin{align}
\lim_{G \rightarrow \infty} \frac{N_\mathbf{w}(i)}{G_i} = P_\mathbf{w},
\label{CLT}
\end{align}
which does not depend on $i$ as the MC is ergodic, and then we have approximately, $$E_\mathbf{w}(i)=G_i \frac{P_{^{-}\mathbf{w}}P_{\mathbf{w}^{-}}}{P_{^-\mathbf{w}^-}} \mbox{ and }$$
$$E_\mathbf{w}^R = \sum_{i} r_i E_\mathbf{w}(i) = \sum_{i} r_i G_i \frac{P_{^{-}\mathbf{w}}P_{\mathbf{w}^{-}}}{P_{^-\mathbf{w}^-}}, \ i=1,2, \dots, .$$
Thus
\begin{eqnarray*}
\lim_{G \rightarrow \infty} \frac{E_\mathbf{w}^R(i)}{E_\mathbf{w}^R} = \frac{r_i f_i}{\sum_{j} r_j f_j}; \\
\lim_{G \rightarrow \infty} \frac{1 - N_{^{-}\mathbf{w}}^R(i)/N_{^{-}\mathbf{w}^{-}}^R(i)}{ 1 - N_{^{-}\mathbf{w}}^R/N_{^{-}\mathbf{w}^{-}}^R}  = 1; \\
\lim_{G \rightarrow \infty} \frac{1 - N_{\mathbf{w}^{-}}^R(i)/N_{^{-}\mathbf{w}^{-}}^R(i)}{ 1 - N_{\mathbf{w}^{-}}^R/N_{^{-}\mathbf{w}^{-}}^R}  = 1.
\end{eqnarray*}
From the above three equations, we have
\[ \lim_{G \rightarrow \infty} \frac{\hat{\sigma}_\mathbf{w}^R(i)}{\hat{\sigma}_\mathbf{w}^R} = \sqrt{\frac{r_i f_i}{\sum_{j} r_j f_j}}. \]
Therefore
\begin{align*}
\frac{N_\mathbf{w}^R(i) - E_\mathbf{w}^R(i)}{\hat{\sigma}_\mathbf{w}^R}
&= \left(  \frac{r_i N_\mathbf{w} (i) - r_i E_\mathbf{w}(i)}{\hat{\sigma}_\mathbf{w}^R (i)} \right) \frac{\hat{\sigma}_\mathbf{w}^R (i)}{\hat{\sigma}_\mathbf{w}^R}  \\
&\rightarrow N \left (0, \frac{r_i^2 f_i}{\sum_j r_j f_j} \right).
\end{align*}

{{ For Part b) of this lemma, note that
$$ 0 \le  N_\mathbf{w}^R - \sum_{j=1}^B N^R_\mathbf{w}(j) \le 2 M k $$
as the only differences in the counts occur due to not counting occurrences at the boundaries of the regions and there are $M$ reads. As $N_\mathbf{w} \sim P_\mathbf{w}  G$ and as $\lim_{G \rightarrow \infty}\frac{Mk}{\sqrt{G}} \rightarrow 0$, the second assertion follows.
}}

\hfill  \text{$\square$}  \\


From Lemma \ref{lemma:expectation_difference}, we can easily show the first assertion in Theorem 1 by noting that
\begin{align*}
 Z_\mathbf{w}^R & =  \frac{N_\mathbf{w}^R - E_\mathbf{w}^R}{\hat{\sigma}_\mathbf{w}^R}  =  \sum_{i} \frac{N_\mathbf{w}^R (i) - E_\mathbf{w}^R(i)}{\hat{\sigma}_\mathbf{w}^R} + \frac{\sum_{i} E_\mathbf{w}^R(i) - E_\mathbf{w}^R}{\hat{\sigma}_\mathbf{w}^R}.
 \end{align*}
 The {{last summand}} tends to 0 by \eqref{eq:expectation_difference}. {{When $G_i$ is large then the}}  $i$-th term {{is close to}} a normal distribution with mean 0 and variance $\frac{r_j^2 f_j}{\sum_i r_i f_i}$. Since the dependence between the segments is {{weak}}, we can treat the terms in the first summand as independent. Part a) of Theorem 1 is proved.


Now we prove the part b) of Theorem 1. Suppose there are only two regions with coverage $r_i$ and region length $G_i$, $i=1,2$.
With \eqref{eq:expectation_difference}, $S_k^R$ has approximately the same distribution as
\begin{align*}
S_k^{R, *}
&= \sum_\mathbf{w} \frac{ \left( N_\mathbf{w}^R - E_\mathbf{w}^R(1) - E_\mathbf{w}^R(2) \right)^2}{E_\mathbf{w}^R}\\
&= \sum_\mathbf{w} \left( \sum_{i} \frac{N_\mathbf{w}^R(i) - E_\mathbf{w}^R(i)}{\sqrt{E_\mathbf{w}^R}}   \right)^2  \\
&= \sum_\mathbf{w} \left( \sum_{i} \frac{N_\mathbf{w}(i) - E_\mathbf{w}(i)}{\sqrt{E_\mathbf{w}(i)}} \sqrt{\frac{r_i E_\mathbf{w}^R(i)}{E_\mathbf{w}^R}} \right)^2  \\
&= \sum_\mathbf{w} \left( \sum_{i} W_i \frac{N_\mathbf{w}(i) - E_\mathbf{w}(i)}{\sqrt{E_\mathbf{w}(i)}}   \right)^2,
\end{align*}
where $$W_i = \sqrt{\frac{r_i E_\mathbf{w}^R(i)}{E_\mathbf{w}^R}} \approx \sqrt{ \frac{r_i^2 f_i}{\sum_j r_{j}f_j }}, i=1,2. $$

 Generally suppose that the reads come from $B$ regions with each region having the same coverage. Let $r_i$ be the coverage and $G_i$ be the genomic length of the $i$-th region. Using the same idea as  above, we can approximate $S_k^R$ by
\begin{align}
\label{S_k^R_W}
S_k^{R,*} = \sum_\mathbf{w} \left( \sum_{i=1}^B W_i \frac{N_\mathbf{w}(i) - E_\mathbf{w}(i)}{\sqrt{E_\mathbf{w}(i)}} \right)^2.
\end{align}

Note that from Section 6.6.1 in \cite{reinert2005statistics}, the vector
$${\bf{\tilde{N}}}(i) :=\left( \frac{N_\mathbf{w}(i) - E_\mathbf{w}(i)}{\sqrt{E_\mathbf{w}(i)}} \right)_{\mathbf{w} \in \mathit{A}^k} \rightarrow \mathit{N}\left(0, \Sigma^2_{L^k \times L^k} \right), $$ in distribution, where $\Sigma^2_{L^k \times L^k}$ is the covariance matrix with rank $df_k$. Hence,  in distribution,
$$ \sum_{\mathbf{w} \in \mathit{A}^k} \left( \frac{N_\mathbf{w}(i) - E_\mathbf{w}(i)}{\sqrt{E_\mathbf{w}(i)}} \right)^2 \rightarrow \chi^2_{(df_k)} ,$$  where $\chi^2_{(df_k)}$ is a $\chi^2$ random variable with $df_k=(L-1)^2 L^{k-2}$ degrees of freedom. {{On the other hand,}} we can find a $L^k \times df_k$ matrix $M$ such that
$ \Sigma^2_{L^k \times L^k}= M M^T .$
Let $M^{-1}$ be the pseudo-inverse of $M$, then we have
$$M^{-1} {\bf{\tilde{N}}}(i)  = \begin{pmatrix}
Z_1(i)
\\ ...
\\ Z_{df_k}(i)
\end{pmatrix}  = Z(i) $$
such that approximately $Z(i) \sim \mathit{N}(0, I_{df_k}). $
As  ${\bf{\tilde{N}}}(i) = M Z(i)$ {{we obtain that}}
$$
{\bf{\tilde{N}}}(i) ^T \cdot{\bf{\tilde{N}}}(i)
=  Z(i)^T M^T M Z(i).
$$
Since the left hand side has   approximately a $\chi^2$ distribution with $df_k$ degrees of freedom, the right hand side should be in distribution close to $Z(i)^t Z(i)$. Thus {{approximately,}} $M^T M = I_{df_k \times df_k}.$
Also note that $M$ does not depends on $i$ because the correlation structure of  ${\bf{\tilde{N}}}(i) $ is the same across the regions under the null model.
Then (\ref{S_k^R_W}) for $S_k^{R,*}$ can be written as
\begin{align*}
S_k^{R,*}
&= \left( \sum_{i=1}^B W_i {\bf{\tilde{N}}}(i)  \right)^T \left( \sum_{i=1}^B W_i {\bf{\tilde{N}}}(i)  \right) \\
&= \left( \sum_{i=1}^B W_i Z(i) \right)^T M^T M \left( \sum_{i=1}^B W_i Z(i)  \right) \\
&= \left( \sum_{i=1}^B W_i Z(i) \right)^T \left( \sum_{i=1}^B W_i Z(i)  \right) \\
&= \sum_{k=1}^{df_k} \left( \sum_{i=1}^B W_i Z_k(i) \right)^2.
\end{align*}
Since $Z_k(i)$ are all  approximately  i.i.d $\mathit{N}(0, 1)$ random variables and $ \sum_{i=1}^B W_i^2 \approx  \sum_{i=1}^B \frac{r_i^2 f_i}{\sum_j r_{j}f_j }$,
we obtain that, {{in distribution,}}
 $$Y_k = \sum_{i=1}^B W_i Z_k(i)  \rightarrow \mathit{N}\left(0, \sum_{i=1}^B W_i^2 \right) \mbox{ and } \frac{Y_k}{\sqrt{\sum_{i=1}^B W_i^2}}  \rightarrow \mathit{N}(0, 1) . $$
Hence
\begin{align*}
S_k^R =  \left(\sum_{i=1}^B W_i^2  \right)  \sum_{k=1}^{df_k}\left( \frac{Y_k}{\sqrt{\sum_{i=1}^B W_i^2}} \right)^2
\rightarrow  \left( \frac{ \sum_{i=1}^B r_i^2 f_i}{\sum_{j=1}^B r_{j}f_j } \right) \chi^2(df_k).   \hfill \text{$\square$}
\end{align*}

\section{Methods}


\subsection{Estimating the order of a Markov chain based on Theorem 1}

{{We  assume that  $k \ge 2$ and that the assumptions of  Theorem 1 are satisfied. Moreover we assume for now that $d$ is known; in}}
practice, the effective coverage $d$ is replaced by the estimated
value $\hat{d}$.
In addition to our estimator
\begin{equation} \label{eq:r-Sk-estimate}
\hat{r}_{S_k}= \mathrm{argmin}_k \left \{\frac{S^R_{k+1}}{ S_{k}^R} \right \} - 1
\end{equation}
 we  define four related estimators based on Theorem 1; they are all analogs of corresponding established estimators for the order of a Markov chain.

\begin{enumerate}
\item  Instead of using $S_k^R$ directly, we can calculate the p-value
\footnotesize
\[ p_k = P \left( S^R_k \geq s^R_k \right) = P \left( S^R_k/d \geq s^R_k/d \right) = P \left( \chi^2_{df_k} \geq s^R_k/d \right), \]
\normalsize
where $s^R_k$ is the observed value of $S^R_k$ based on the short read data. We expect $p_k$ to be small for $k < r+2$ while $p_k$ will not be that small for $k \geq r+2$. Therefore, we expect $\log(p_{k+1})/\log(p_k)$ to be the smallest when $k = r+1$. Thus we can also estimate the order of a MC by
\begin{equation}
\hat{r}_{p_k} = \mathrm{argmin}_k \left\{ \frac{ \log(p_{k+1})}{\log(p_k)} \right\} - 1.
\label{eq:r-pk-estimate}
\end{equation}

\item For a given significance level $\alpha$, we check consecutively if $p_{k+1} < \alpha$ and stop when $p_{k+1} \geq \alpha$. We estimate $r$ by
\begin{equation}
\hat{r}_{h} = \mathrm{argmin}_k \{p_{k+1} \geq \alpha\} - 1, \text{for a given significant level $\alpha$.}
\label{eq:r-hypothesis-estimate}
\end{equation}
To avoid early stopping, we can also require that  both $p_k$ and $p_{k+1}$ are larger than $\alpha$.

\item If $k \geq r+2$, then $Z^R_{\mathbf{w}}/\sqrt{d}$  is approximately  standard normal $N(0,1)$ and thus $(Z^R_{\mathbf{w}})^2/d$ has approximately a $\chi^2$-distribution with one degree of freedom. If $k < r+2$, for some $k$-word $\mathbf{w}$, $(Z^R_{\mathbf{w}})^2/d$ is generally larger than a $\chi^2$-distributed random variable with one degree of freedom. Therefore
we would expect $Z_{\mathrm{max}}^R(k)$ to be large when $k < r+2$ and $Z_{\mathrm{max}}^R(k)$ to be relatively small for $k \geq r+2$. As $Z_{\mathrm{max}}^R$ is the maximum value over $L^k$ variables, we should divide $Z_{\mathrm{max}}^R(k)$ by $L^k$. Therefore, we estimate the order of the MC $r$ by
\begin{equation}\label{eq:r_Z-estimate}
\hat{r}_{Z_k}
= \mathrm{argmin}_k \left \{\frac{Z_{\mathrm{max}}^R (k+1)}{Z_{\mathrm{max}}^R (k)} \right \} - 1,
\end{equation}
where  $Z_{\mathrm{max}}^R(k) = \max_{\mathbf{w}, |\mathbf{w}| = k }|Z_\mathbf{w}^R|.$

\item
{{Extending the method by
\cite{morvai2005order} and
\cite{peres2005two}, for}}  a set of short reads and a $(k-1)$-word $\mathbf{v}=v_1 \cdots v_{k-1}$, define
$$
 \triangle^{k-1}(\mathbf{v}) =  \max_{a \in {\mathcal A}} \left| N^R_{\mathbf{v} a} - \frac{N^R_{^-\mathbf{v} a} N^R_{\mathbf{v}}}{N^R_{^-\mathbf{v}}} \right|;
$$
then
$$
\triangle^k = \max_{\mathbf{v} \in {\mathcal A}^{k-1}} \{  \triangle^{k-1}(\mathbf{v}) \} = \max_{\mathbf{w} \in {\mathcal A}^{k}} \left| N^R_{\mathbf{w}}- \frac{N^R_{^- \mathbf{w}}N^R_{\mathbf{w}^-}}{N^R_{^- \mathbf{w}^-}}  \right|
$$
is the maximum difference between the number of occurrences of a $k$-word and its estimated expectation under the  $(k-2)$-th order MC. {{Our}} Peres-Shields{{-type}} estimator is
\begin{equation} \label{rps}
 \hat{r}_{\mathrm{PS}}(x) = \mathrm{argmax}_{k} \left\{ \frac{\triangle^{k}}{\triangle^{k+1}} \right\} - 1.
\end{equation}

\end{enumerate}

\subsection{Estimating the order of the MC based on modified AIC and BIC }
Several methods based on the Akaike information criterion (AIC) and the Bayesian information criterion (BIC) have been proposed to estimate the order of MC based on long sequences \citep{narlikar2013one,tong1975determination,katz1981some,hurvich1989regression,zhao2001determination}.
The AIC and BIC for long sequences are defined by
\begin{align}
AIC(k) &= - 2 \log(\text{Sequence-Likelihood}, {\cal M}_k) + 2 |{\cal M}_k|,
\label{eq:AIC} \\
AICc(k) &= AIC(k) + \frac{ 2 |{\cal M}_k| \left( |{\cal M}_k| + 1 \right) }{ \left( |{\cal S}|_k - |{\cal M}_k - 1  \right)}, \label{eq:AICc}  \text{and} \\
BIC(k) &= - 2 \log(\text{Sequence-Likelihood}, {\cal M}_k) + |{\cal M}_k|  \log|{\cal S}|_k, \label{eq:BIC}
\end{align}
where ${\cal M}_k$ indicates the $k$-th order Markov chain, $|{\cal M}_k|$ is the number of parameters for the $k$-th order Markov model, i.e. $|{\cal M}_k| = (L-1)L^k$, and $|{\cal S}|_k$ is the size of the data. For a long sequence of length $G$, we have that $|{\cal S}|_k = G-k$ is the number of (k+1) tuples.

However, no formulas have been defined for AIC and BIC of Markov models based on NGS data. In the following we modify the definitions of AIC and BIC given in (\ref{eq:AIC}), (\ref{eq:AICc}) and (\ref{eq:BIC}) respectively so that they are applicable to NGS data.
For a NGS short read sample, we define the log-pseudo-likelihood under the $k$-th order  MC model as,
$$\log(_p LH_k^R)=\sum_{\mathbf{w} \in L^{k+1}} N_{\mathbf{w}}^R  \ \log\frac{N_{\mathbf{w}}^R}{N_{\mathbf{w}^-}^{R}},$$
by replacing $N_{\mathbf{w}}$ with $N_{\mathbf{w}}^R$ in the log-likelihood of a long sequence.
We think of the factor $d$ as the effective coverage of the reads along the genome. So the pseudo-likelihood of the NGS data under the $k$-th order of MC model is approximately the likelihood of the effectivele covered region along the genome to the power of $d$, i.e. $_p LH_k^R \approx (\text{likelihood of the covered genomic region})^d$. Thus, the log-likelihood of the covered genomic region is approximately $\log(_p LH_k^R)/d$. Based on the idea of AIC for long sequences, we minimize $ -2 \log(_p LH_k^R)/d + 2 (L-1)L^k$ with respect to $k$. Therefore, we propose the AIC for k=1, 2, \dots, based on NGS data as
\begin{equation*}
 AIC^R(k) = -2 \log(_p LH_k^R) + 2 d (L-1)L^k.
\end{equation*}

To define BIC and AICc for NGS data, we also need to find a suitable analog of the data size $|{\cal S}|_k$ in (\ref{eq:AICc}) and  (\ref{eq:BIC}). A naive substitute is the total effective read length $M (\beta-k)$, or in other words, the number of $(k+1)$-tuples used to estimate the likelihood under the $k$-th order MC model. However, the reads can overlap and adjustments are needed. Using the effective coverage factor $d$ as a normalization factor from the NGS to long sequence, the length of the effective covered genomic region is $\frac{M (\beta-k)}{d}$. Then we define AICc and BIC for the NGS read data as
\begin{align*}
AICc^R(k) &= AIC^R(k) + \frac{2(L-1)L^k \left( (L-1)L^k + 1 \right)}{ \left( \frac{M * (\beta-k)}{d} - (L-1)L^k - 1 \right)}, \text{and} \\
BIC^R(k) &= -2 \log(_p LH_k^R) + d (L-1)L^k \log \frac{M * (\beta-k)}{d}.
\end{align*}
{{We define}} the estimators of the order of a Markov model based on AIC, AICc and BIC for the NGS read data by
\begin{align}
\hat{r}_{AIC^R}& = \mathrm{argmin}_k{AIC^R(k)} \label{eq:AIC-R}  \\
\hat{r}_{AICc^R} &= \mathrm{argmin}_k{AICc^R(k)} \label{eq:AICc-R}  \\ \text{and} \ \
\hat{r}_{BIC^R} &= \mathrm{argmin}_k{BIC^R(k)} \label{eq:BIC-R}
\end{align}
In order to see the effect of the normalization by the effective coverage factor $d$,  we denote by $\hat{r}_{AIC}$, $\hat{r}_{AICc}$ and $\hat{r}_{BIC}$ the naive estimators, which are obtained by simply substituting the log-sequence-likelihood and the size of the data by the log-pseudo-likelihood and the number of (k+1)-tuples in the NGS read data without normalization by $d$.

\section{Results}

\subsection{Simulation studies for the validation of the theoretical results}

{{ For the simulation studies}} we first generate MCs of different orders. {{For realistic parameter values,}}  the transition probability matrices of the MCs are based on real cis-regulatory module (CRM) DNA sequences in mouse forebrain from \cite{blow2010chip}. We start the sequence from the stationary distribution.
 In addition to the first order MC model described above, we also simulate a second order MC. Table \ref{transition-matrix} shows the transition probability of a) the first and b) the second order MC matrices we used in the simulations.

\begin{table}[!htbp]
\centering
\setlength{\tabcolsep}{18pt}
\subtable[The transition probability matrix of the first order MC]{
\begin{tabular}{ c | c c c c }
& A & C & G & T  \\ \hline
A & 0.39 & 0.15 & 0.21 & 0.25 \\
C & 0.31 & 0.19 & 0.12 & 0.38 \\
G & 0.31 & 0.23 & 0.20 & 0.26 \\
T & 0.26 & 0.18 & 0.21 & 0.35 \\
\end{tabular}} \\
\subtable[The transition probability matrix of the second order MC]{
\begin{tabular}{ c | c c c c }
 & A & C & G & T \\ \hline
AA & 0.39 & 0.19 & 0.19 & 0.24 \\
AC & 0.25 & 0.26 & 0.21 & 0.27 \\
AG & 0.38 & 0.17 & 0.29 & 0.15 \\
AT & 0.30 & 0.24 & 0.22 & 0.24 \\
CA & 0.46 & 0.20 & 0.15 & 0.18 \\
CC & 0.26 & 0.40 & 0.11 & 0.22 \\
CG & 0.23 & 0.16 & 0.14 & 0.48 \\
CT & 0.29 & 0.20 & 0.15 & 0.37 \\
GA & 0.26 & 0.25 & 0.23 & 0.26 \\
GC & 0.25 & 0.17 & 0.33 & 0.25 \\
GG & 0.26 & 0.30 & 0.28 & 0.17 \\
GT & 0.29 & 0.12 & 0.22 & 0.37 \\
TA & 0.19 & 0.19 & 0.21 & 0.41 \\
TC & 0.12 & 0.29 & 0.24 & 0.35 \\
TG & 0.27 & 0.17 & 0.25 & 0.32 \\
TT & 0.11 & 0.18 & 0.24 & 0.47 \\
\end{tabular}
}
\caption{The transition probability matrices of a) the first and b) the second order Markov chain in our simulation studies.}
\label{transition-matrix}
\end{table}

We generate MCs with length of $G=10^5$ and $2\times10^5$.
We simulate NGS {{data}} by sampling {{a varying}} number of reads of different lengths from the MC{{ as follows}}.
{{We}} use the Lander-Waterman model for physical mapping
\citep{lander1988genomic} to sample NGS reads homogeneously with read length $\beta
= 100, 200, 300, 400, 500$ and the number of reads $M =  500, 1000$.
The coverage $c$ is calculated as $c = M \beta/G$ and the effective
coverage based on the first $\beta-k+1$ positions is $d = c_{\mbox{eff}} + 1$, where $c_{\mbox{eff}} = M (\beta-k+1)/(G-k+1) $.
Under each combination of $(G, M, \beta)$, we calculate the values of
$Z_\mathbf{w}^R$ for each word $\mathbf{w}$ and $S_k^R$ based on the NGS read data.
Then we repeat the processes 2000 times to obtain
the empirical distribution of $Z_\mathbf{w}^R$ and $S_k^R$.
Finally, we compare the mean and
variance of $S_k^R$ with their corresponding theoretical
approximations and test the fit of the data to the theoretical
approximate distribution using the Kolmogorov-Smirnov (KS) test.

With the current NGS technologies,
the reads are generally not homogeneously sampled from the genome. In
order to see the effects of inhomogeneous sampling on the
approximate distributions of $S_k^R$ and $Z_\mathbf{w}^R$, similarly as in \cite{song2013alignment} we implement
the following simulation.  We divide the long sequences into 100
blocks, $b_1, \dots, b_t, \dots, b_{100}$. For each block $b_t$,
the sampling probability
$\lambda_i$ for each position in this block  is proportional to a random number which is drawn independently from
 the gamma
distribution $\Gamma(1,20)$ (one number per block).

It has been observed that the probability that a fragment is sequenced in NGS depends on its nucleotide content. Empirical studies showed that the dependency on GC content is unimodal and we use the empirical unimodal distribution on GC curve shown in the software developed by \cite{benjamini2012summarizing} as an example to generate the reads. The other parameters are the same as before.

Sequencing error is another factor that reduces the data quality. Currently, Illumina sequencing has an error rate about 0.1\% and 454 sequencing has an error rate about 1\% \citep{glenn2011field}.
In order to see the effect of sequencing errors on the distribution of $S_k^R$ and $Z_\mathbf{w}^R$, in each position of a read, we randomly replace the letter in that position with one of the other three letters with equal probability 0.005; the different letter is drawn with equal probability. The remaining simulation steps stay the same as before.

Next we evaluate the proposed estimators of the order of MCs, $\hat{r}_{S_k}$, $\hat{r}_{p_k}$, $\hat{r}_h$, $\hat{r}_{Z_k}$ and $\hat{r}_{PS}$ 
and the estimator for the effective coverage $d$ based on equation (7) in the main text, respectively.
{{For estimating}} the effective coverage $d$ by equation (7) in the main text, we using 3-tuples for a first order MC,  and 4-tuples for a second order MC.
For evaluating the proposed estimators, we let $k$ range from 2 to 6, i.e. we choose the model from 1st, 2nd, $\cdots$, 5th order MC. The performance of the estimator is measure by the precision rate, i.e. the percentage of times in 1000 repeats the estimator gives the true order.
In order to see the effect of genome length and sequencing depth on the estimators, we take the genome length 5000, 7500, 10000, 20000, 50000, while fixing the read coverage to be 1; we take the read coverage 0.05, 0.1, 0.25, 0.5, 0.75, 1, while fixing the genome length to be $10^5$.
We also set the sequencing error rate at 10\%, which is relatively high compared to the true sequencing error rate in real sequencing, 
as to distinguish clearly between the estimators with respect to 
their robustness to sequencing errors. For the estimator $\hat{r}_h$, we set $\alpha=0.05$.

When the sequence length is short or the coverage is low, it is possible that the numbers of occurrences of some $k$-words are close to zero and the expected numbers of occurrences are very low; the estimated expected number of occurrences may turn out to be 0 for some $k$-words. In order to overcome the issue, we add a pseudo count of 1 to the number of occurrence of all words, which is a common procedure to avoid division by zero and is equivalent to incorporating a flat prior during the parameter estimation in terms of Bayesian statistics, see \cite{narlikar2013one,strelioff2007inferring}.

\subsubsection{Simulation results for Theorem 1: Homogeneous sampling of the reads}
 We use simulations to validate the theoretical results in Theorem 1. In our simulation study, here we first generate sequences with genome length of $G = 1\times 10^5$ and $2\times 10^5$ bps under the first order MC model as shown in Table S1(a).
$S_3^R$ with their corresponding theoretical approximations in Theorem 1  are given in Table S2. It can be seen from the table
that the approximate mean and variance of $S_3^R$ are very close to
their simulated values  except in the case of a large number ($M=1000$) of very short reads  ($\beta=100$) in a small genome ($G=10^5$ bps).

\begin{table}[!htbp]
\centering
\setlength{\tabcolsep}{3pt}
\subtable[Genome length $G=1 \times 10^5$ ]{
    \begin{tabular}{c|ccccc|ccccc}\hline
 & \multicolumn{5}{c|}{$M=500$} & \multicolumn{5}{c}{$M=1000$}\\\hline
$\beta$ & $\hat{\text{mean}}$ & mean & $\hat{\text{var}}$ &
var & p-value & $\hat{\text{mean}}$ & mean & $\hat{\text{var}}$ &
var & p-value \\\hline
$100$ & 53.4 & 54.0 & 160.0 & 162.0 & 0.06 & 70.8 & 72.0 & 280.0 & 288.0 & 0.002 \\
$200$ & 71.6 & 72.0 & 277.4 & 288.0 & 0.99 & 106.9 & 108.0 & 660.5 & 648.0 & 0.15 \\
$300$ & 89.8 & 90.0 & 482.9 & 450.0 & 0.39 & 143.9 & 144.0 & 1139.5 & 1152.0 & 0.15 \\
$400$ & 107.6 & 108.0 & 655.0 & 648.0 & 0.84 & 179.8 & 180.0 & 1901.3 & 1800.0 & 0.20 \\
$500$ & 126.0 & 126.0 & 905.2 & 882.0 & 0.86 & 216.6 & 216.0 &
2675.3 & 2592.0 & 0.26
\\\hline
    \end{tabular}
} \subtable[Genome length $G=2 \times 10^5$ ]{
    \begin{tabular}{c|ccccc|ccccc}\hline
 & \multicolumn{5}{c|}{$M=500$} & \multicolumn{5}{c}{$M=1000$}\\\hline
$\beta$ & $\hat{\text{mean}}$ & mean & $\hat{\text{var}}$ &
var & p-value & $\hat{\text{mean}}$ & mean & $\hat{\text{var}}$ &
var & p-value \\\hline
$100$ & 44.6 & 45.0 & 110.9 & 112.5 & 0.12 & 53.7 & 54.0 & 142.7 & 162.0 & 0.61 \\
$200$ & 53.5 & 54.0 & 159.6 & 162.0 & 0.13 & 71.8 & 72.0 & 278.4 & 288.0 & 0.64 \\
$300$ & 62.7 & 63.0 & 222.5 & 220.5 & 0.95 & 89.0 & 90.0 & 440.1 & 450.0 & 0.41 \\
$400$ & 71.7 & 72.0 & 286.0 & 288.0 & 0.51 & 107.4 & 108.0 & 645.7 & 648.0 & 0.26 \\
$500$ & 80.9 & 81.0 & 357.5 & 364.5 & 0.69 & 124.7 & 126.0 &
821.3 & 882.0 & 0.99
\\\hline
    \end{tabular}
}
\caption{Comparison of mean and variance of
$S_3^R$ with their corresponding theoretical approximations under a first order MC model and the fit of the data to the theoretical approximate distribution
using the KS test. The simulation process was repeated 2000 times for each combination of $(G, M, \beta)$.
The columns $\hat{\text{mean}}$ and $\hat{\text{var}}$ are the simulated mean and variance; the columns mean and var are the theoretical mean and
variance.}
\label{estimate_homo_Sk}
\end{table}

\begin{figure}
\centering
\scriptsize
\tiny
\begin{subfigure}[homogeneous, without error]{
\includegraphics[height=6.5cm,width=6.5cm,angle=0]{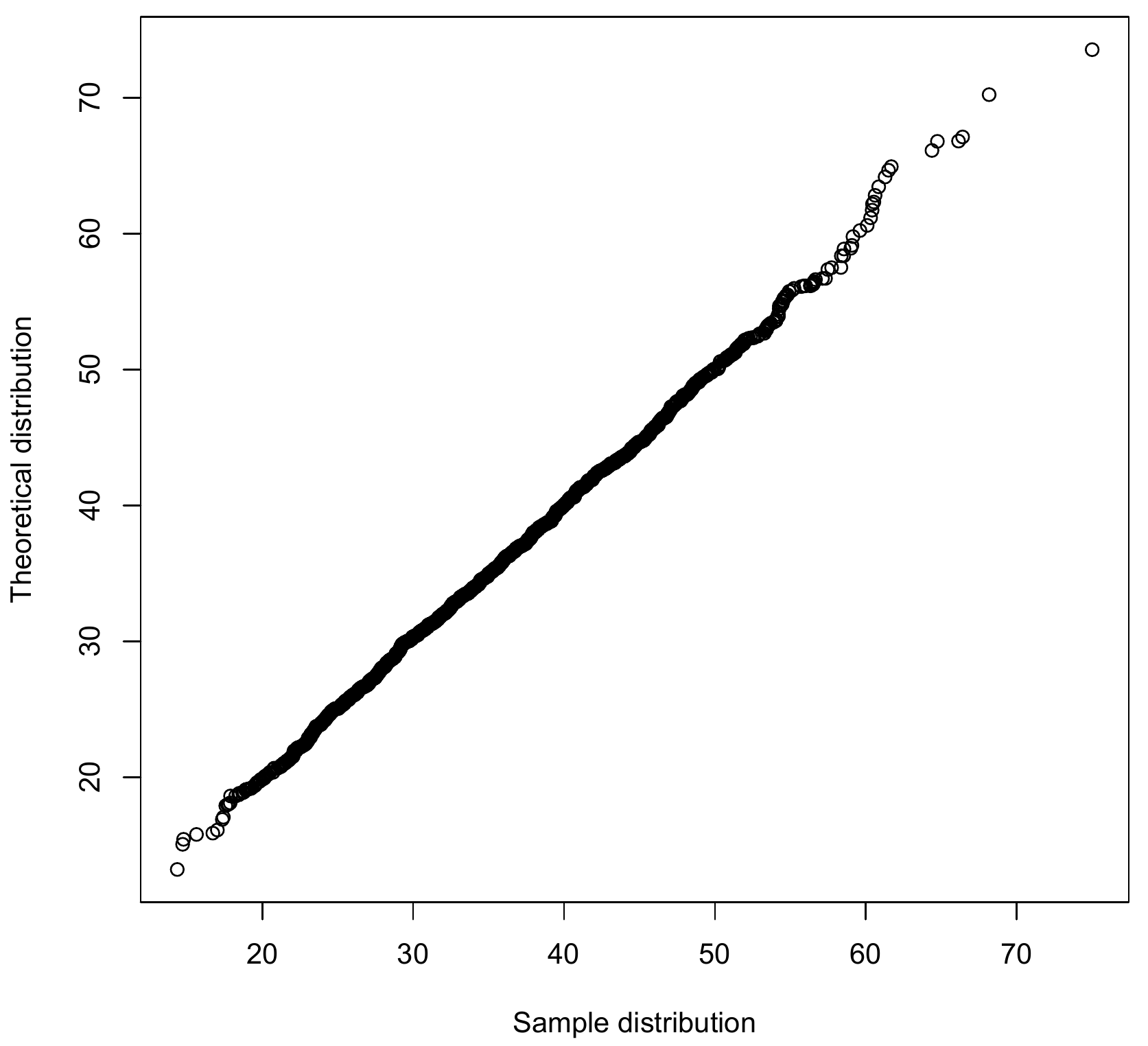}
}\end{subfigure}
\begin{subfigure}[inhomogeneous, without error]{
\includegraphics[height=6.5cm,width=6.5cm,angle=0]{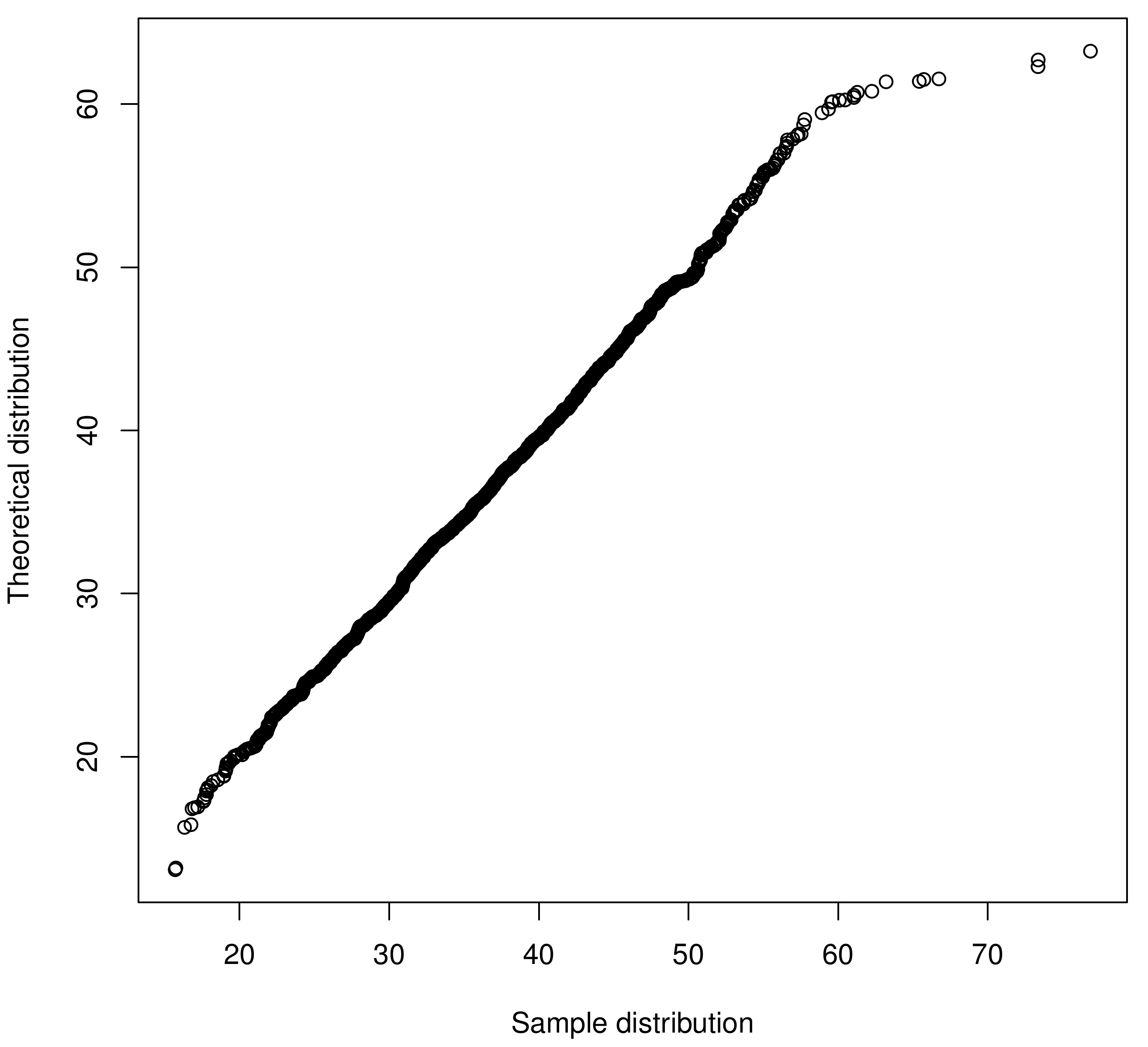}
}\end{subfigure}
\begin{subfigure}[inhomogeneous on GC content]{
\includegraphics[height=6.5cm,width=6.5cm,angle=0]{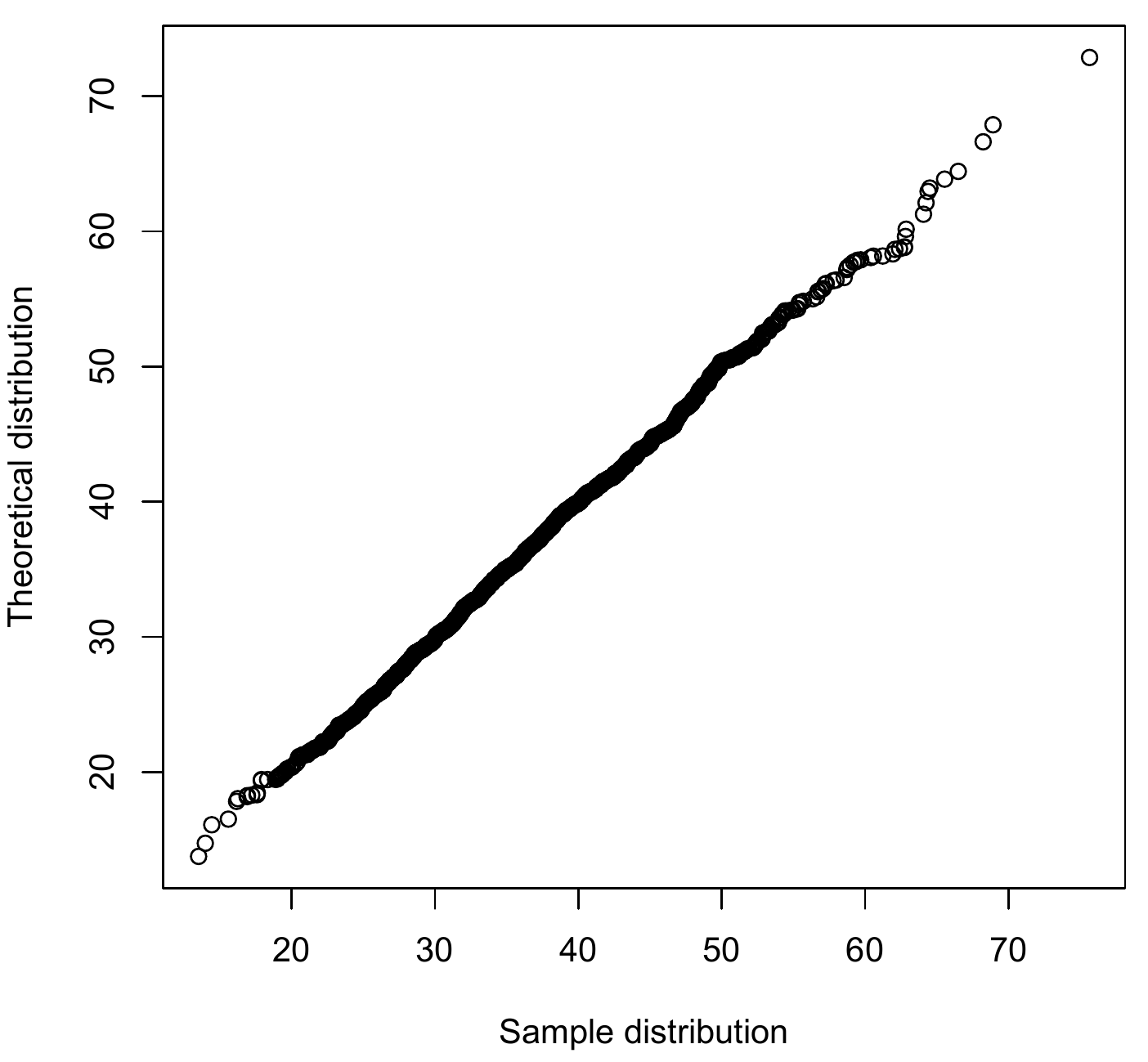}
}\end{subfigure}
\begin{subfigure}[homogeneous, with error]{
\includegraphics[height=6.5cm,width=6.5cm,angle=0]{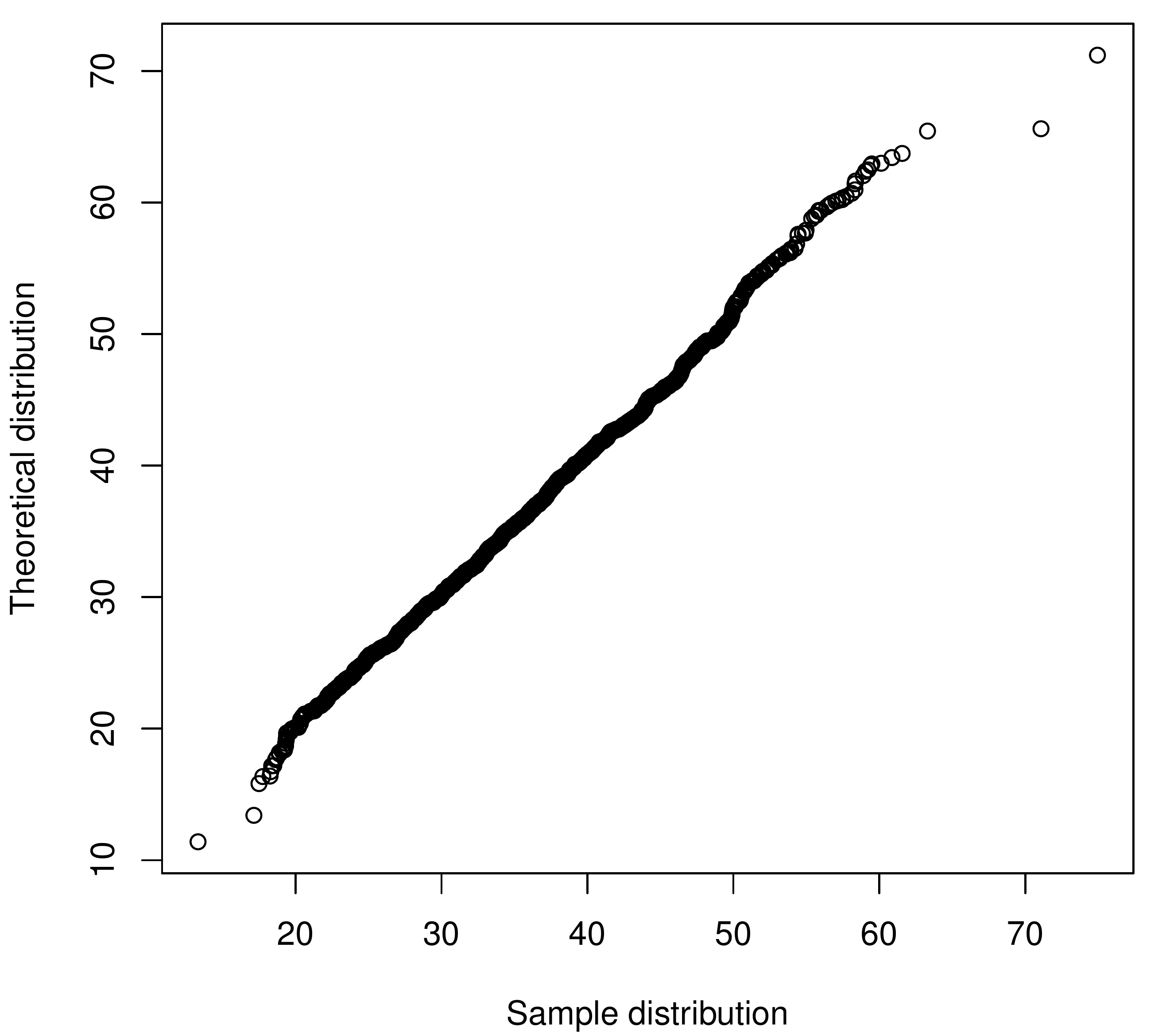}
}\end{subfigure}
\caption{Q-Q plots of the 2000 $S_3^R/d$ scores v.s. 2000 scores sampled from a
$\chi^2_{36}$ distribution; $(G, \beta, M) = (10^5, \ 200, \ 1000)$. {a): homogeneous sampling without error, b): inhomogeneous sampling without error, c): inhomogeneous sampling with sampling rate depending on GC content,  d): homogeneous sampling with error.}}
 \label{qq_inhomo_Sk}
\end{figure}

\begin{figure}
\centering
\scriptsize
\begin{subfigure}[homogeneous, without error]{
\includegraphics[height=6.5cm,width=6.5cm,angle=0]{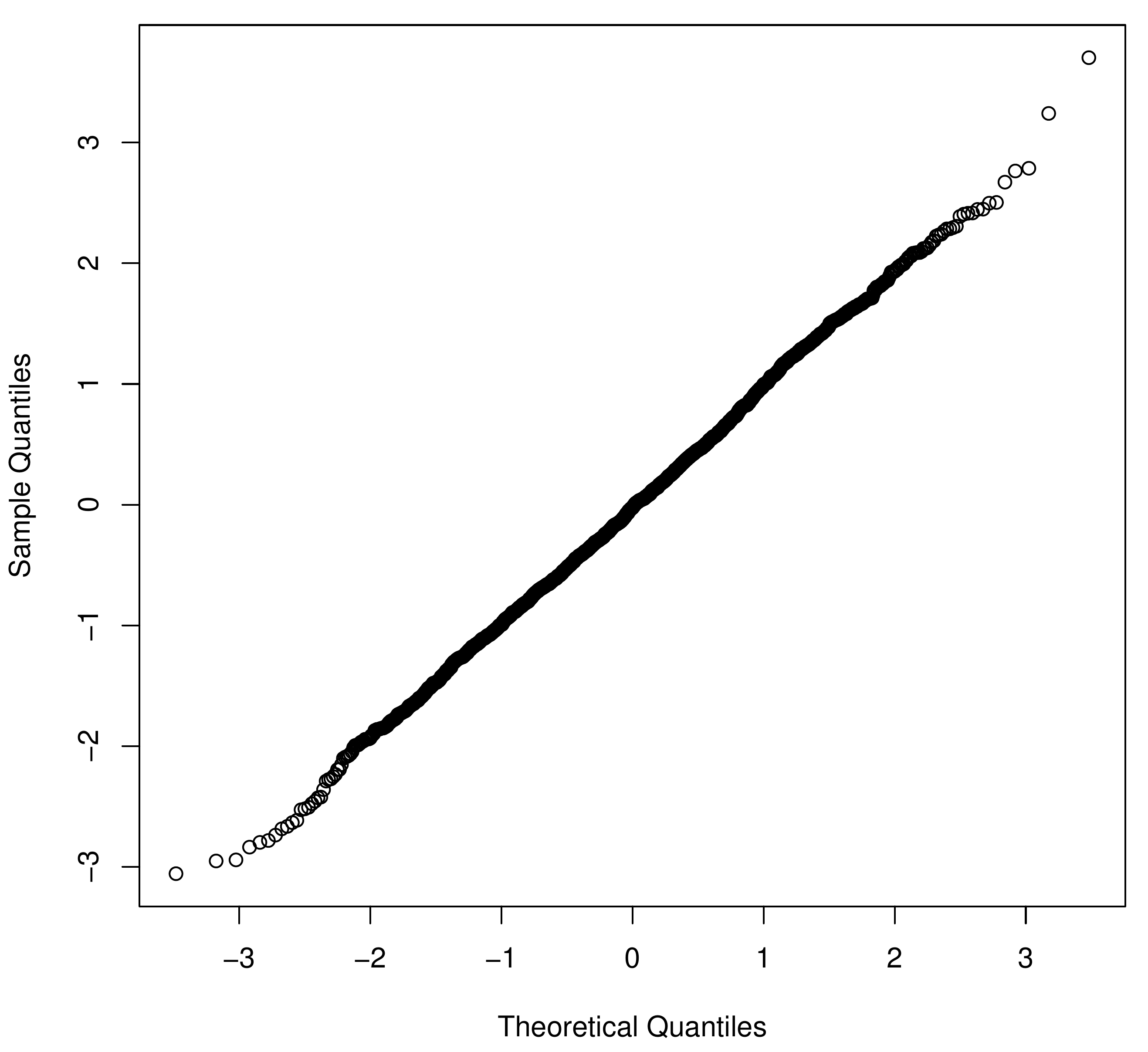}
}\end{subfigure}
\begin{subfigure}[inhomogeneous, without error]{
\includegraphics[height=6.5cm,width=6.5cm,angle=0]{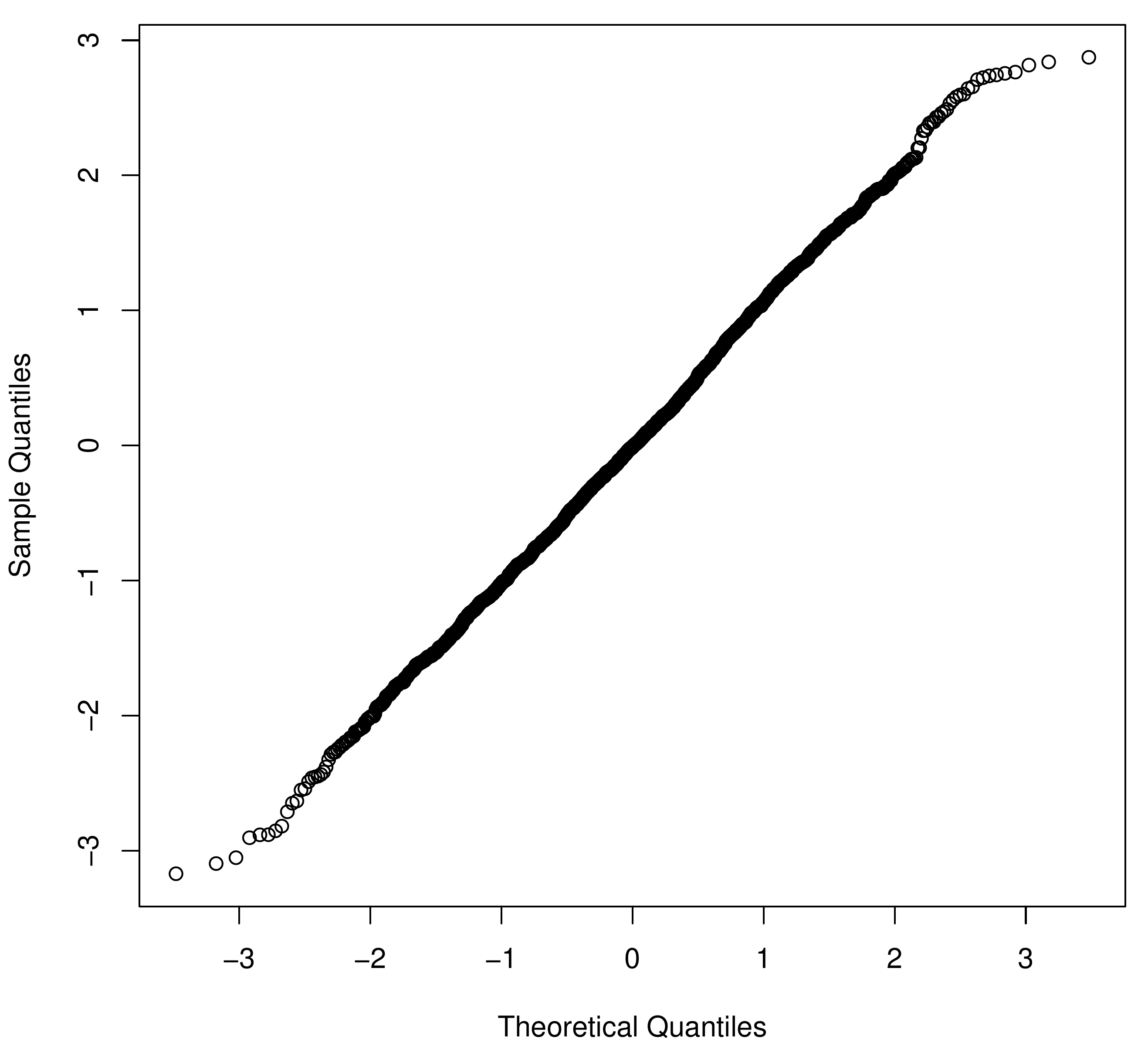}
}\end{subfigure}
\begin{subfigure}[inhomogeneous on GC content]{
\includegraphics[height=6.5cm,width=6.5cm,angle=0]{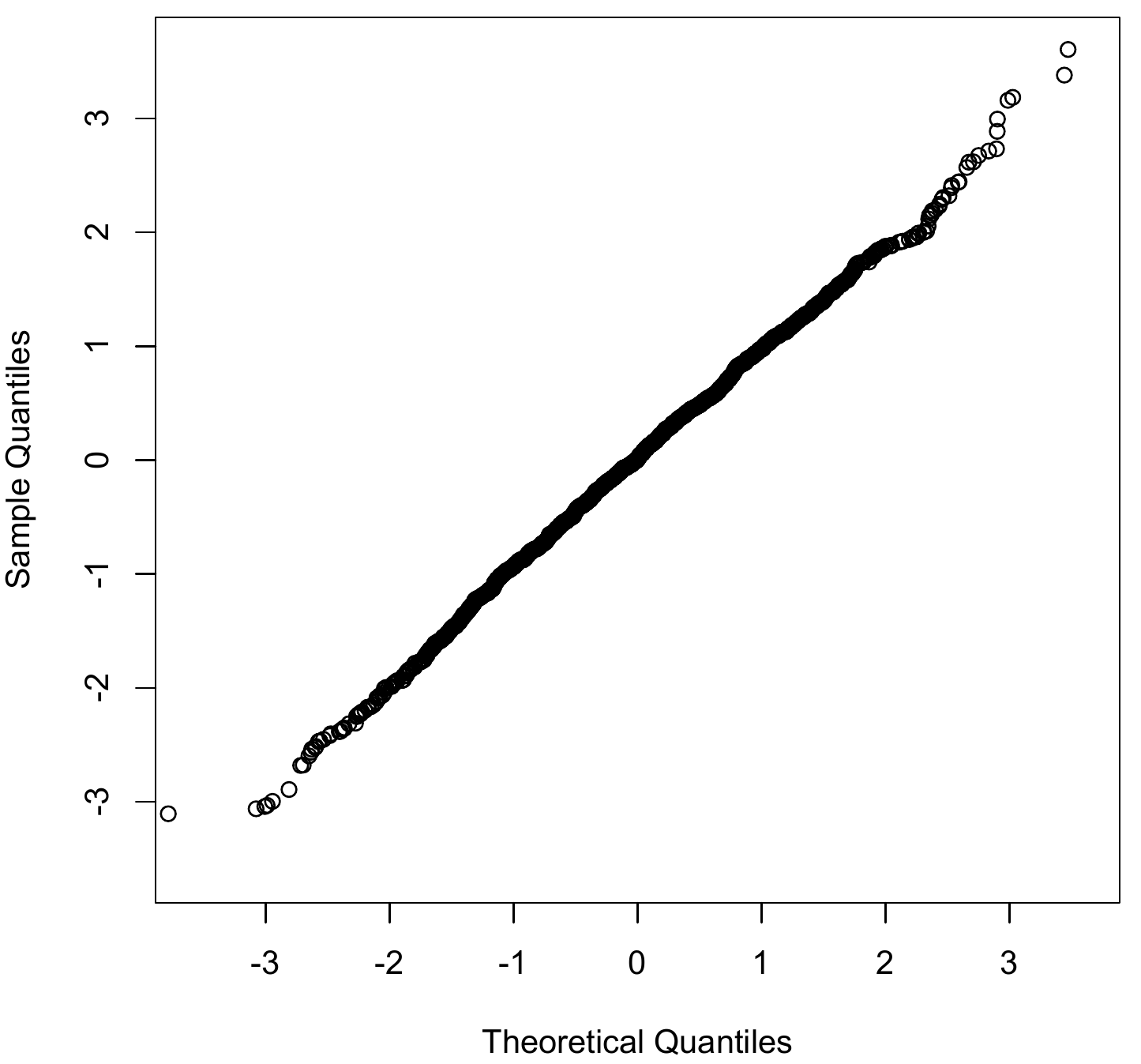}
}\end{subfigure}
\begin{subfigure}[homogeneous, with error]{
\includegraphics[height=6.5cm,width=6.5cm,angle=0]{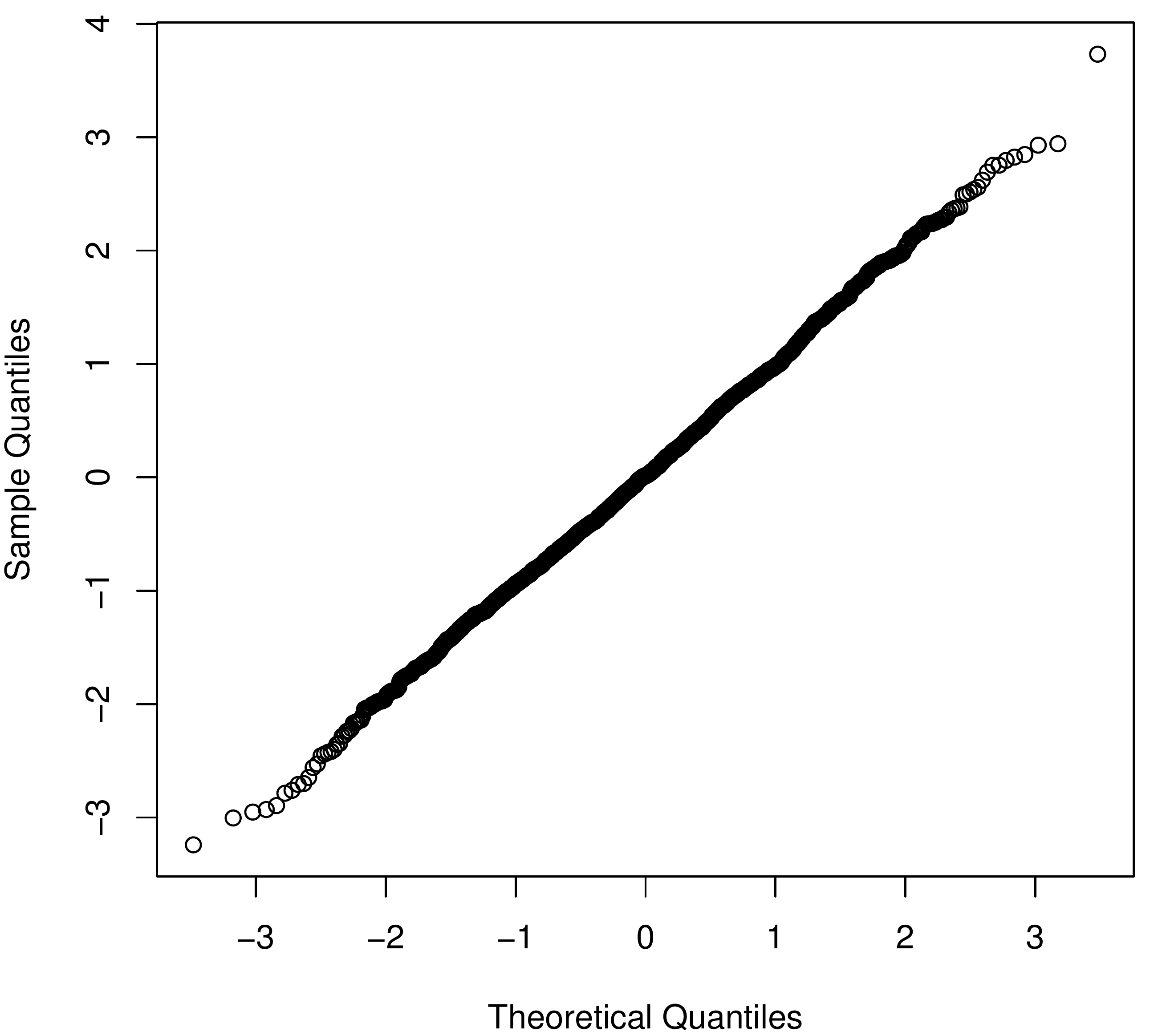}
}\end{subfigure}
\caption{ Q-Q plots of the 2000 $Z^R_{ACT}/\sqrt{d}$ scores of the word $ACT$ v.s.
the 2000 scores sampled from the standard normal distribution; $(G, \beta, M) = (10^5, \ 200, \ 1000)$. {a): homogeneous sampling without error, b): inhomogeneous sampling without error, c): inhomogeneous sampling with sampling rate depending on GC content,  d): homogeneous sampling with error.}}
\label{qq_inhomo_Zk}
\end{figure}

Moreover, Figure \ref{qq_inhomo_Sk}(a) shows typical Q-Q (Quantile-Quantile) plots
of $S_3^R/d$ versus the distribution of $\chi^2_{36}$,
where the subscript 36 indicates the degree of freedom of the
$\chi^2$ distribution, under different models of sampling reads with/without sequencing errors. Similarly, Figure \ref{qq_inhomo_Zk}(a) show the Q-Q plots of
$Z_{ACT}^R/\sqrt{d}$ versus the standard normal distribution under different scenarios; here, the word under consideration is  ${\mathbf{w}}=ACT $. The theoretical approximations work well in this situation.

Simulation studies under the higher order MC model and on genomes inserted with repeated regions are carried out in a similar {{fashion (data not shown)}}. The same conclusion that the theoretical approximate distributions of $S_k^R$ and $Z_\mathbf{w}^R$ fit their simulated distributions well holds.

\subsubsection{Simulation results for Theorem 1: The effect of inhomogeneous sampling and of sequencing error}
\label{subsection:inhomogeneous}

We use Q-Q plots to show the effects of inhomogeneous sampling and sequencing errors on the distribution of $S_k^R$ and $Z_\mathbf{w}^R$ for $(G, \beta, M) = (10^5, \ 200, \ 1000)$.
Figure S1 (a, b, c, d) shows the Q-Q plots of the 2000 $S_3^R/d$ scores from (a) homogeneous sampling, (b) inhomogeneous sampling with rate not depending on GC content, (c) inhomogeneous sampling with rate depending on GC content, and  (d) homogeneous sampling with sequencing errors v.s. 2000 scores sampled from $\chi^2_{36}$ distribution.  The factor $d$ is $c_{\mbox{eff}}+1$ in homogeneous sampling; and  $d$ is calculated from the exact distribution of the sampled reads along the sequence in the inhomogeneous sampling situation. Figure S2 gives the Q-Q plots for $Z_{ACT}^R/\sqrt{d}$, showing the effect of inhomogeneous sampling and sequencing error on the distribution of $Z_{ACT}^R/\sqrt{d}$.
All Q-Q plots show a satisfactory fit, confirming that the theoretical results from Theorem 1 hold even when the assumptions are not necessarily satisfied.




\subsubsection{Simulation results on estimating the order of MCs based on simulated NGS reads}

\begin{figure}
\centering
\begin{subfigure}[a first order MC, $c=1$, homogeneous]
{
\includegraphics[height=6.5cm,width=6.5cm,angle=0]{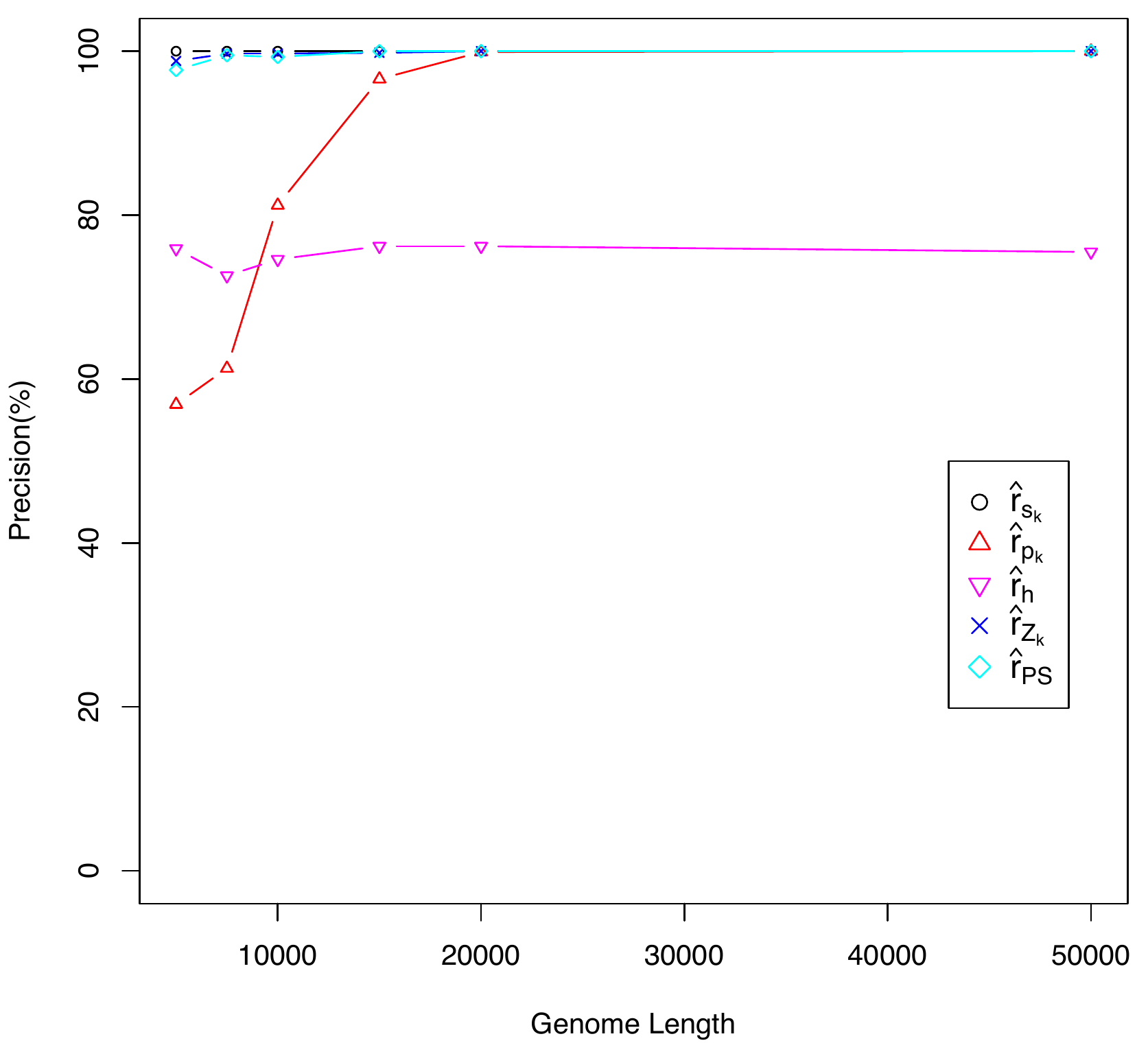}
 }%
\end{subfigure}
\begin{subfigure}[a first order MC, $G=10^5$, homogeneous]
{
\includegraphics[height=6.5cm,width=6.5cm,angle=0]{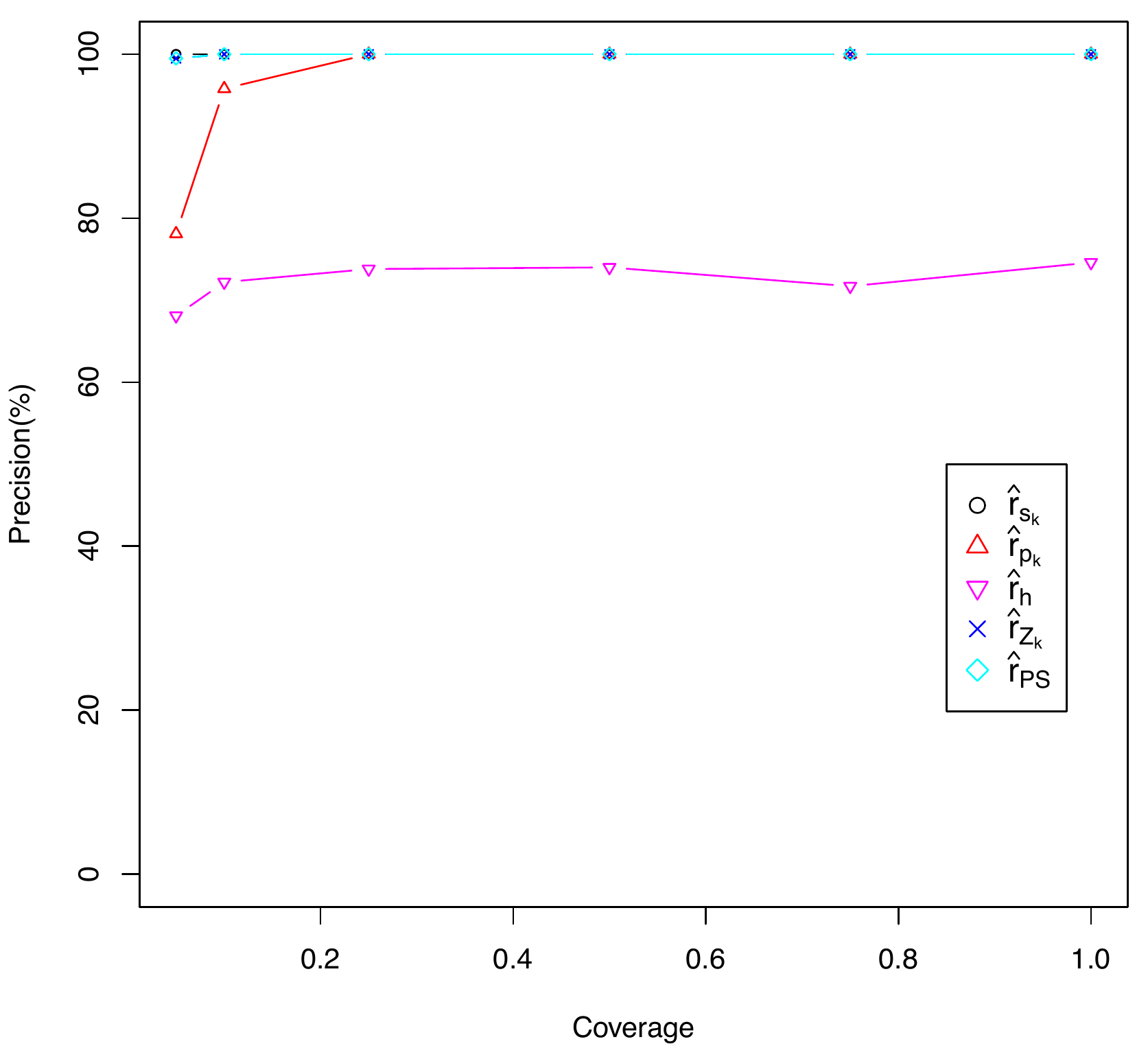}
 }%
\end{subfigure}
\begin{subfigure}[a second order MC, $c=1$, homogeneous]
{
\includegraphics[height=6.5cm,width=6.5cm,angle=0]{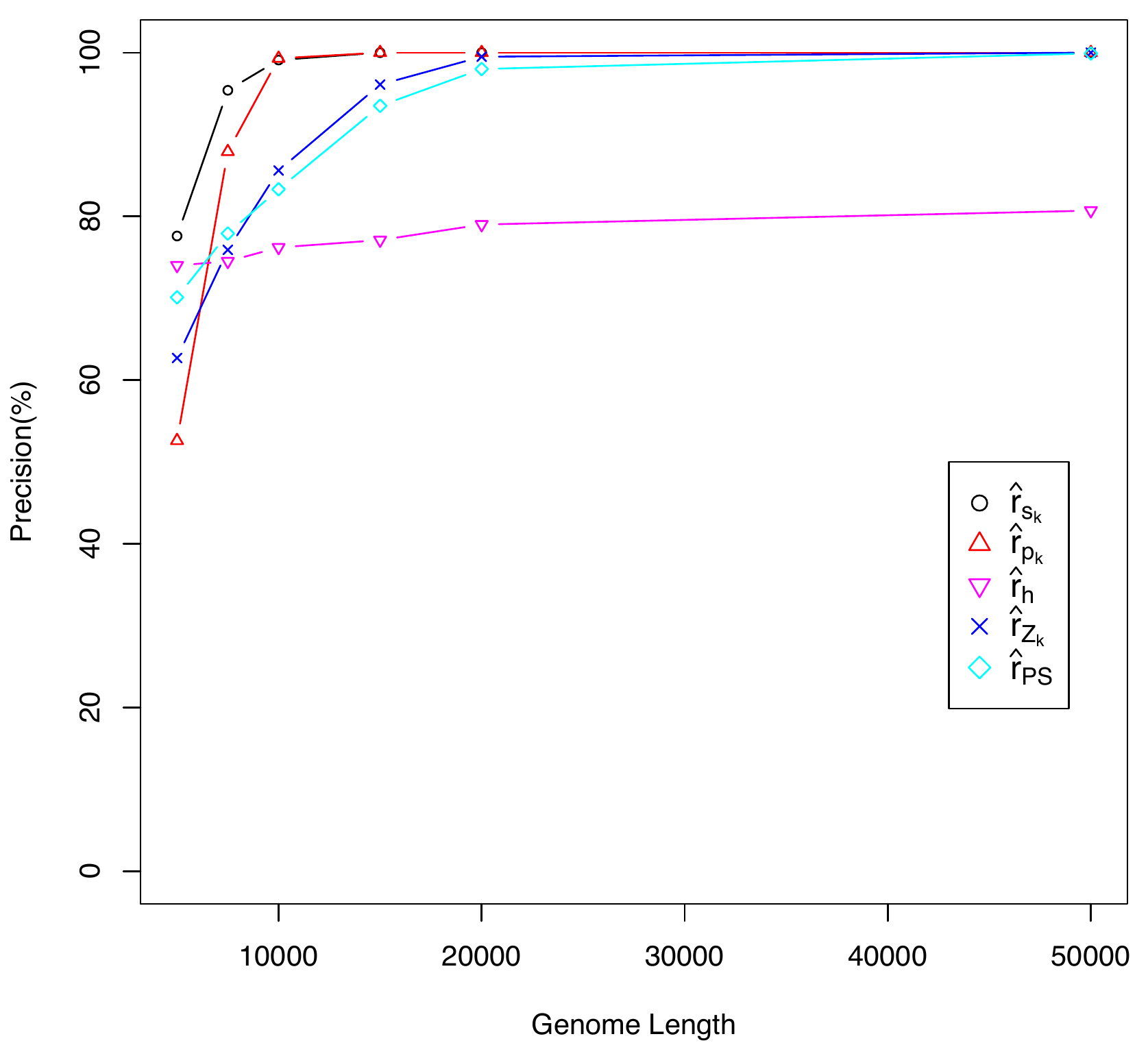}
 }%
\end{subfigure}
\begin{subfigure}[a second order MC, $G=10^5$, homogeneous]
{
\includegraphics[height=6.5cm,width=6.5cm,angle=0]{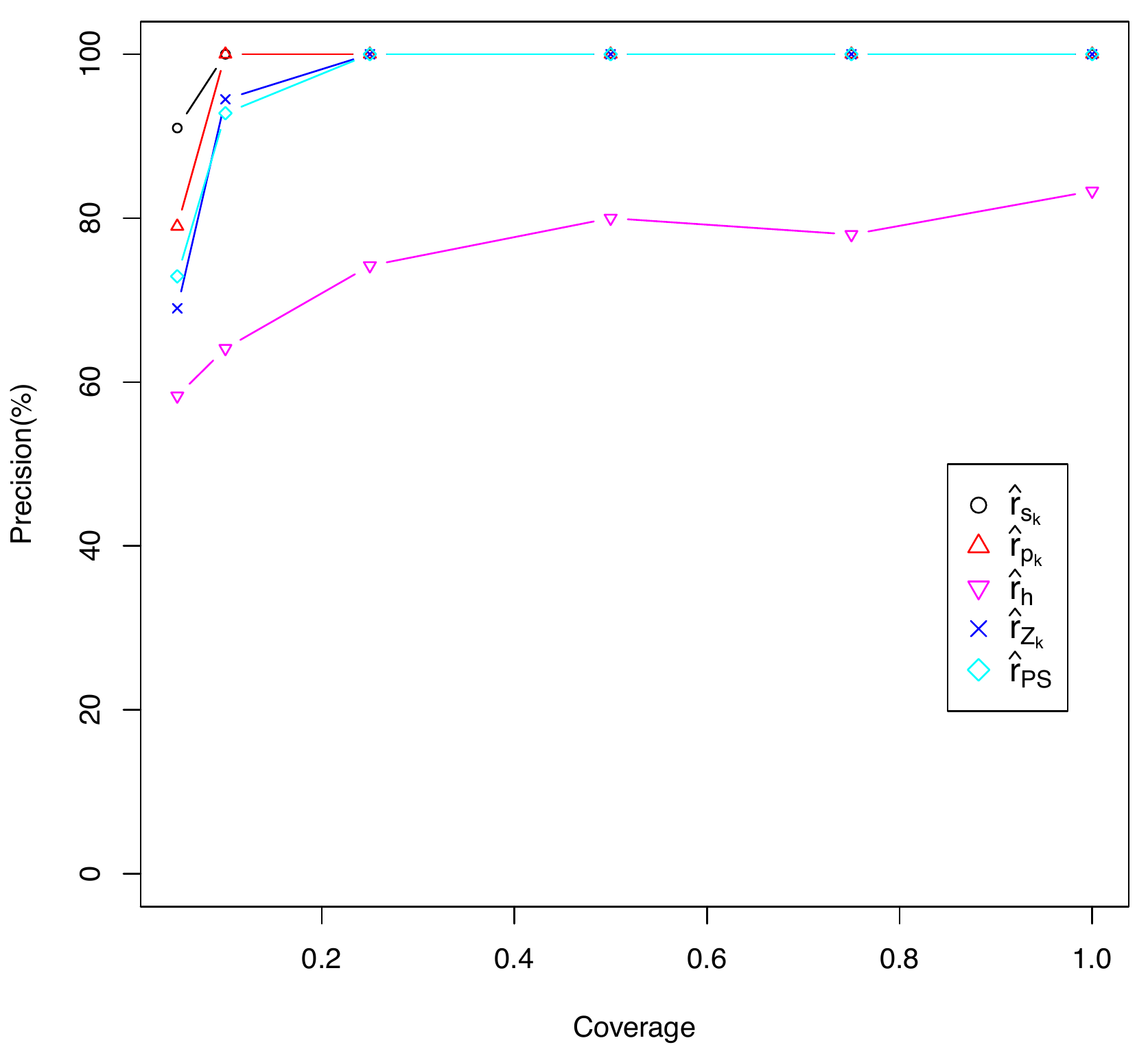}
 }%
\end{subfigure}
\caption{The precision rates of the estimators for the order of a MC under a first and a second order MC, with homogeneous sampling and sequencing error rate 10\%. (a,b): The effects of genome length and read coverage to the precision rates under a first order MC. (c,d): The effects of genome length and read coverage on the precision rates under a second order MC.}
\label{5stat_homo}
\end{figure}

Figure \ref{5stat_homo} shows the effects of sequence length and read coverage on the precision of the estimators under a first and second order MCs, with homogeneous sampling and sequencing error rate 10\%.
It can be seen that all the five estimators perform reasonably well. In particular, the precision rate of estimators $\hat{r}_{S_k}$, $\hat{r}_{p_k}$, $\hat{r}_{Z_k}$ and $\hat{r}_{PS}$ reach 100\% when the genome length is larger than 20000 bps and the read coverage is greater than 0.2. The estimator $\hat{r}_h$ performs slightly worse than the other four estimators. Since the estimator $\hat{r}_h$ is based on hypothesis testing with a given significant level $\alpha$, the precision rate is not able to reach 100\%. It is also possible that no $k$ from 2 to 6 satisfies $p_{k-1}<\alpha$ and both $p_k$ and $p_{k+1}$ are larger than $\alpha$ such that the estimator $\hat{r}_h$ fails to give an estimation.
We observe that the precision of $\hat{r}_h$ is sensitive to the accuracy of the estimation of the effective coverage $d$. In the simulation, if we take the underlying true value of $d$ in place of the estimated value of $\hat{d}$ in the computation, the precision rate of  $\hat{r}_h$ goes up to above 90\%.

\begin{figure}
\centering
\begin{subfigure}[a first order MC, $c=1$, inhomogeneous]{
\includegraphics[height=6.5cm,width=6.5cm,angle=0]{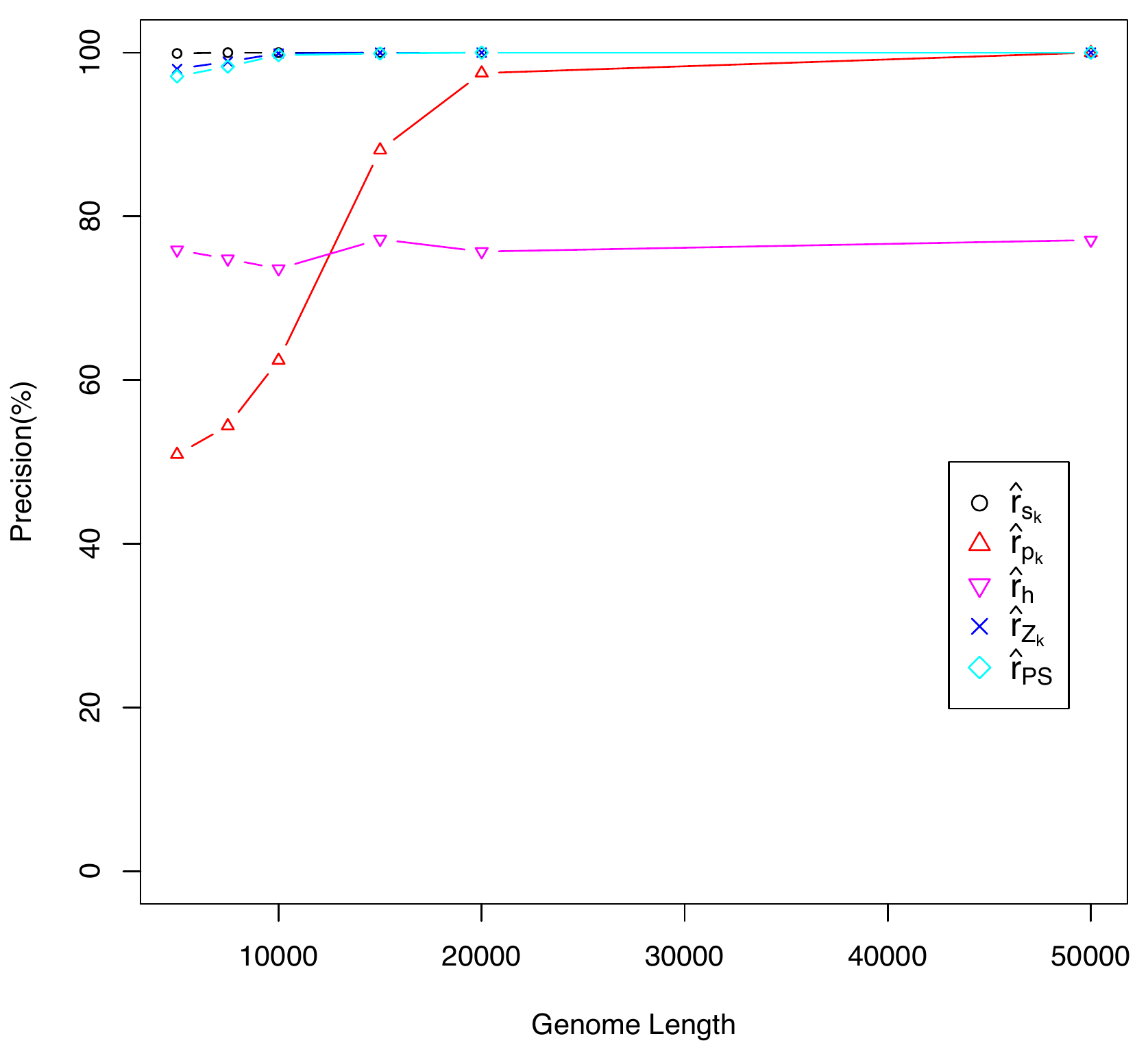}}
\end{subfigure}
\begin{subfigure}[a first order MC, $G=10^5$, inhomogeneous]{
\includegraphics[height=6.5cm,width=6.5cm,angle=0]{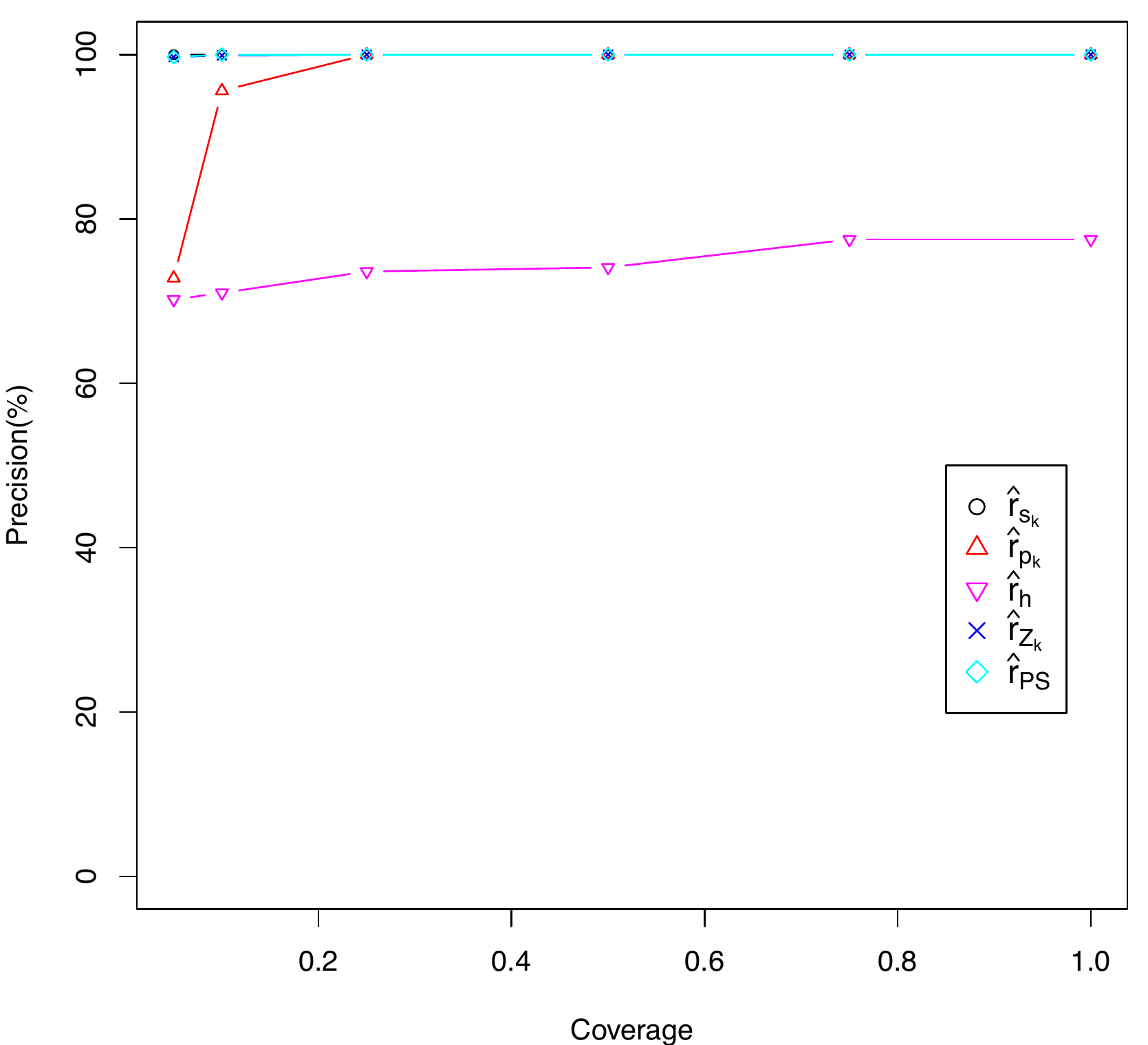}}
\end{subfigure}
\begin{subfigure}[a secomd order MC, $c=1$, inhomogeneous]{
\includegraphics[height=6.5cm,width=6.5cm,angle=0]{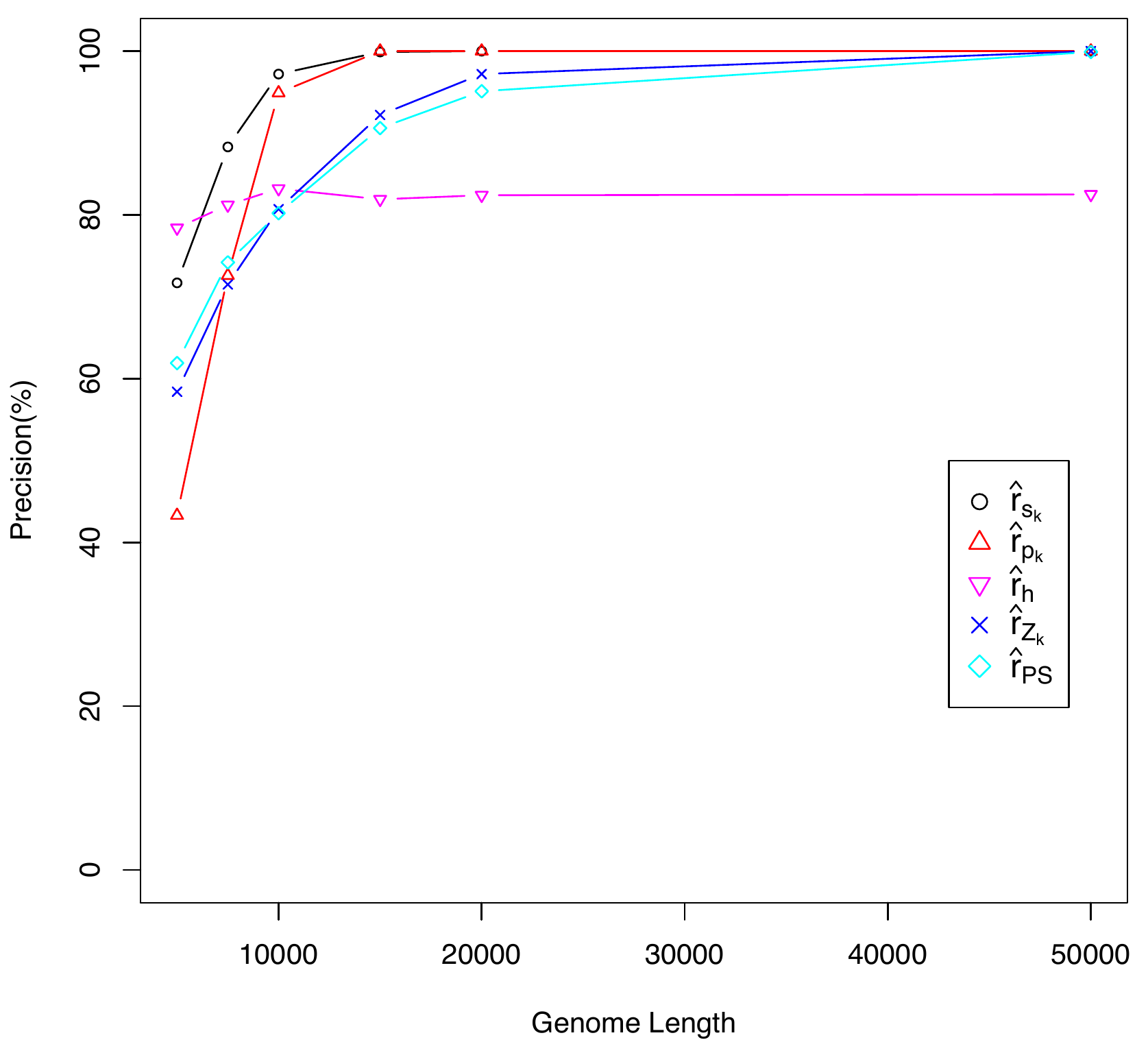}}
\end{subfigure}
\begin{subfigure}[a second order MC, $G=10^5$, inhomogeneous]{
\includegraphics[height=6.5cm,width=6.5cm,angle=0]{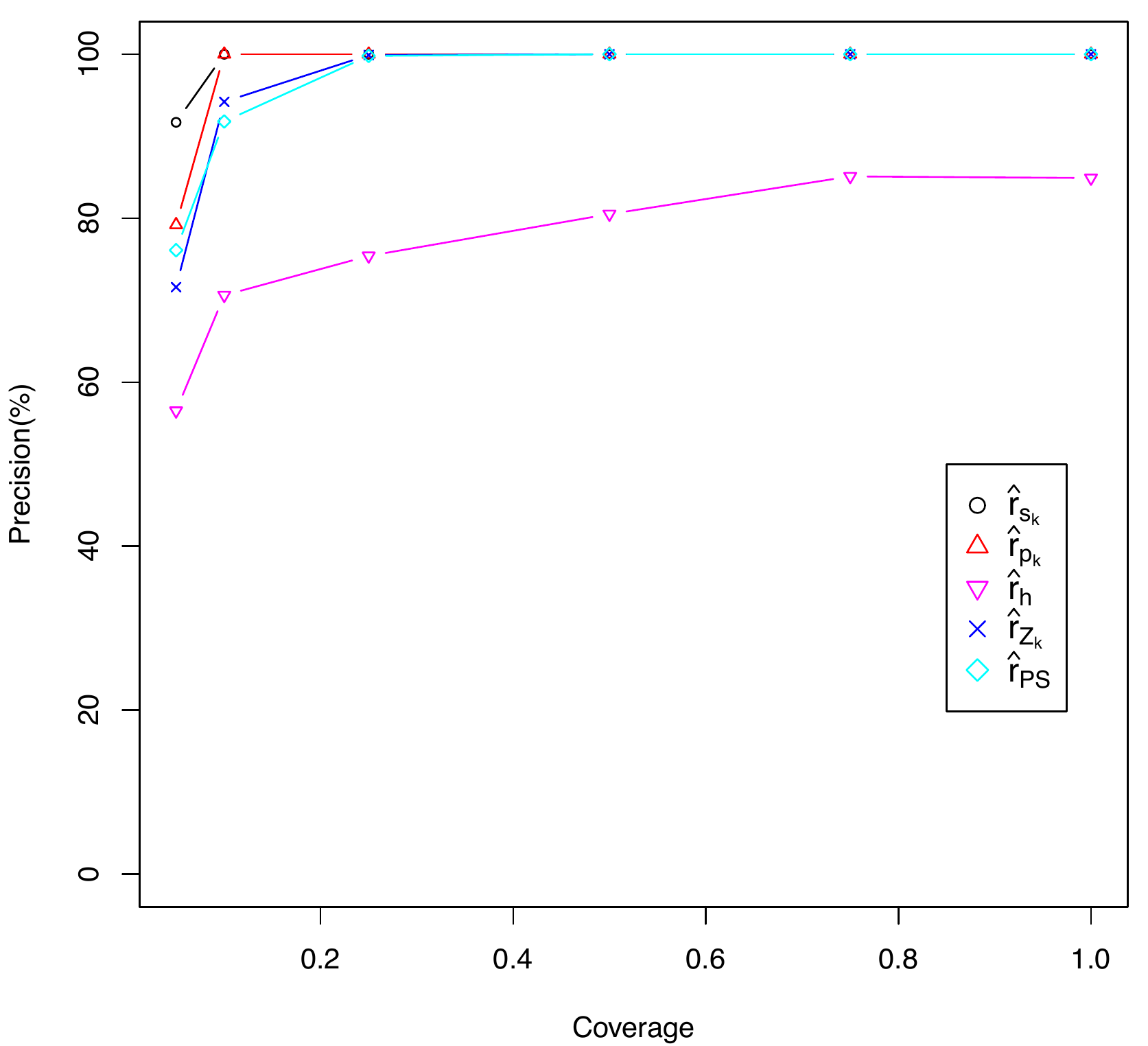}}
\end{subfigure}
\caption{The precision rates of the estimators for the order of a MC under a first and a second order MC, with inhomogeneous sampling and sequencing error rate 10\%. (a,b): The effects of genome length and read coverage to the precision rates under a first order MC. (c,d): The effects of genome length and read coverage on the precision rates under a  second order MC.}
\label{5stat_inhomo}
\end{figure}

For inhomogeneous sampling, Figure \ref{5stat_inhomo} shows the effects of sequence length and read coverage on the precision of the estimators. We can see that the precision rates under the inhomogeneous sampling start at slightly lower values than those under the homogeneous case. As the genome length and the read coverage increase, the estimators perform as well as they perform under the homogeneous sampling.

\begin{figure}
\centering
\begin{subfigure}[a first order MC, $c=1$, homogeneous]{
\includegraphics[height=6.5cm,width=6.5cm,angle=0]{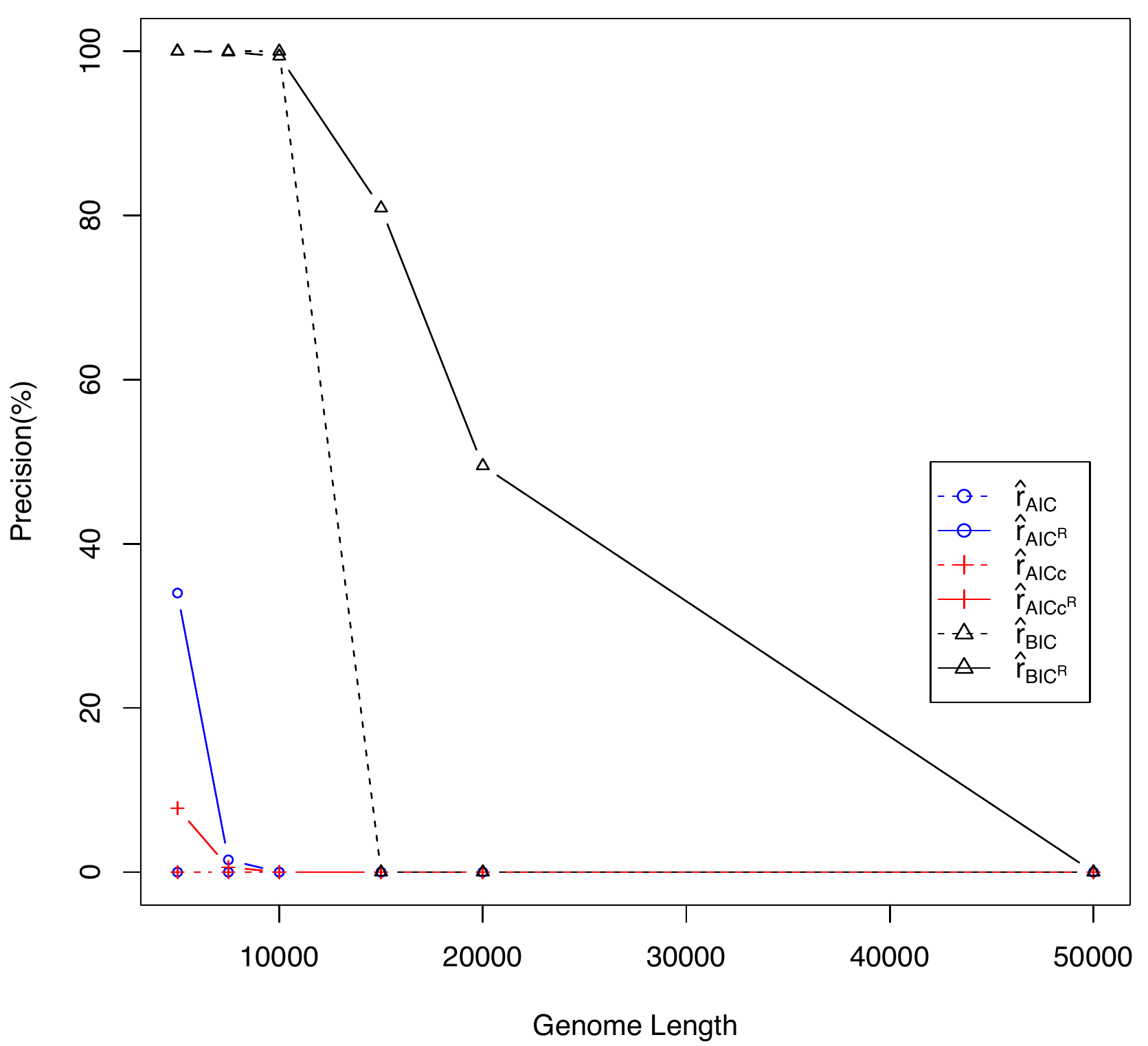}
}\end{subfigure}
\begin{subfigure}[a first order MC, $G=10^5$, homogeneous]{
\includegraphics[height=6.5cm,width=6.5cm,angle=0]{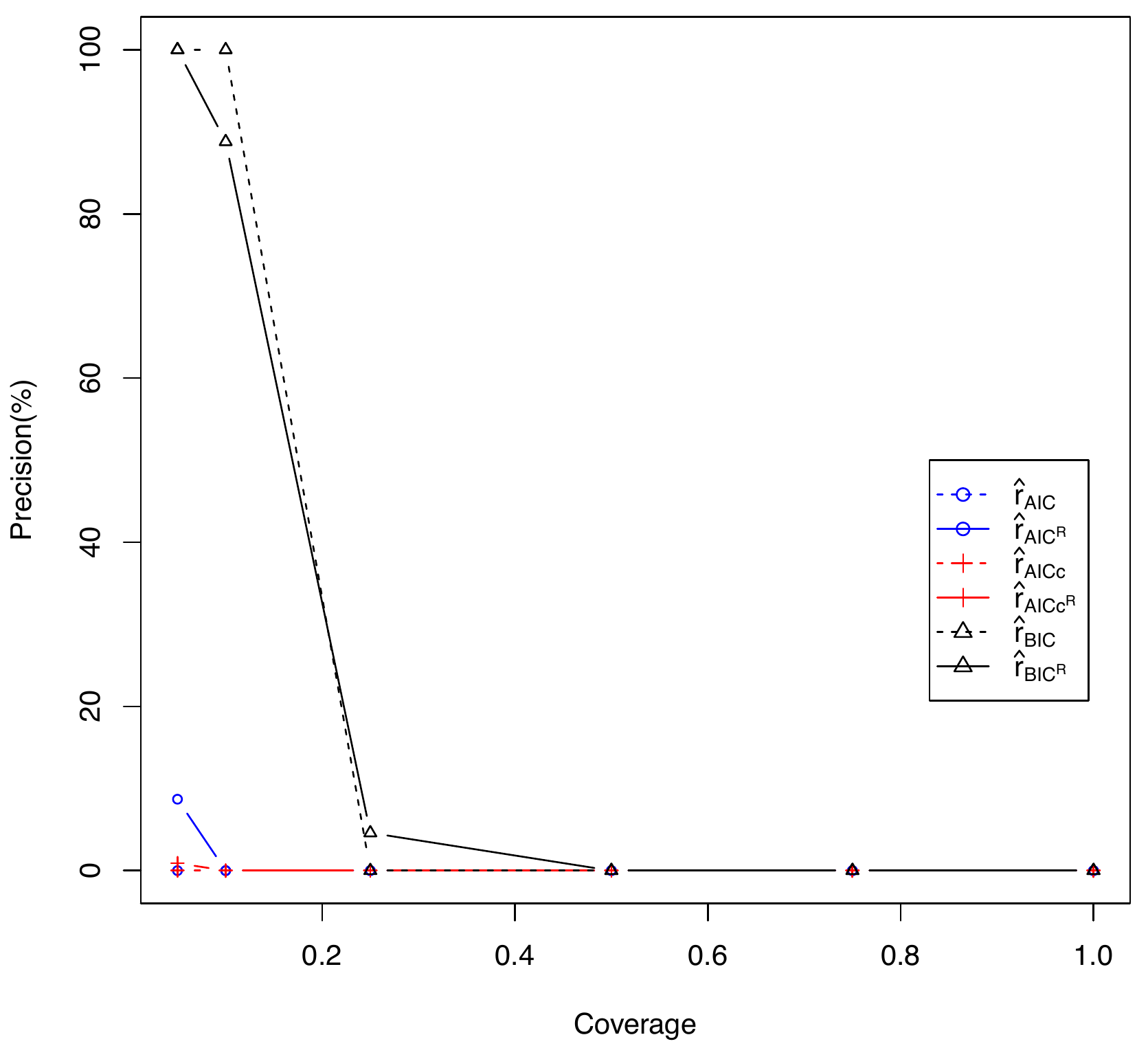}
}\end{subfigure}
\begin{subfigure}[a second order MC, $c=1$, homogeneous]{
\includegraphics[height=6.5cm,width=6.5cm,angle=0]{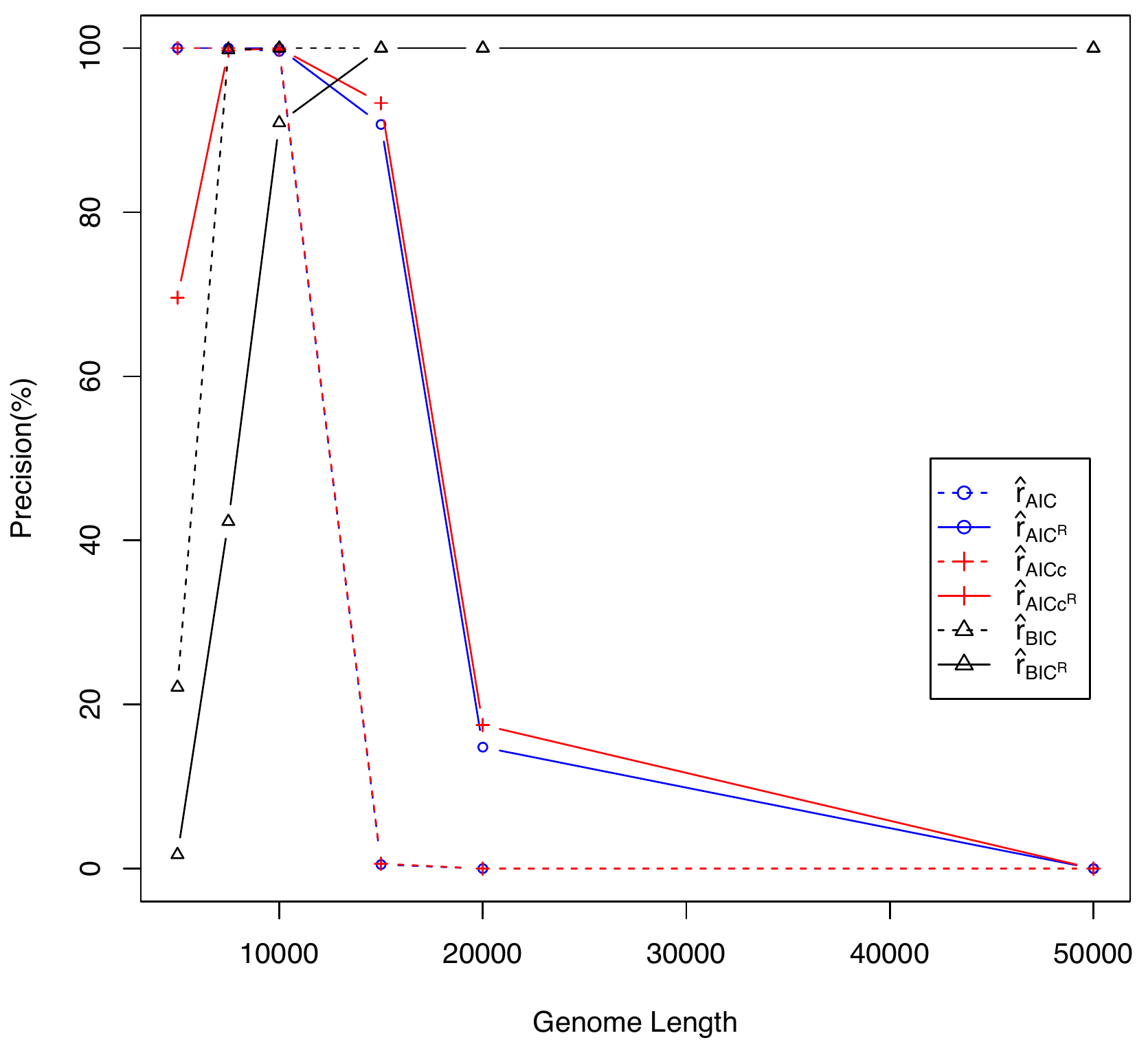}
}\end{subfigure}
\begin{subfigure}[a second order MC, $G=10^5$, homogeneous]{
\includegraphics[height=6.5cm,width=6.5cm,angle=0]{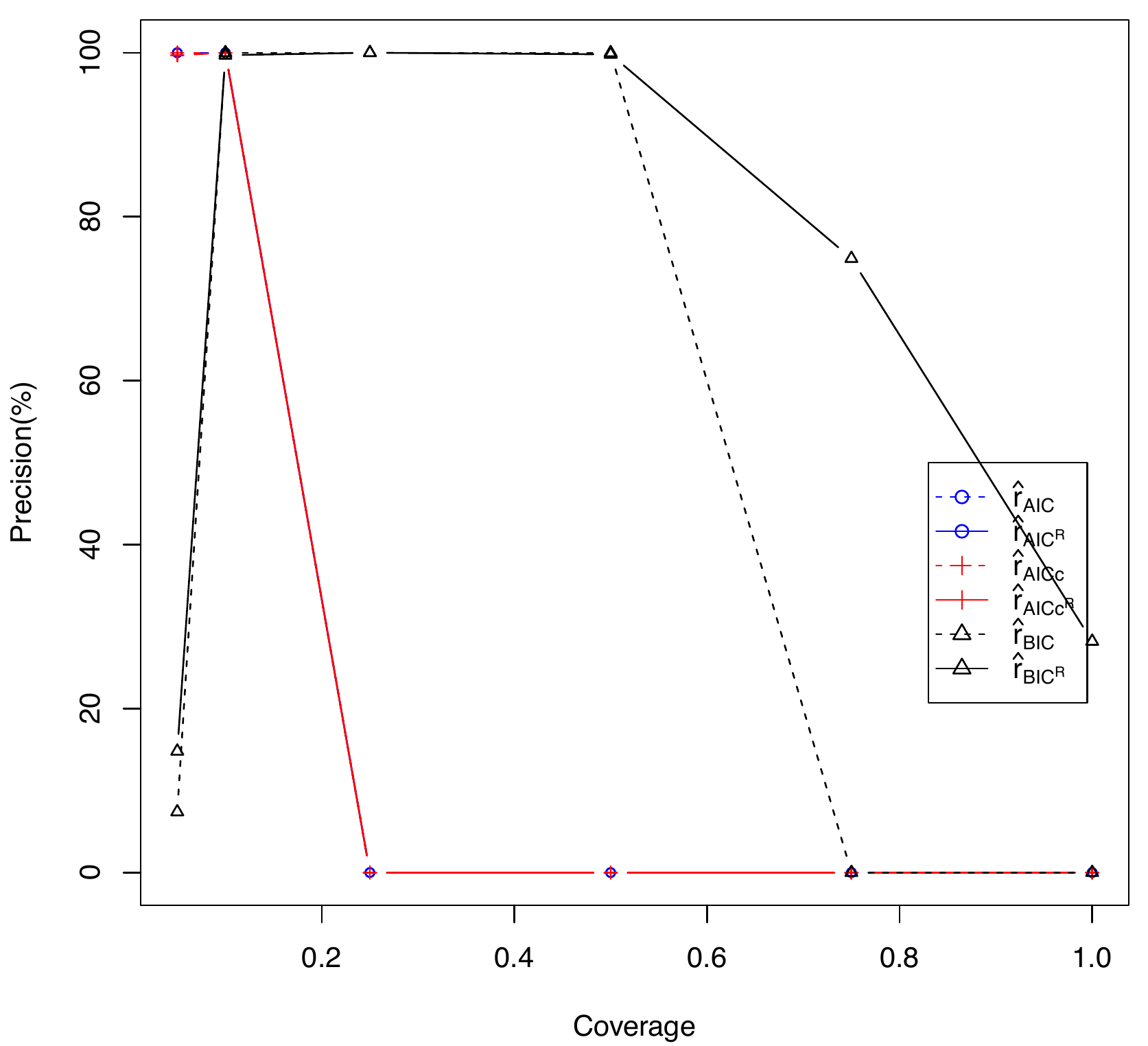}
}\end{subfigure}
\caption{The precision rates of the AIC and BIC based estimators for the order of MC under a first and a second order MC, with homogeneous sampling and sequencing error rate 10\%. (a,b): The effects of genome length and read coverage to the precision rates under a first order MC. (c,d): The effects of genome length and read coverage on the precision rates under a second order MC.}
\label{AICBICstat_homo}
\end{figure}

\begin{figure}
\centering
\begin{subfigure}[a first order MC, $c=1$, inhomogeneous]{
\includegraphics[height=6.5cm,width=6.5cm,angle=0]{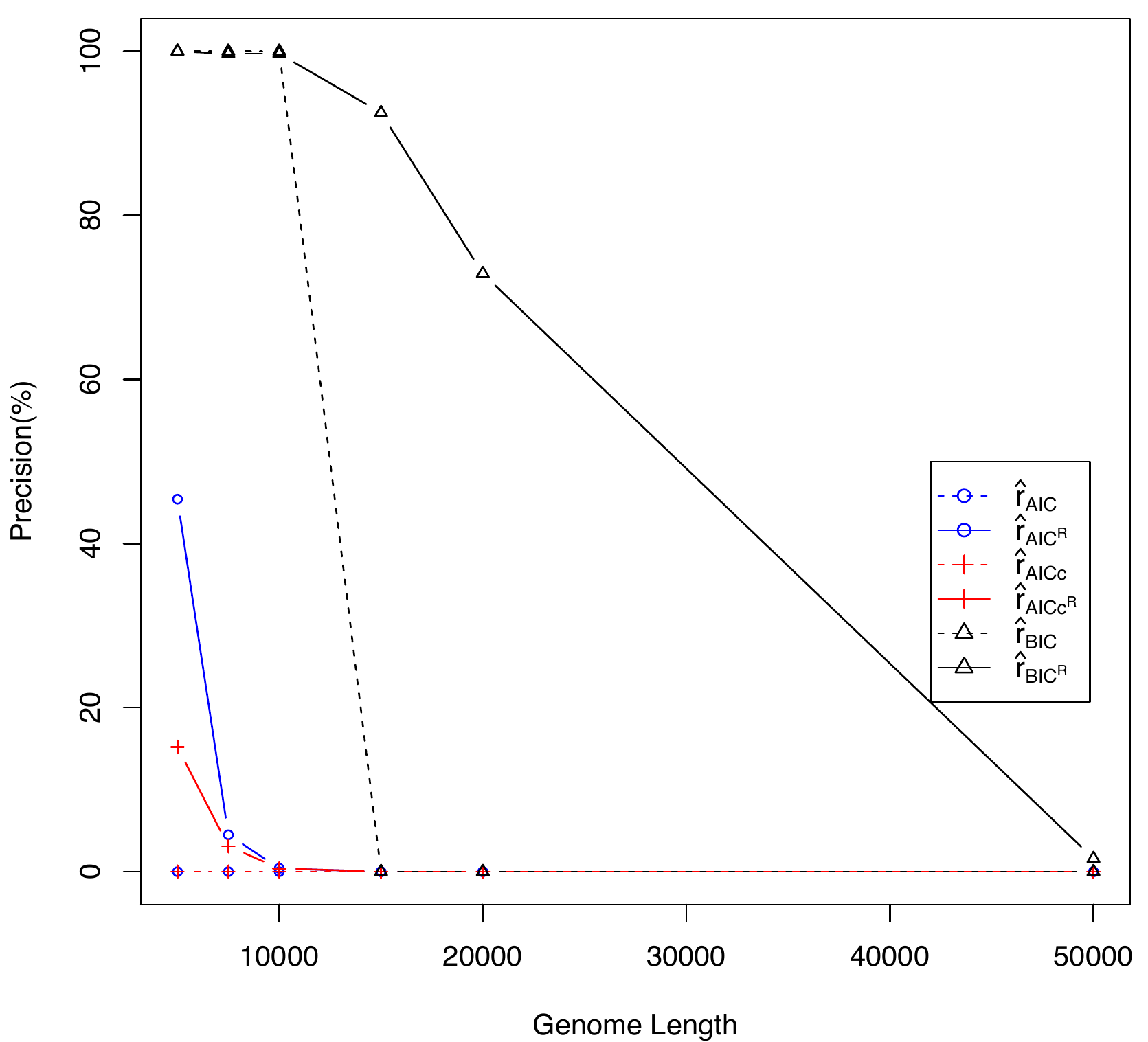}
}\end{subfigure}
\begin{subfigure}[a first order MC, $G=10^5$, inhomogeneous]{
\includegraphics[height=6.5cm,width=6.5cm,angle=0]{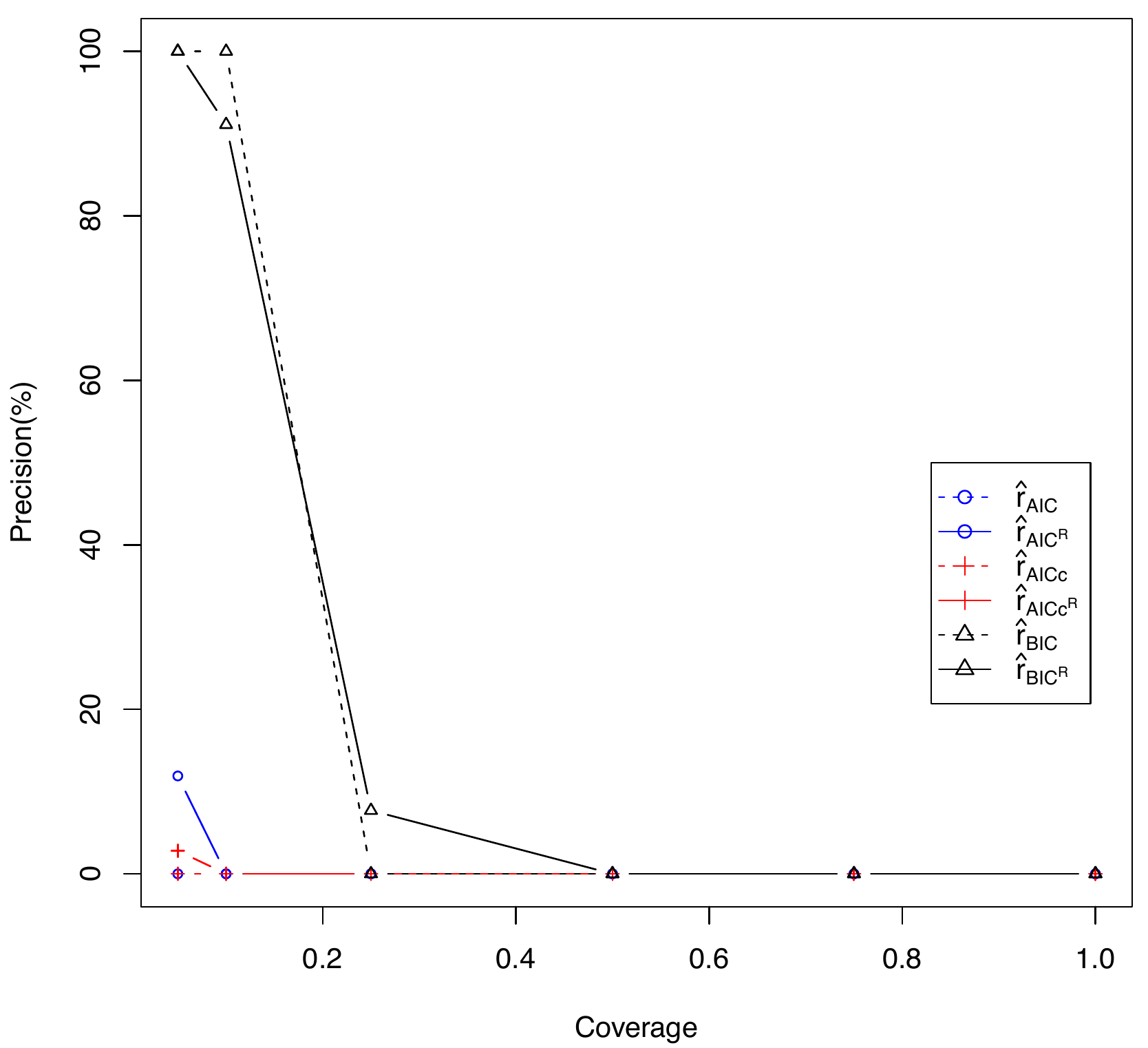}
}\end{subfigure}
\begin{subfigure}[a second order MC, $c=1$, inhomogeneous]{
\includegraphics[height=6.5cm,width=6.5cm,angle=0]{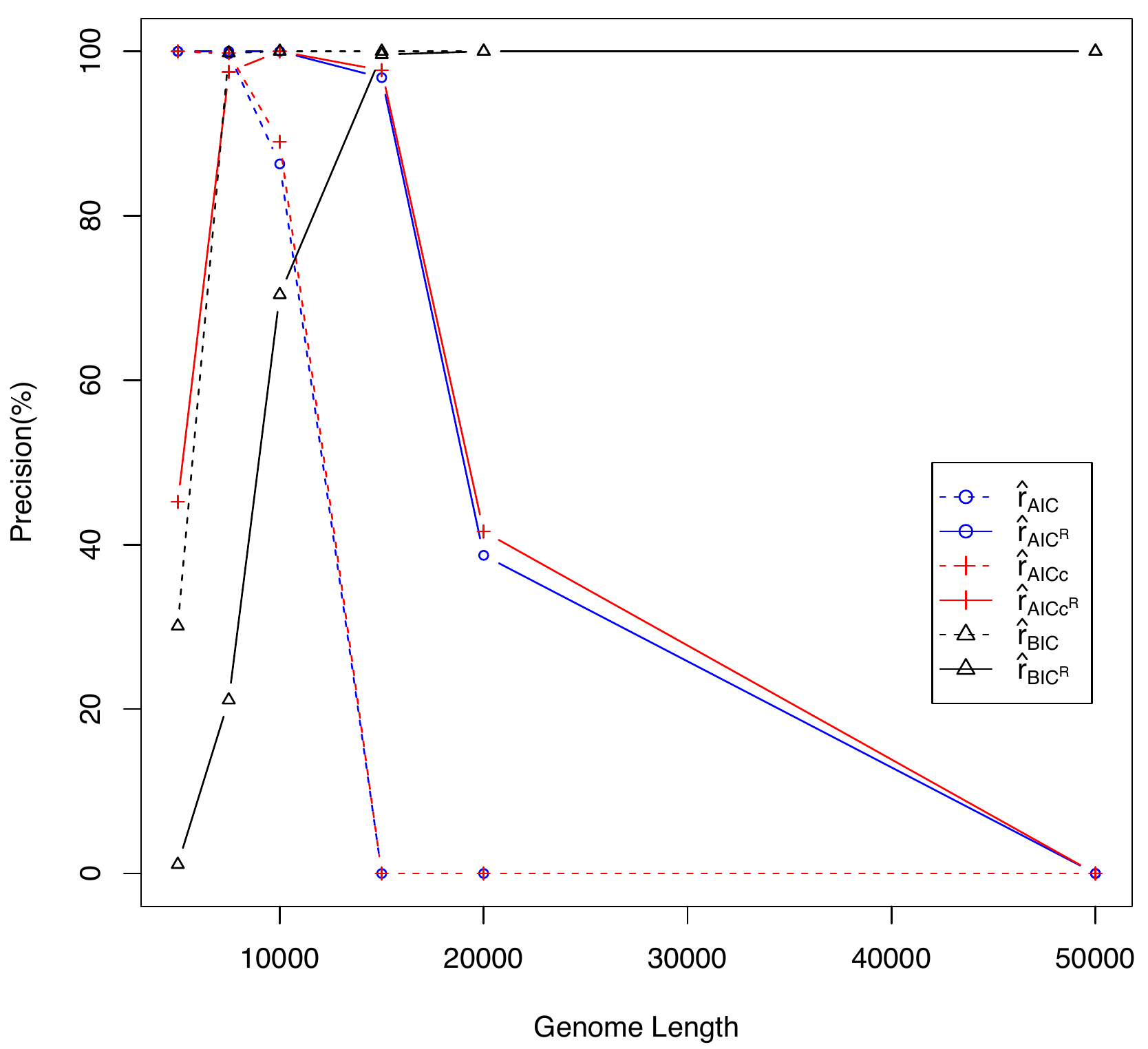}
}\end{subfigure}
\begin{subfigure}[a second order MC, $G=10^5$, inhomogeneous]{
\includegraphics[height=6.5cm,width=6.5cm,angle=0]{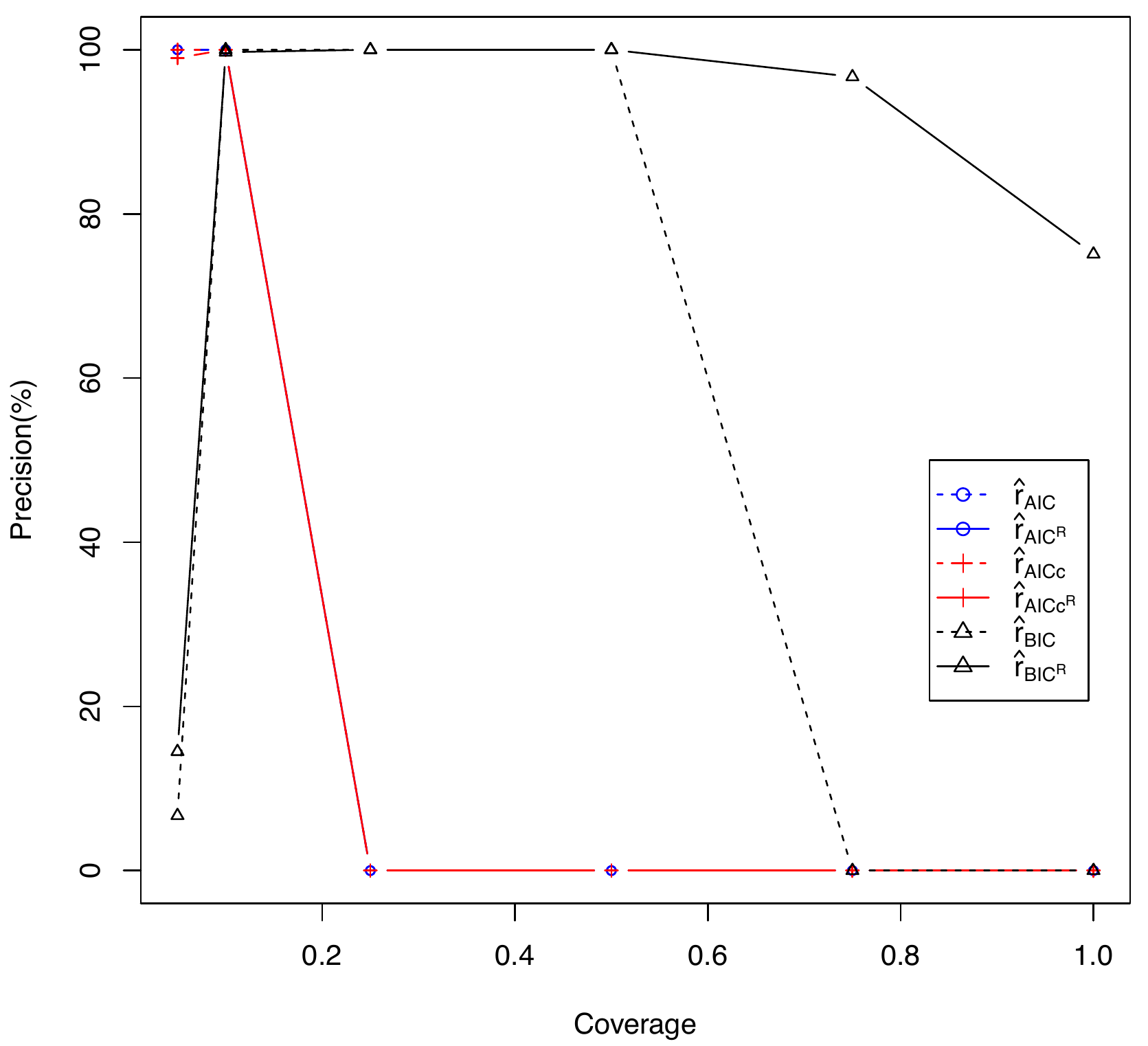}
}\end{subfigure}
\caption{The precision rates of the AIC and BIC based estimators for the order of MC under a first and a second order MC, with inhomogeneous sampling and sequencing error rate 10\%. (a,b): The effects of genome length and read coverage to the precision rates under a first order MC. (c,d): The effects of genome length and read coverage on the precision rates under a  second order MC.}
\label{AICBICstat_inhomo}
\end{figure}

Figure \ref{AICBICstat_homo} and \ref{AICBICstat_inhomo} show the {{disappointing}} precision rates of the AIC and BIC based estimators of the order of a MC under a first and a second order MC, with homogeneous (Figure \ref{AICBICstat_homo}) and inhomogeneous sampling (Figure \ref{AICBICstat_inhomo}). The sequencing error rate is 10\% and read length is 100.
It can be seen that the normalized estimators $\hat{r}_{AIC^R}$, $\hat{r}_{AICc^R}$ and $\hat{r}_{BIC^R}$ (in solid lines) have better performance than the their corresponding naive estimators $\hat{r}_{AIC}$, $\hat{r}_{AICc}$ and $\hat{r}_{BIC}$ (in dotted lines). The BIC based estimator, $\hat{r}_{BIC^R}$, has generally the best performance among all the AIC and BIC based statistics. However, the precision of $\hat{r}_{BIC^R}$ is low compared to the five estimators $\hat{r}_{S_k}$, $\hat{r}_{p_k}$, $\hat{r}_{h}$, $\hat{r}_{Z_k}$, and $\hat{r}_{PS}$. With the increase of the genome length and read coverage, although $\hat{r}_{BIC^R}$ is able to reach a very high precision rate at some point, the precision finally drops with further increase of the genome length and the read coverage, and it fails to give a consistent estimation.
The AIC and AICc based estimators, $\hat{r}_{AIC}$ and $\hat{r}_{AICc}$, do not differ much in the precision rate.

\subsubsection{Simulation results on estimating the effective coverage  $d$}
\label{subsection:d-estimation}
For inhomogeneous sampling of the reads, it is not clear how we estimate the parameter $d$ in Theorem 1 based purely on the naive read coverage $c$. In equation (7) in the main text, we propose a method to estimate $d$ based on the statistics $(Z_\mathbf{w}^R)^2, \mathbf{w} \in {\mathcal A}^k$.  We assess its accuracy using the average relative error, defined by $\mathrm{MRE} = \frac{1}{T} \sum_{i}^T \frac{|\hat{d}_i - d|}{d}$, where $\hat{d}_i$ is the estimated value of $d$ in the $i$-th simulation out of the total $T$ repeats, and by the  root-mean-relative-squared-error (RMRSR)  defined as
$ \mathrm{RMRSR} = \frac{1}{d} \sqrt{\sum_i^T (\hat{d}_i - d)^2/T}. $

The values of $d$ for the four combinations of $(\beta, M)$ and the average of their estimations are given in the first and second rows of Table \ref{estimate-d}. The MRE and RMRSR for the different combinations of $(\beta, M)$ are given in the third and fourth rows of Table \ref{estimate-d}. Table \ref{estimate-d} shows the results for (a) the homogeneous and (b) the inhomogeneous sampling of the reads using $k=3$. In general, although the estimation of the effective coverage $d$ becomes slightly inaccurate as the read coverage increases, the estimation is reasonably good. Under the inhomogeneous sampling, the relative errors are slightly higher than those under the homogeneous case, which reflects the greater randomness the inhomogeneous sampling brings into the model; while the MRE error seems unaffected by the choice of $d$, the RMRSR shows a tendency to increase with increasing $d$.

We only consider $(\beta, M) = (100, 2000)$ for inhomogeneous sampling with sampling rate depending on GC content as before. In this case, the value of $d$ is 2.08. The mean value of  estimated $d$ is 2.16, the MRE is 0.273, and the RMRSR is 0.358 indicating that the value of $d$ can be accurately estimated.

\begin{table}[!ht]
\centering
\setlength{\tabcolsep}{3pt}
 \begin{tabular}{ccccc}\hline
 & $(\beta,M)=(100,1000)$ & $(100,2000)$ & $(500,1000)$ & $(500,2000)$ \\\hline
  \multicolumn{5}{c}{(a) homogeneous sampling of reads} \\ \hline
$d$ & 2 & 3 & 6 & 11 \\
$\bar{d}$ & 2.05 & 3.08 & 6.16 & 11.35 \\
$\mathrm{MRE}$ & 0.278 & 0.283 & 0.281 & 0.279 \\
$\mathrm{RMRSR}$ & 0.328 & 0.328 & 0.339 & 0.338 \\\hline
 \multicolumn{5}{c}{(b) inhomogeneous sampling of reads}  \\ \hline
$d$ & 2.80 & 4.60 & 9.56 & 18.15 \\
$\bar{d}$ & 2.85 & 4.92 & 10.03 & 18.86 \\
$\mathrm{MRE}$ & 0.283 & 0.281 & 0.281 & 0.292 \\
$\mathrm{RMRSR}$ & 0.321 & 0.348 & 0.353 & 0.359 \\ \hline
\end{tabular}
\caption{The estimation of the effective coverage $d$ for  a) the
homogeneous and  b) inhomogeneous sampling of reads with $G=10^5$ bps, $\beta=100, 500$ bps, and
$M=1000, 2000$;  $\bar{d}=\sum_{i=1}^{T}\hat{d}_{i}/T$; $T$=2000.}
\label{estimate-d}
\end{table}

\swallow{

\section{Example: estimating the order of MCs for human and mouse genome using NGS data}

We downloaded a real NGS sample of human (ERX226056) and a sample of mouse (ERX288029) from NCBI Sequence Read Archive (SRA). We estimate the effective coverage $d$ and the order of the MC for each of the samples.

 To estimate $d$ using $k$-words by equation (7), the word length $k$ must be at least $r + 2$, where $r$ is the order of the MC for the sequence.
Since $r$ is unknown, we let $k$ to be relatively large and see when the estimated effective coverage becomes stable. In our estimation, we only consider $k$-words with counts at least 10 so that $(Z_{\mathbf{w}}^R)^2/d$ is approximately $\chi^2$-distributed with one degree of freedom. Table 4(a) shows the estimated values of $d$ for human and mouse with various $k$.

The estimation of $d$ in depends on the choice of the word length $k$. It can be seen that the estimated $d$ decreases first, then becomes stable and then increases as $k$ gets greater. Note that, in our simulation, we only consider $k$-words with counts at least 10, so one potential reason for the increase of $d$ could be that when $k$ is too large, there are not enough words for estimating the median of $Z_{\mathbf{w}}^R$. Table 4(b) shows the proportion of $k$-words with count at least 10; it confirms that for mouse, using words of length $k=16$ only takes 2\% of all words into account, and hence it is no surprise that the estimated $d$ appears not to be stable any more.

For human, the estimated values of $d$ is 90.76, 26.68, 13.41, 9.57, 7.87 and 9.01 for $k$ from 11 to 16. The estimated value become stable at $k=$15 or 16. We take the average of the estimated values of $d$ based on these $k$ as the estimated $d$, i.e. the $\hat{d}_{\text{human}}=8.44$. Similarly, the estimated value of $d$ stabilizes at $k=$ 14 or 15 for mouse, so $\hat{d}_{\text{mouse}}=5.31$.

We note that the coverage of the reads calculated using $c = M\beta/G$ is 4.78 and 2.84 for human and mouse, respectively, where $M$ is the number of reads, $\beta$ is the read length and $G$ is the genome length. Thus, the estimated value of $d$ is somewhat larger than $c + 1$ due to the unevenness of the sampling of the reads.

Then based on the estimated values $\hat{d}_{\text{human}}=8.44$ and $\hat{d}_{\text{mouse}}=5.31$,  the fraction of $k$-words $(Z_{\mathbf{w}}^R)^2/\hat{d}$ within $\chi^2(1)$ distribution with type I error 0.01 is shown in Table 4(c).

\begin{table}[!ht]
\scriptsize
\centering
\setlength{\tabcolsep}{2pt}
\begin{subfigure}[The estimated values of effective coverage $d$ for human and mouse using different values of word length $k$.]{
 \begin{tabular}{cccccccccc}  \hline
   & \multicolumn{6}{c}{word length $k$}  \\ \hline
species & 11 & 12 & 13 & 14 & 15 & 16  \\  \hline
human  & 90.76 &  26.68 & 13.41 & 9.57 & 7.87 & 9.01   \\
mouse  & 27.42 & 10.60 & 6.53 & 5.36 & 5.26 & 8.9   \\ \hline \\
\end{tabular}
\label{mouse-human-d-estimator}
}\end{subfigure}\\
\begin{subfigure}[The proportion of $k$-words with count at least 10, for different values of word length $k$.]{
 \begin{tabular}{ccccccccc}  \hline
   & \multicolumn{4}{c}{word length $k$}  \\ \hline
species & 13 & 14 & 15 & 16   \\  \hline
human  & 0.89 & 0.64 & 0.37 & 0.12   \\
mouse  & 0.72 & 0.43 &  0.15 & 0.02   \\ \hline \\
\end{tabular}
\label{mouse-human-d-prop}
}\end{subfigure}
\begin{subfigure}[The fraction of $k$-words $(Z_{\mathbf{w}}^R)^2/\hat{d}$ fall into $\chi^2(1)$ distribution with type I error 0.01 for different values of word length $k$. ]{
 \begin{tabular}{ccccccccc}  \hline
   & \multicolumn{5}{c}{word length $k$}  \\ \hline
species & 9 & 10 & 11 & 12 & 13   \\  \hline
human  & 0.17 &  0.32 & 0.55 & 0.76 & 0.89   \\
mouse  & 0.24 & 0.44 & 0.67 & 0.84 & 0.93   \\  \hline \\
\end{tabular}
\label{mouse-human-fraction}
}\end{subfigure}\\
\caption{The results on inferring the order of MCs for human and mouse genome using NGS data. a) The estimated values of effective coverage $d$ for human and mouse using different values of word length $k$. b) The proportion of $k$-words with count at least 10, for different values of word length $k$. c) The fraction of $k$-words $(Z_{\mathbf{w}}^R)^2/\hat{d}$ fall into $\chi^2(1)$ distribution with type I error 0.01, using the $\hat{d}_{\text{human}}=8.44$ and the $\hat{d}_{\text{mouse}}=5.31$, for different values of word length $k$.}
\label{mouse-human}
\end{table}

 Note that for mouse, both the estimated orders we obtain from the real NGS sample and the MetaSim simulated NGS sample are 10, while for human, we obtain 11 from the real NGS sample and 10 from the MetaSim sample. The small difference may be caused by the heterogenous sampling versus homogeneous sampling, and variation between the genome of ITU Indian Telugu in the UK(ERX226056) and the standard human genome in UCSC.

}

\subsection{The relationship among 28 vertebrate species}

Table \ref{28order} shows the estimated orders of MC for the 28 genomes of vertebrate species based on their NGS samples. For each of the 28 species, we compute the fraction of the k-tuple words that have $(Z_{\mathbf{w}}^R)^2/\hat{d}$ within the 99\% of a $\chi^2$ distribution with one degree of freedom, for $k=8, 9, \dots, 14$. Using 80\% as a threshold, we estimate the orders of MC for each species to be the smallest $k-2$ under which the fraction of words that can be explained by the $(k-2)$-th order MC is greater than the threshold. The last two columns of Table \ref{28order} are the AIC-predicted optimal and BIC-predicted optimal orders obtained in \cite{narlikar2013one}.

We cluster the NGS datasets of the 28 species listed in the Table \ref{28order}, using  $d_2^*$ and $d_2^S$ under MC models of varying order.
Figure S1 in  \cite{miller200728} gives the phylogenetic tree of the 28 species based on alignment methods, with branch lengths noted on it. We obtain the distance matrix of the 28 species by computing pairwise disstance between each pair of the 28 species from the phylogenetic tree; the matrix is given as Table \ref{28distance}. {{We compare}} the dissimilarity matrices using $d_2^*$ and $d_2^S$  under MC models with different  order {{to the matrix  Table \ref{28distance} which we use as underlying truth}}.

\subsection{The relationship among 13 tropical tree species with unknown reference genomes}

We also apply our method to the 13 tree species (8 \textit{Fagaceae} and 5 \textit{Moraceae}) based on the NGS shotgun read data sets in \cite{cannon2010assembly}. The reference genome sequences for the 13 tree species are unknown.
 Figure 1 in the main text shows the clustering results using $d_2^*$ under MCs of order 0, 4, 8 and 9. For the clustering results using $d_2^S$, see Figure \ref{13tree_d2shepp}.

\swallow{
Figure \ref{13tree_d2star} shows the clustering results using $d_2^*$ under MCs of order 0, 4, 8 and 9. The trees are built based on all the reads. The number on the branch refers to the frequency of the branch occurring among the 30 clusterings based on random sampled 10\% reads.
From the results we can see,
under the i.i.d model, Lithocarpus mixes up with Castanopsis; Trigonobalanus can not be separated from the rest of Fagaceae. While under the MC of order greater than 4,  Trigonobalanus successfully gets separated from the rest of the Fagaceaes.
Besides, within the Moraceae group, Ficus fistulas and Ficus langkokensis forms a subgroup under the i.i.d model, and they are separated under the MC with order greater than 4.  At the meanwhile, Ficus langkokensis is the closest Maraceae to the Fagaceae under 4th order MC, Ficus fistulosa becomes the closest species to the Fagaceaes under 8th and 9th order MCs. It also can be seen that three branches of the tree under MC of order 9 are frequencies of occurrence less than 30. When using the MC of a very high order, the tree becomes unstable.

\begin{figure}[ht]
\centering
\begin{subfigure}[k11, order0, $d_2^*$]{\includegraphics[height=6cm,width=15cm,angle=0]{13treeimage/k11_order0_longseq_d2star_consensus_tree1.pdf}
\label{fig:subfig1}
 }%
\end{subfigure}
 \begin{subfigure}[k11, order4, $d_2^*$]{\includegraphics[height=6cm,width=15cm,angle=0]{13treeimage/k11_order4_longseq_d2star_consensus_tree1.pdf}
   \label{fig:subfig2}
 }%
\end{subfigure}
 \begin{subfigure}[k11, order8, $d_2^*$]{\includegraphics[height=6cm,width=15cm,angle=0]{13treeimage/k11_order8_longseq_d2star_consensus_tree1.pdf}
   \label{fig:subfig3}
 }%
\end{subfigure}
 \begin{subfigure}[k11, order9, $d_2^*$]{\includegraphics[height=6cm,width=15cm,angle=0]{13treeimage/k11_order9_longseq_d2star_consensus_tree1.pdf}
   \label{fig:subfig3}
 }%
\end{subfigure}
\caption{The clustering of the 13 tropical tree species using $d_2^*$ under MC with order of 0(i.i.d), 4, 8 and 9.}
\label{13tree_d2star}
\end{figure}

}

\bibliography{document_sun}
\bibliographystyle{natbib}

\begin{landscape}
\begin{table}[h]
\setlength{\tabcolsep}{4pt}
\centering
\begin{tabular}{cc|ccccccc|c|cc} \hline
    &       &   \multicolumn{6}{c}{word length $k$}   &               &               \\
genome & common name &  8                  & 9           & 10      & 11   & 12   & 13   & 14   &  estimated order & APO & BPO \\ \hline
hg38                 & human       & 0.12    & 0.25 & 0.45 & 0.68 & 0.85 & 0.9  & 0.78 & 10                                & 12            & 10            \\
panTro4              & chimp       & 0.14    & 0.28 & 0.49 & 0.71 & 0.87 & 0.91 & 0.78 & 10                                & 12            & 10            \\
rheMac3              & macaque     & 0.16    & 0.32 & 0.55 & 0.76 & 0.9  & 0.92 & 0.78 & 10                                & 12            & 10            \\
otoGar3              & bushbaby    & 0.18    & 0.37 & 0.58 & 0.78 & 0.9  & 0.92 & 0.77 & 10                                & na            & na            \\
tupBel1              & tree shrew  & 0.17    & 0.34 & 0.57 & 0.78 & 0.9  & 0.9  & 0.69 & 10                                & na            & na            \\
rn5                  & rat         & 0.17    & 0.33 & 0.55 & 0.75 & 0.88 & 0.92 & 0.79 & 10                                & 12            & 10            \\
mm10                 & mouse       & 0.16    & 0.29 & 0.49 & 0.69 & 0.84 & 0.89 & 0.76 & 10                                & 12            & 10            \\
cavPor3              & guinea pig  & 0.17    & 0.32 & 0.52 & 0.72 & 0.86 & 0.9  & 0.77 & 10                                & 13            & 10            \\
oryCun2              & rabbit      & 0.12    & 0.25 & 0.45 & 0.7  & 0.87 & 0.91 & 0.78 & 10                                & na            & na            \\
sorAra2              & shrew       & 0.13    & 0.27 & 0.47 & 0.71 & 0.87 & 0.91 & 0.76 & 10                                & na            & na            \\
eriEur2              & hedgehog    & 0.12    & 0.24 & 0.44 & 0.66 & 0.83 & 0.86 & 0.70 & 10                                & na            & na            \\
canFam3              & dog         & 0.15    & 0.32 & 0.56 & 0.79 & 0.91 & 0.93 & 0.77 & 10                                & 12            & 10            \\
felCat5              & cat         & 0.16    & 0.33 & 0.57 & 0.79 & 0.92 & 0.94 & 0.81 & 10                                & 12            & 10            \\
equCab2              & horse       & 0.17    & 0.34 & 0.56 & 0.78 & 0.9  & 0.93 & 0.81 & 10                                & 12            & 10            \\
bosTau7              & cow         & 0.12    & 0.24 & 0.43 & 0.67 & 0.84 & 0.89 & 0.76 & 10                                & 12            & 10            \\
dasNov3              & armadillo   & 0.12    & 0.24 & 0.41 & 0.63 & 0.81 & 0.88 & 0.78 & 10                                & na            & na            \\
loxAfr3              & elephant    & 0.12    & 0.21 & 0.37 & 0.59 & 0.78 & 0.88 & 0.79 & 11                                & na            & na            \\
echTel2              & tenrec      & 0.16    & 0.3  & 0.49 & 0.7  & 0.86 & 0.92 & 0.83 & 10                                & na            & na            \\
monDom5              & opossum     & 0.14    & 0.26 & 0.44 & 0.64 & 0.81 & 0.86 & 0.73 & 10                                & 13            & 11            \\
ornAna1              & platypus    & 0.13    & 0.29 & 0.51 & 0.75 & 0.89 & 0.91 & 0.69 & 10                                & 12            & 10            \\
galGal4              & chicken     & 0.4     & 0.67 & 0.83 & 0.94 & 0.97 & 0.89 & 0.64 & 8                                 & 11            & 8             \\
anoCar2              & lizard      & 0.15    & 0.25 & 0.42 & 0.63 & 0.81 & 0.86 & 0.69 & 10                                & 12            & 10            \\
xenTro3              & frog        & 0.17    & 0.3  & 0.49 & 0.71 & 0.86 & 0.88 & 0.67 & 10                                & 12            & 10            \\
tetNig2              & tetraodon   & 0.54    & 0.78 & 0.91 & 0.97 & 0.98 & 0.82 & 0.40 & 8                                 & na            & na            \\
fr3                  & fugu        & 0.57    & 0.8  & 0.92 & 0.97 & 0.98 & 0.83 & 0.43 & 7                                 & 11            & 8             \\
gasAcu1              & stickleback & 0.48    & 0.7  & 0.85 & 0.94 & 0.97 & 0.9  & 0.54 & 8                                 & 11            & 8             \\
oryLat2              & medaka      & 0.34    & 0.5  & 0.69 & 0.85 & 0.94 & 0.88 & 0.58 & 9                                 & na            & na            \\
danRer7              & zebrafish   & 0.17    & 0.27 & 0.42 & 0.62 & 0.81 & 0.88 & 0.68 & 10                                & 13            & 10     \\ \hline  \\
\end{tabular}
\caption{Estimating the order of the MC for 28 species based on NGS read samples. The 'genome' column is the scientific names shown in UCSC; 'estimated order' is the minimum order of MC $r$ that can explain greater than 80\% of the $(r+2)$-tuple word at type I error 0.01; 'APO' and 'BPO' columns show the AIC-predicted optimal and BIC-predicted optimal orders obtained in \cite{narlikar2013one}; the middle columns with order $k=8, 9, \dots, 14$ are the fraction of words $\mathbf{w}$ with length $k$ that have $(Z_{\mathbf{w}}^R)^2/\hat{d}$ within the 99\% of a $\chi^2$ distribution with one degree of freedom. 'na' means the numbers are not provided in \cite{narlikar2013one}.}
\label{28order}
\end{table}
\end{landscape}

\begin{landscape}
\begin{table}[p]
\footnotesize
\resizebox{1.3\textwidth}{!}{%
\setlength{\tabcolsep}{4pt}
\begin{tabular}{l|llllllllllllllllllllllllllll}
& human  & chimp & macaque & bushbaby & tree shrew & rat  & mouse & guinea pig & rabbit & shrew & hedgehog & dog  & cat  & horse & cow  & armadillo & elephant & tenrec & opossum & platypus & chicken & lizard & frog & tetraodon & fugu & stickleback & medaka & zebrafish     \\ \hline
human       & 0.00  & 0.02    & 0.07     & 0.26       & 0.29 & 0.46  & 0.46       & 0.38   & 0.36  & 0.48     & 0.45 & 0.35 & 0.35  & 0.30 & 0.35      & 0.33     & 0.35   & 0.48    & 0.72     & 0.98    & 1.10   & 1.21 & 1.54      & 1.85 & 1.89        & 1.82   & 2.01      & 1.83 \\
chimp       &       & 0.00    & 0.07     & 0.26       & 0.29 & 0.46  & 0.46       & 0.38   & 0.36  & 0.48     & 0.45 & 0.35 & 0.35  & 0.30 & 0.35      & 0.33     & 0.35   & 0.48    & 0.72     & 0.98    & 1.10   & 1.21 & 1.54      & 1.85 & 1.89        & 1.82   & 2.01      & 1.83 \\
macaque     &       &         & 0.00     & 0.25       & 0.28 & 0.45  & 0.45       & 0.37   & 0.35  & 0.47     & 0.44 & 0.34 & 0.34  & 0.29 & 0.34      & 0.32     & 0.34   & 0.47    & 0.71     & 0.97    & 1.09   & 1.20 & 1.53      & 1.84 & 1.88        & 1.81   & 2.00      & 1.82 \\
bushbaby    &       &         &          & 0.00       & 0.33 & 0.50  & 0.50       & 0.42   & 0.40  & 0.52     & 0.49 & 0.39 & 0.39  & 0.34 & 0.39      & 0.37     & 0.39   & 0.52    & 0.76     & 1.02    & 1.14   & 1.25 & 1.58      & 1.89 & 1.93        & 1.86   & 2.05      & 1.87 \\
tree shrew  &       &         &          &            & 0.00 & 0.49  & 0.49       & 0.41   & 0.39  & 0.51     & 0.48 & 0.38 & 0.38  & 0.33 & 0.38      & 0.36     & 0.38   & 0.51    & 0.75     & 1.01    & 1.13   & 1.24 & 1.57      & 1.88 & 1.92        & 1.85   & 2.04      & 1.86 \\
rat         &       &         &          &            &      & 0.00  & 0.16       & 0.48   & 0.52  & 0.66     & 0.63 & 0.53 & 0.53  & 0.48 & 0.53      & 0.51     & 0.53   & 0.66    & 0.90     & 1.16    & 1.28   & 1.39 & 1.72      & 2.03 & 2.07        & 2.00   & 2.19      & 2.01 \\
mouse       &       &         &          &            &      &       & 0.00       & 0.48   & 0.52  & 0.66     & 0.63 & 0.53 & 0.53  & 0.48 & 0.53      & 0.51     & 0.53   & 0.66    & 0.90     & 1.16    & 1.28   & 1.39 & 1.72      & 2.03 & 2.07        & 2.00   & 2.19      & 2.01 \\
guinea pig  &       &         &          &            &      &       &            & 0.00   & 0.44  & 0.58     & 0.55 & 0.45 & 0.45  & 0.40 & 0.45      & 0.43     & 0.45   & 0.58    & 0.82     & 1.08    & 1.20   & 1.31 & 1.64      & 1.95 & 1.99        & 1.92   & 2.11      & 1.93 \\
rabbit      &       &         &          &            &      &       &            &        & 0.00  & 0.56     & 0.53 & 0.43 & 0.43  & 0.38 & 0.43      & 0.41     & 0.43   & 0.56    & 0.80     & 1.06    & 1.18   & 1.29 & 1.62      & 1.93 & 1.97        & 1.90   & 2.09      & 1.91 \\
shrew       &       &         &          &            &      &       &            &        &       & 0.00     & 0.47 & 0.47 & 0.47  & 0.42 & 0.47      & 0.49     & 0.51   & 0.64    & 0.88     & 1.14    & 1.26   & 1.37 & 1.70      & 2.01 & 2.05        & 1.98   & 2.17      & 1.99 \\
hedgehog    &       &         &          &            &      &       &            &        &       &          & 0.00 & 0.44 & 0.44  & 0.39 & 0.44      & 0.46     & 0.48   & 0.61    & 0.85     & 1.11    & 1.23   & 1.34 & 1.67      & 1.98 & 2.02        & 1.95   & 2.14      & 1.96 \\
dog         &       &         &          &            &      &       &            &        &       &          &      & 0.00 & 0.20  & 0.25 & 0.32      & 0.36     & 0.38   & 0.51    & 0.75     & 1.01    & 1.13   & 1.24 & 1.57      & 1.88 & 1.92        & 1.85   & 2.04      & 1.86 \\
cat         &       &         &          &            &      &       &            &        &       &          &      &      & 0.00  & 0.25 & 0.32      & 0.36     & 0.38   & 0.51    & 0.75     & 1.01    & 1.13   & 1.24 & 1.57      & 1.88 & 1.92        & 1.85   & 2.04      & 1.86 \\
horse       &       &         &          &            &      &       &            &        &       &          &      &      &       & 0.00 & 0.27      & 0.31     & 0.33   & 0.46    & 0.70     & 0.96    & 1.08   & 1.19 & 1.52      & 1.83 & 1.87        & 1.80   & 1.99      & 1.81 \\
cow         &       &         &          &            &      &       &            &        &       &          &      &      &       &      & 0.00      & 0.36     & 0.38   & 0.51    & 0.75     & 1.01    & 1.13   & 1.24 & 1.57      & 1.88 & 1.92        & 1.85   & 2.04      & 1.86 \\
armadillo   &       &         &          &            &      &       &            &        &       &          &      &      &       &      &           & 0.00     & 0.28   & 0.41    & 0.67     & 0.93    & 1.05   & 1.16 & 1.49      & 1.80 & 1.84        & 1.77   & 1.96      & 1.78 \\
elephant    &       &         &          &            &      &       &            &        &       &          &      &      &       &      &           &          & 0.00   & 0.33    & 0.69     & 0.95    & 1.07   & 1.18 & 1.51      & 1.82 & 1.86        & 1.79   & 1.98      & 1.80 \\
tenrec      &       &         &          &            &      &       &            &        &       &          &      &      &       &      &           &          &        & 0.00    & 0.82     & 1.08    & 1.20   & 1.31 & 1.64      & 1.95 & 1.99        & 1.92   & 2.11      & 1.93 \\
opossum     &       &         &          &            &      &       &            &        &       &          &      &      &       &      &           &          &        &         & 0.00     & 0.90    & 1.02   & 1.13 & 1.46      & 1.77 & 1.81        & 1.74   & 1.93      & 1.75 \\
platypus    &       &         &          &            &      &       &            &        &       &          &      &      &       &      &           &          &        &         &          & 0.00    & 1.10   & 1.21 & 1.54      & 1.85 & 1.89        & 1.82   & 2.01      & 1.83 \\
chicken     &       &         &          &            &      &       &            &        &       &          &      &      &       &      &           &          &        &         &          &         & 0.00   & 0.91 & 1.42      & 1.73 & 1.77        & 1.70   & 1.89      & 1.71 \\
lizard      &       &         &          &            &      &       &            &        &       &          &      &      &       &      &           &          &        &         &          &         &        & 0.00 & 1.53      & 1.84 & 1.88        & 1.81   & 2.00      & 1.82 \\
frog        &       &         &          &            &      &       &            &        &       &          &      &      &       &      &           &          &        &         &          &         &        &      & 0.00      & 1.87 & 1.91        & 1.84   & 2.03      & 1.85 \\
tetraodon   &       &         &          &            &      &       &            &        &       &          &      &      &       &      &           &          &        &         &          &         &        &      &           & 0.00 & 0.44        & 0.77   & 0.96      & 1.48 \\
fugu        &       &         &          &            &      &       &            &        &       &          &      &      &       &      &           &          &        &         &          &         &        &      &           &      & 0.00        & 0.81   & 1.00      & 1.52 \\
stickleback &       &         &          &            &      &       &            &        &       &          &      &      &       &      &           &          &        &         &          &         &        &      &           &      &             & 0.00   & 0.81      & 1.45 \\
medaka      &       &         &          &            &      &       &            &        &       &          &      &      &       &      &           &          &        &         &          &         &        &      &           &      &             &        & 0.00      & 1.64 \\
zebrafish   &       &         &          &            &      &       &            &        &       &          &      &      &       &      &           &          &        &         &          &         &        &      &           &      &             &        &           & 0.00 \\
\end{tabular}
} \\
\caption{The pairwise distance matrix obtain from the Figure S1 in \cite{miller200728}.  Figure S1 in \cite{miller200728} shows the phylogenetic tree of the 28 species based on alignment methods with branch lengths noted on it.}
\label{28distance}
\end{table}
\end{landscape}


\begin{figure}[ht]
\centering
\begin{subfigure}[k11, order0, $d_2^S$]{\includegraphics[height=4.5cm,width=9cm,angle=0]{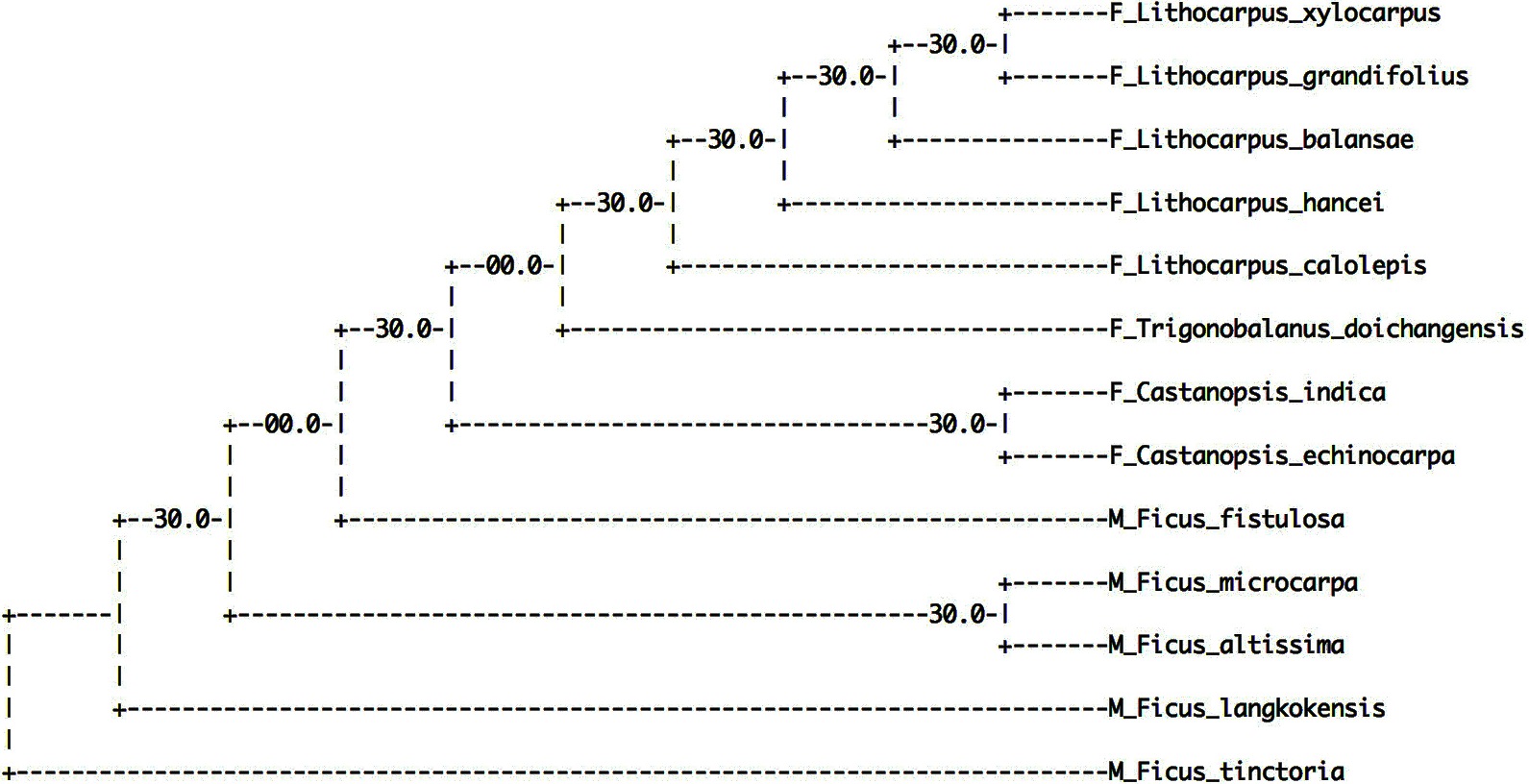}
\label{fig:subfig21}
 }%
\end{subfigure}
 \begin{subfigure}[k11, order4, $d_2^S$]{\includegraphics[height=4.5cm,width=9cm,angle=0]{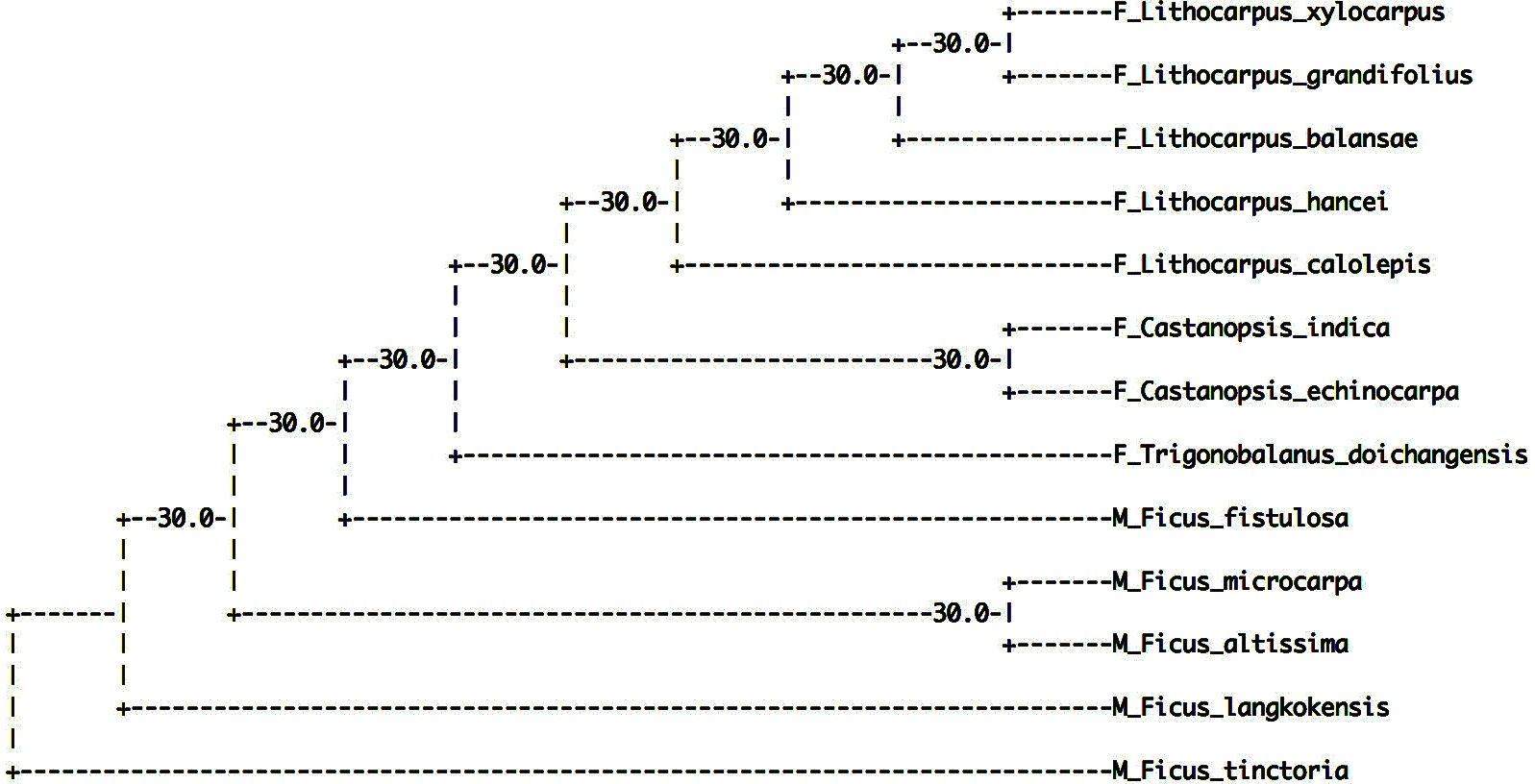}
   \label{fig:subfig22}
 }%
\end{subfigure}
 \begin{subfigure}[k11, order8, $d_2^S$]{\includegraphics[height=4.5cm,width=9cm,angle=0]{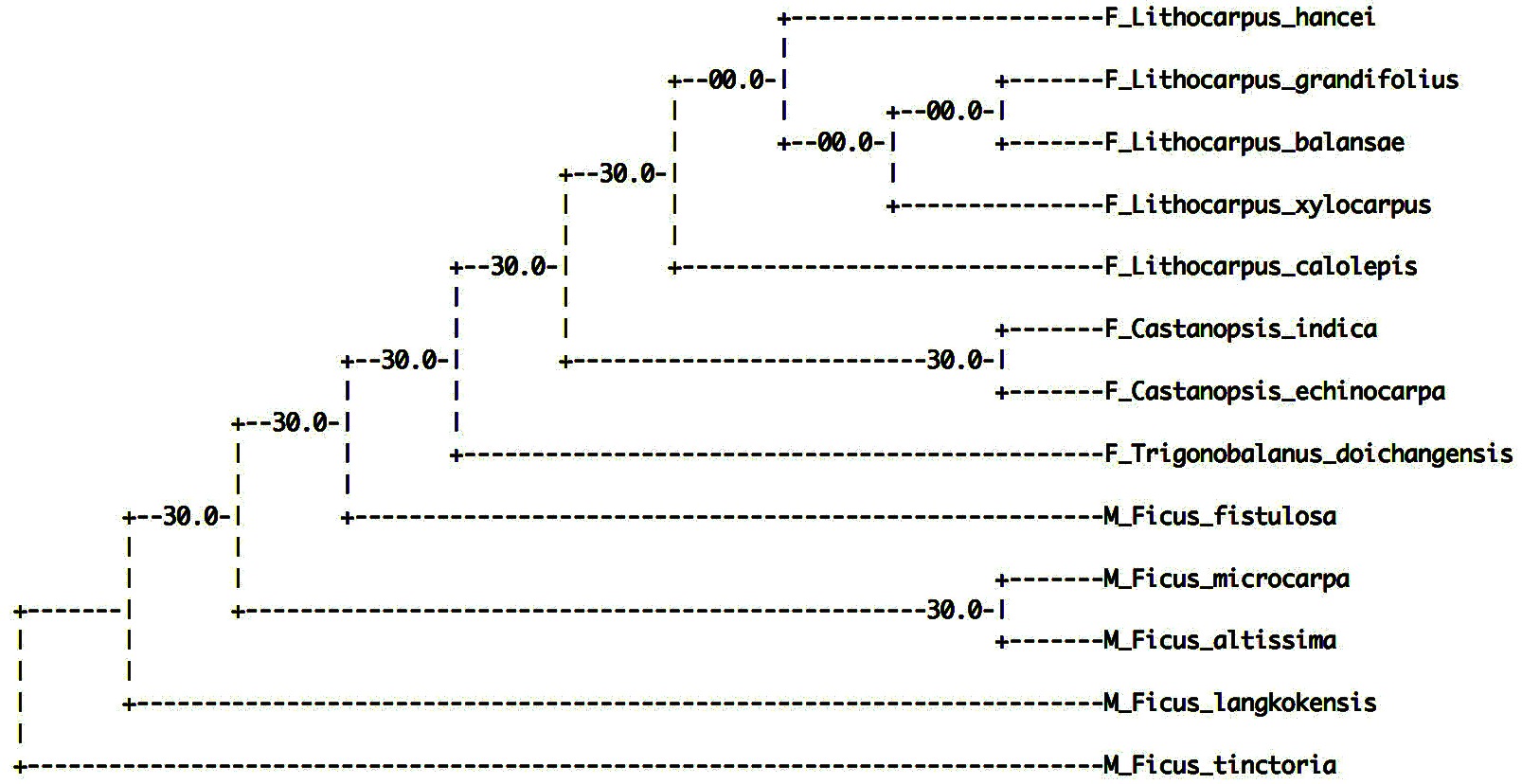}
   \label{fig:subfig23}
 }%
\end{subfigure}
 \begin{subfigure}[k11, order9, $d_2^S$]{\includegraphics[height=4.5cm,width=9cm,angle=0]{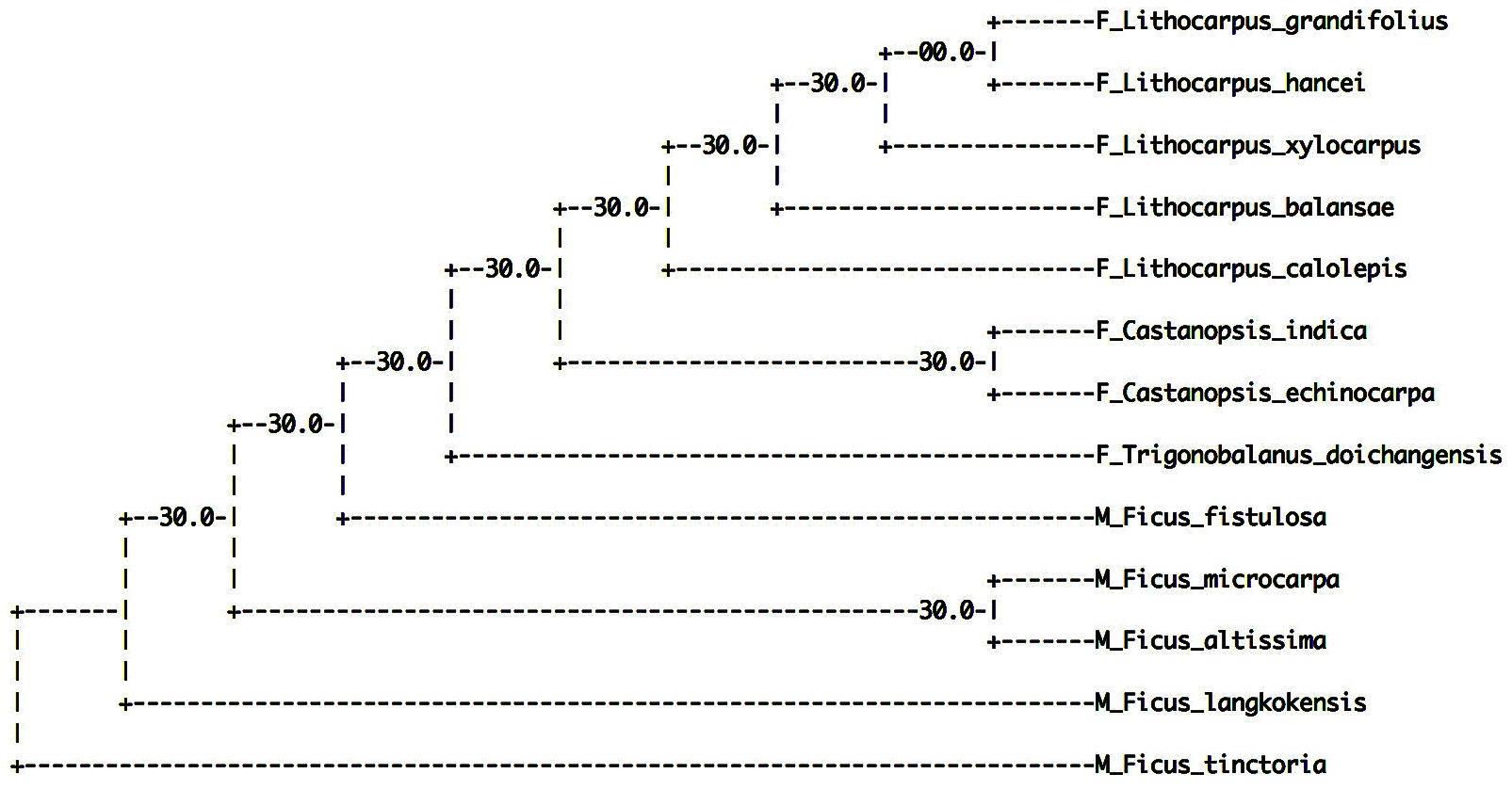}
   \label{fig:subfig24}
 }%
\end{subfigure}
\caption{The clustering of the 13 tropical tree species using $d_2^S$ under MC with order of 0(i.i.d), 4, 8 and 9. The number on the branch refers to the frequency of the branch occurring among the 30 clusterings based on random sampled 10\% reads. {The letter `F' at the beginning of the names represents \textit{Fagaceae}; similarly the letter `M' represents \textit{Maraceae}. }}
\label{13tree_d2shepp}

\end{figure}

\end{document}